\documentclass[aps,prd,preprintnumbers,reprint,superscriptaddress]{revtex4-1}
\usepackage[utf8]{inputenc}
\usepackage{graphicx}
\usepackage{amssymb}
\usepackage{amsmath}
\begin{document}
\title{Pion and kaon parton distribution functions from basis light front quantization and QCD evolution}
\author{Jiangshan Lan}
\email{jiangshanlan@impcas.ac.cn}
\affiliation{Institute of Modern Physics, Chinese Academy of Sciences, Lanzhou 730000, China}
\affiliation{School of Nuclear Science and Technology, University of Chinese Academy of Sciences, Beijing 100049, China}
\affiliation{Lanzhou University, Lanzhou 730000, China}
\author{Chandan Mondal}
\email{mondal@impcas.ac.cn} \affiliation{Institute of Modern Physics, Chinese Academy of Sciences, Lanzhou 730000, China}
\affiliation{School of Nuclear Science and Technology, University of Chinese Academy of Sciences, Beijing 100049, China}
\author{Shaoyang Jia}
\email{sjia@iastate.edu} \affiliation{Department of Physics and Astronomy, Iowa State University, Ames, IA 50011, USA}
\author{Xingbo Zhao}
\email{xbzhao@impcas.ac.cn} \affiliation{Institute of Modern Physics, Chinese Academy of Sciences, Lanzhou 730000, China}
\affiliation{School of Nuclear Science and Technology, University of Chinese Academy of Sciences, Beijing 100049, China}
\author{James P. Vary}
\email{jvary@iastate.edu} \affiliation{Department of Physics and Astronomy, Iowa State University, Ames, IA 50011, USA}
\collaboration{BLFQ Collaboration}
\begin{abstract}
We investigate the parton distribution functions (PDFs) of the pion and the kaon by combining quantum chromodynamics (QCD) evolution with the basis light front quantization. The initial PDFs result from the light front wave functions obtained by diagonalizing the effective Hamiltonian consisting of the holographic QCD confinement potential, a complementary longitudinal confinement potential, and  the color-singlet Nambu--Jona-Lasinio interactions. The valence-quark PDF of the pion, after QCD evolution, is consistent with the result from the E-0615 experiment at Fermilab. Meanwhile, the pion structure function calculated from the PDFs agrees with the ZEUS and the H1 experiments at DESY-HERA for large $x$. Additionally, the ratio of the up quark PDF of the kaon to that of the pion is in agreement with the NA-003 experiment at CERN. We also present the cross section for the pion-nucleus induced Drell-Yan process with the obtained pion PDFs supplemented by the PDFs of the target nuclei. 
\end{abstract}
\maketitle
\section{Introduction}\label{SecI}
Parton distribution functions (PDFs) encode the nonperturbative structure of a hadron by specifying how its longitudinal momentum is distributed to quarks and gluons. The determination of PDFs from the analysis of hard scattering processes is one of the main topics of hadron physics 
 \cite{Bordalo:1987cs,Freudenreich:1990mu,Sutton:1991ay,Gluck:1999xe,Wijesooriya:2005ir, Conway:1989fs,Badier:1983mj,Aicher:2010cb,Watanabe:2017pvl,Hecht:2000xa,Nam:2012vm,Detmold:2003tm,Holt:2010vj,Pumplin:2002vw,Ball:2017nwa,Alekhin:2017kpj,Dulat:2015mca,Harland-Lang:2014zoa,Bednar:2018mtf}. The structure of hadrons including their PDFs is expected to be described by quantum chromodynamics (QCD) in the low energy region where quarks are confined. 
In addition to color confinement, the explicit and the dynamical breaking of the global chiral dynamics leads to pions having a small mass when compared to other hadrons, taking the role of the pseudoscalar Goldstone bosons. In a chiral perturbation theory, the dynamics of which preserves local chiral symmetry, the pions dress the constituent quarks of an isolated nucleon~\cite{Thomas:2007bc,Theberge:1980ye,Thomas:1982kv}. Meanwhile, the pseudoscalar kaons are the counterparts of the pions with one strange valence quark, the structure of which is crucial to our understanding of CP symmetry violation~\cite{Woods:1988za,Barr:1993rx,Gibbons:1993zq}. In this article, we are interested in explaining the partonic structure of the pions and the kaons in terms of their PDFs.

One of the available experiments with access to the pion PDFs is the Drell-Yan dilepton production in $\pi^-$-tungsten reactions~\cite{Bordalo:1987cs,Freudenreich:1990mu,Sutton:1991ay}. Several next-to-leading order (NLO) analyses of this Drell-Yan process have been performed by Refs.~\cite{Sutton:1991ay,Gluck:1999xe,Wijesooriya:2005ir}. The subsequent determination of the nucleon and the light meson PDFs with associated uncertainties from the experiment is available in Refs.~\cite{Aicher:2010cb,Watanabe:2017pvl,Hecht:2000xa,Nam:2012vm,Detmold:2003tm,Holt:2010vj,Pumplin:2002vw,Ball:2017nwa,Alekhin:2017kpj,Dulat:2015mca,Harland-Lang:2014zoa,Barry:2018ort}. The pion PDF has also been the subject of detailed analyses in the phenomenological models in Refs.~\cite{Frederico:1994dx,Shigetani:1993dx,Shigetani:1993dx}, also including the chiral quark model~\cite{Broniowski:2007si} and anti-de Sitter (AdS)/QCD models~\cite{Gutsche:2014zua,Gutsche:2013zia,Ahmady:2018muv,deTeramond:2018ecg}. The pion PDFs have also been investigated within lattice QCD~\cite{Brommel:2006zz,Martinelli:1987bh,Detmold:2003tm,Abdel-Rehim:2015owa,Oehm:2018jvm,Sufian:2019bol}. See Ref. \cite{Lin:2017snn} for the corresponding review of lattice QCD results. Additionally, the first global fit analysis of PDFs in the pion has been performed in Ref.~\cite{Barry:2018ort}.

Although meson PDFs are expected to be universal, tension exists regarding the behavior of the pion valence PDF. On the one hand, from the analyses of the Drell-Yan data \cite{Sutton:1991ay,Wijesooriya:2005ir}, the large-$x$ behavior of the pion valence PDF is expected to fall off linearly or slightly faster, which is supported by the constituent quark models~\cite{Frederico:1994dx,Shigetani:1993dx}, the Nambu--Jona-Lasinio (NJL) model~\cite{Shigetani:1993dx}, and duality arguments~\cite{Melnitchouk:2002gh}. This observation disagrees with perturbative QCD where the behavior of the same function has been predicted to be ${(1-x)^2}$~\cite{Farrar:1979aw,Berger:1979du,Brodsky:2006hj,Yuan:2003fs}, a behavior further supported by the Bethe-Salpeter equation (BSE) approach~\cite{Hecht:2000xa,Ding:2019lwe}.
However, the reanalysis of the data for the Drell-Yan process~\cite{Aicher:2010cb} including the next-to-leading logarithmic threshold resummation effects shows a considerably softer valence PDF at high $x$  when compared to the NLO analysis \cite{Sutton:1991ay,Wijesooriya:2005ir}. 

Information from experiments on the light-quark PDF of the kaons exists in the form of the ratio of the up (u) quark valence PDF in the kaon to that in the pion \cite{Badier:1983mj, Conway:1989fs}. Theoretically, the kaon's valence PDF from the BSE approach has been investigated in Ref.~\cite{Nguyen:2011jy}. A more recent study of the pion and kaon valence PDFs from the BSE with a beyond-rainbow-ladder truncation of QCD shows a good agreement with the experimental valence PDF of the pion~\cite{Shi:2018mcb}. The kaon's PDF has also been studied in several quark models such as the gauge-invariant nonlocal chiral-quark model \cite{Nam:2012vm}, the dressed quark model \cite{Chen:2016sno}, and the NJL model \cite{Hutauruk:2016sug,Davidson:2001cc}. Meanwhile, the quasi-PDFs for the pion and the kaon have been given in Refs.~\cite{Xu:2018eii,Broniowski:2017wbr,Radyushkin:2017gjd}.

Our theoretical calculation of meson structures is based on the basis light front quantization (BLFQ) approach, which has been developed for solving many-body bound state problems in quantum field theories~\cite{Vary:2009gt,Wiecki:2014ola,Li:2015zda}. BLFQ is a Hamiltonian-based formalism incorporating the light front dynamics \cite{Brodsky:1997de}. This formalism has been successfully applied to the quantum electrodynamics (QED) systems including the electron self-energy~\cite{Zhao:2014xaa} and positronium with strong coupling \cite{Wiecki:2014ola}. It has also been applied to heavy quarkonia~\cite{Li:2017mlw} and $B_c$ mesons~\cite{Tang:2018myz} both as QCD bound states. Recently, the BLFQ approach using a Hamiltonian that includes the color-singlet NJL interaction to account for the chiral dynamics has been applied to the light mesons \cite{Jia:2018ary}. Furthermore, the BLFQ formalism has been extended to time-dependent strong external field problems such as those in non-linear Compton scattering~\cite{Zhao:2013jia}. (For the reviews of BLFQ and its application, see Refs.~\cite{Leitao:2017esb,Adhikari:2016idg,Li:2017mlw,Zhao:2014xaa,Zhao:2013cma,Chen:2016dlk,Wiecki:2014ola,Li:2018uif,Li:2015zda,Li:2017uug,Vary:2009gt,Zhao:2013jia,Chakrabarti:2014cwa,Adhikari:2018umb}.) With light front kinematics, the PDFs can also be calculated using the microcanonical ensemble~\cite{Jia:2018hxd}.

In this work, we elaborate on Ref.~\cite{Lan:2019vui} in the determination of the valence quark PDFs of the pion and the kaon at independent initial scales using the light front wave functions (LFWFs). These wave functions were obtained within the framework of BLFQ by diagonalizing the effective light front Hamiltonian whose interactions include the light front holographic QCD (LFHQCD) confinement potential in the transverse direction~\cite{Brodsky:2014yha}, a longitudinal confinement potential~\cite{Li:2015zda}, and the NJL interactions~\cite{Klimt:1989pm}. These LFWFs have been successfully applied to compute the parton distribution amplitudes and the electromagnetic form factors for the pion and the kaon \cite{Jia:2018ary}.  We then evolve our initial valence quark PDFs of the pion and the kaon utilizing the next-to-next-to-leading order (NNLO) Dokshitzer-Gribov-Lipatov-Altarelli-Parisi (DGLAP) equations~\cite{Dokshitzer:1977sg,Gribov:1972ri,Altarelli:1977zs}  to the relevant scales in order to compare with the result of PDFs from the E-0615 experiment at Fermilab, with the pion structure function from the ZEUS and the H1 experiments at DESY-HERA, and with the ratio $u^K_{\rm v}(x,\mu^2)/u^\pi_{\rm v}(x,\mu^2)$ from the NA-003 experiment at CERN.  Using our pion PDFs in conjunction with the nuclear PDFs from the nuclear Coordinated Theoretical-Experimental Project on QCD (nCTEQ) 2015 global fit~\cite{Kovarik:2015cma}, instead of the collection of free nucleon PDFs used in Ref. \cite{Lan:2019vui}, we further calculate the cross section for the pion-nucleus induced Drell-Yan process, to show that our PDFs consistently describe the measured cross section data from a variety of experiments. 

The paper is organized as follows. Section~\ref{SecI} is the introduction. The valence PDFs for the pion and the kaon from the BLFQ-NJL model are given in Sec.~\ref{sc:BLFQ_NJL}. Section~\ref{sc:PDFs} discusses results of these PDFs following DGLAP evolution. Specifically, in Sec.~\ref{result}, we present the pion and the kaon PDFs at various scales as well as the implied structure function for the pion. Based on these pion PDFs, the cross section for the unpolarized Drell-Yan process is calculated in Sec.~\ref{SecIII}. Section~\ref{SecVI} is the summary.
\section{BLFQ-NJL model for the light mesons}\label{sc:BLFQ_NJL}
\subsection{The light front confinement and NJL interactions for the light mesons}\label{model}
Let us start with an overview of the BLFQ-NJL model for the light mesons following Ref.~\cite{Jia:2018ary}. In the approach of BLFQ, the structures of the bound states are embedded in the LFWFs obtainable as the solutions of the time-independent light front Schr\"{o}dinger equation 
\begin{equation}
H_{\mathrm{eff}}\vert \Psi\rangle=M^2\vert \Psi\rangle,\label{eq:LF_Schrodinger}
\end{equation}
where $H_{\mathrm{eff}}$ is the effective Hamiltonian of the system with the mass squared $M^2$ being the eigenvalue of $\vert \Psi\rangle$. In general, $\vert \Psi\rangle$ is the vector in the Hilbert space spanning into all Fock sectors. In the valence Fock sector, the effective Hamiltonian for the light mesons with non-singlet flavor wave functions is given by~\cite{Jia:2018ary}
\begin{align}
H_\mathrm{eff} &= \frac{\vec k^2_\perp + m_q^2}{x} + \frac{\vec k^2_\perp+m_{\bar q}^2}{1-x}
+ \kappa^4 \vec \zeta_\perp^2 \nonumber\\
& \quad - \frac{\kappa^4}{(m_q+m_{\bar q})^2} \partial_x\big( x(1-x) \partial_x \big)+H^{\rm eff}_{\rm NJL},\label{eqn:Heff}
\end{align}
where $m_q$ ($m_{\bar q}$) is the mass of the quark (antiquark), and $\kappa$ is the strength of the confinement. $\vec \zeta_\perp \equiv \sqrt{x(1-x)} \vec r_\perp$ is the holographic variable~\cite{Brodsky:2014yha}, with $\vec{k}_\perp$ being the conjugate variable of $\vec{r}_\perp$. The $x$-derivative is defined as $\partial_x f(x, \vec\zeta_\perp) = \partial f(x, \vec \zeta_\perp)/\partial x|_{\vec\zeta}$. 
The first two terms in Eq.~\eqref{eqn:Heff} are the light front kinetic energy for the quark and the antiquark. The third and the fourth terms are the confining potential in the transverse direction based on the LFHQCD~\cite{Brodsky:2014yha} and a longitudinal confining potential~\cite{Li:2015zda} that reproduces 3D confinement in the nonrelativistic limit. Additionally, the $H_{\mathrm{NJL}}^{\mathrm{eff}}$ is the color-singlet NJL interaction to account for the chiral dynamics~\cite{Klimt:1989pm}.

The NJL interaction for the positively-charged pion is given by~\cite{Jia:2018ary},
\begin{align}
H_{\mathrm{NJL},\pi}^{\mathrm{eff}} & =G_\pi\, \big\{\bar{u}_{\mathrm{u}s1'}(p_1')u_{\mathrm{u}s1}(p_1)\,\bar{v}_{\mathrm{d}s2}(p_2)v_{\mathrm{d}s2'}(p_2') \nonumber\\
&\quad+ \bar{u}_{\mathrm{u}s1'}(p_1')\gamma_5 u_{\mathrm{u}s1}(p_1)\,\bar{v}_{\mathrm{d}s2}(p_2)\gamma_5 v_{\mathrm{d}s2'}(p_2') \nonumber\\
&\quad+ 2\,\bar{u}_{\mathrm{u}s1'}(p_1')\gamma_5 v_{\mathrm{d}s2'}(p_2')\,\bar{v}_{\mathrm{d}s2}(p_2)\gamma_5 u_{\mathrm{u}s1}(p_1) \big\},\label{eq:H_eff_NJL_pi_ori}
\end{align}
which can be derived from the NJL Lagrangian after the Legendre transform in the two-flavor NJL model~\cite{Klimt:1989pm,Vogl:1989ea,Vogl:1991qt,Klevansky:1992qe}. Here, only the combinations of Dirac bilinears relevant to the valence Fock sector LFWFs of the $\pi^+$ in valence Fock sector are included. For the positively charged kaon, the interaction is given by
\begin{align}
H^{\mathrm{eff}}_{\mathrm{NJL},K}&=G_K\,\big\{- 2\,\bar{u}_{\mathrm{u}s1'}(p_1') v_{\mathrm{s}s2'}(p_2')\,\bar{v}_{\mathrm{s}s2}(p_2) u_{\mathrm{u}s1}(p_1) \nonumber\\
&\quad + 2\,\bar{u}_{\mathrm{u}s1'}(p_1')\gamma_5 v_{\mathrm{s}s2'}(p_2')\,\bar{v}_{\mathrm{s}s2}(p_2)\gamma_5 u_{\mathrm{u}s1}(p_1) \big\},\label{eq:H_eff_NJL_SU_3_ori}
\end{align}
obtained similarly from the Lagrangian of the three-flavor NJL model. Here ${u_{\mathrm{f}s}(p)}$ and ${v_{\mathrm{f}s}(p)}$ are solutions of the free Dirac equation, with the nonitalic subscripts representing the flavors while the italic subscripts designate the spins. Meanwhile, $p_1$ and $p_2$ are the momenta of the valence quark and the valence antiquark, respectively~\cite{Jia:2018ary}.
The coefficients $G_{\pi}$ and $G_{K}$ are independent coupling constants of the theory. We have ignored the instantaneous terms due to the NJL interactions in deriving Eqs.~\eqref{eq:H_eff_NJL_pi_ori}~and~\eqref{eq:H_eff_NJL_SU_3_ori}. Explicit expressions and the detailed calculations of the matrix elements of the NJL interactions in the basis function representation we adopt can be found in Ref.~\cite{Jia:2018ary}.

Parameters in the BLFQ-NJL model are adjusted to reproduce the ground state masses of the pseudoscalar and vector mesons with light-light and light-strange nonsinglet flavor components. Meanwhile, the confining strengths are determined by the experimental charge radii of the $\pi^+$ and the $K^+$~\cite{Jia:2018ary}.
\begin{figure*}
	\centering
	\includegraphics[width=\linewidth]{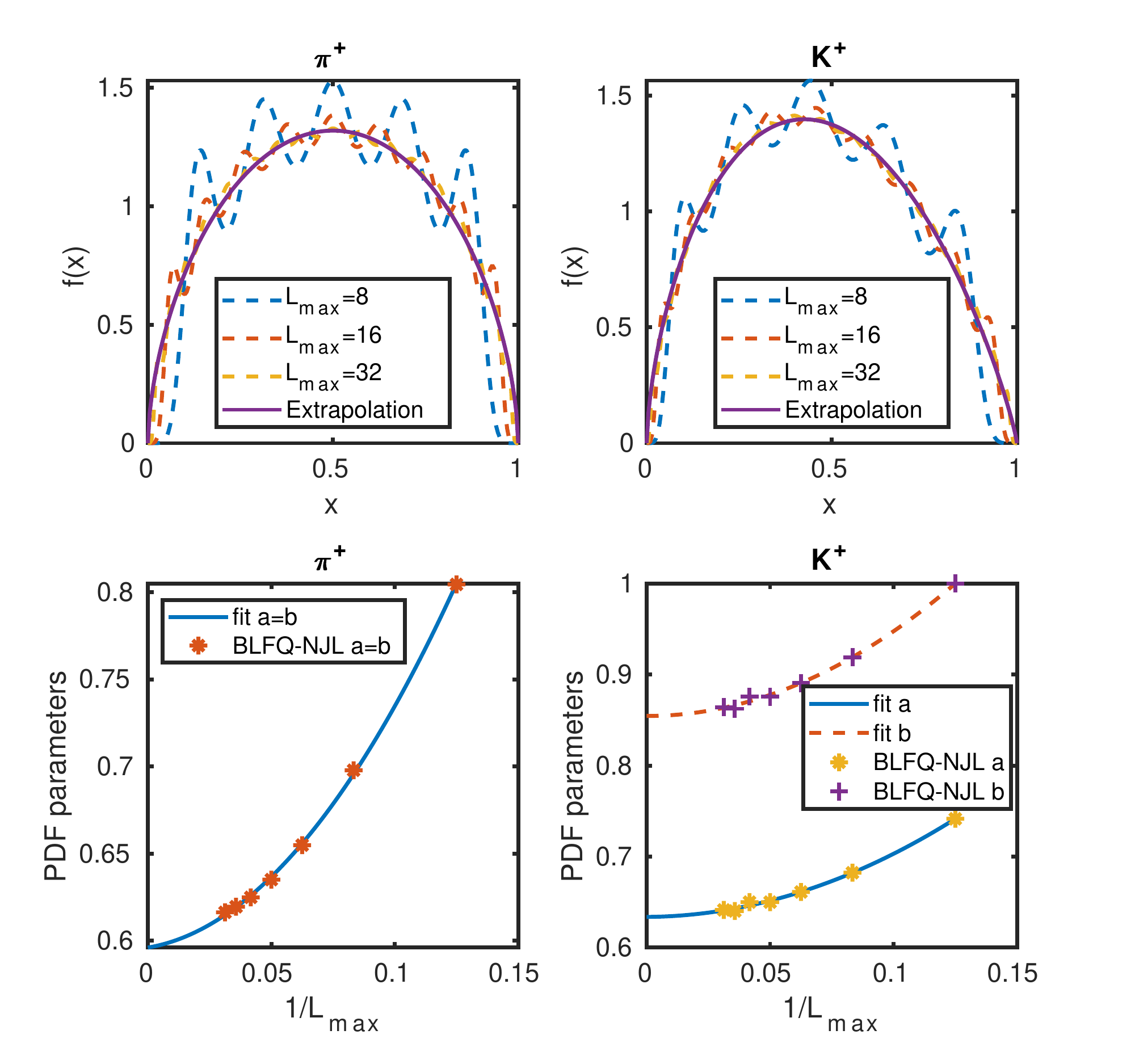}
	\caption{The PDFs for the valence quarks of the $\pi^+$ and $K^+$ mesons. The top-left panel shows the $\pi^+$ valence PDFs calculated from the LFWFs in the BLFQ-NJL model with different $L_{\mathrm{max}}$, together with the extrapolation to $L_{\mathrm{max}}\rightarrow +\infty$. The blue, red, and orange dashed lines correspond to the PDFs obtained from $N_{\mathrm{max}}=8$ and $L_{\mathrm{max}}=8,\,16,$ and $32$, respectively. The purple solid line represents Eq.~\eqref{eq:f_x_parametric} using the extrapolated parameters in Table~\ref{tab:L_max_dep}. The top-right panel presents the corresponding results for the $K^+$. The bottom-left panel shows the extrapolation of the fitting parameters in Eq.~\eqref{eq:f_x_parametric} for the $\pi^+$ valence PDF. Because the u and the d quarks have the same mass in Ref.~\cite{Jia:2018ary}, the parameter $a$ is always identical to $b$ for a fixed $L_{\mathrm{max}}$ for the pion. The bottom-right panel displays the extrapolations of the fitting parameters for the $K^+$ valence PDF. The yellow stars and the purple pluses are the fitting parameters $a$ and $b$ respectively for different $L_{\mathrm{max}}$. The blue solid line and the red dashed line are quadratic functions of $L_{\mathrm{max}}^{-1}$ as the best fits to the data points.}
	\label{fig:pikpdf}
\end{figure*}

\begin{table*}
	\caption{Dependence of the PDF fitting parameters on the longitudinal basis cutoff $L_{\mathrm{max}}$. With $N_{\mathrm{max}}=8$, the extrapolations are carried out by fitting to quadratic functions of $L_{\mathrm{max}}^{-1}$.}\label{tab:L_max_dep}

	\centering
	\begin{tabular}{ccccccccc}
		\hline 		\hline
		$L_{\mathrm{max}}$ & $8$ & $12$ & $16$ & $20$ & $24$ & $28$ & $32$ & Extrapolated to $+\infty$ \\
		\hline
		$\pi^+~a=b$ & $0.8045$  &  $0.6978$  &  $0.6549$  &  $0.6351$  &  $0.6249$  &  $0.6195$  &  $0.6163$ & $0.5961$ \\
		$K^+~a$ &    $0.7415$  &  $0.6823$  &  $0.6611$ &   $0.6500$  &  $0.6500$  &  $0.6403$  &  $0.6414$ & $0.6337$ \\
		$K^+~b$ &    $1.0002$  &  $0.9193$  &  $0.8907$  &  $0.8757$  &  $0.8761$  &  $0.8625$  &  $0.8643$ & $0.8546$ \\
		\hline		\hline
	\end{tabular}
\end{table*}

\begin{figure*}
	\begin{center}
		\includegraphics[width=0.465\textwidth]{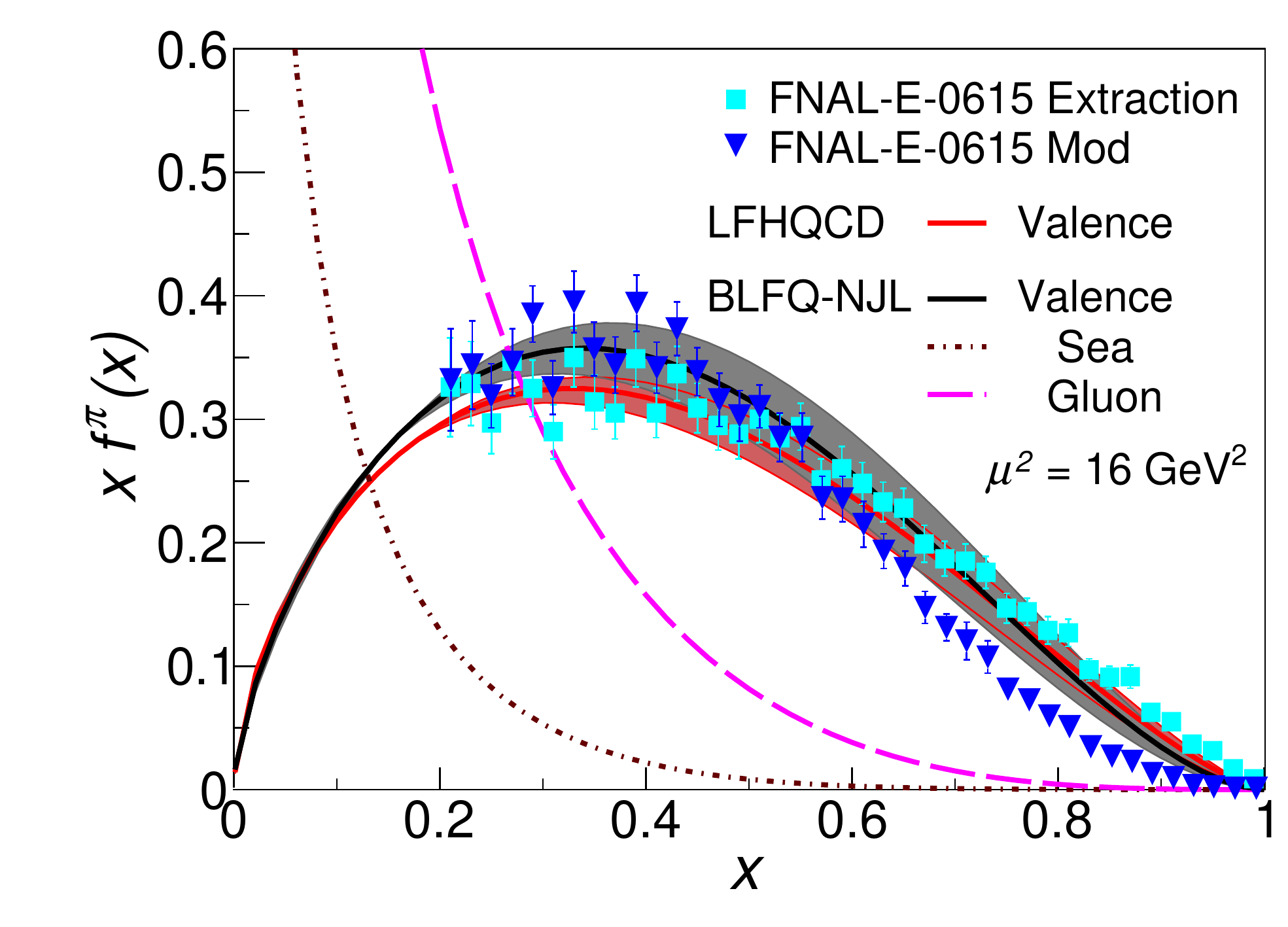}
		\caption{(Color online)~$xf^\pi(x)$ as a function of $x$ for the pion. The grey band corresponds to the pion valence PDF QCD-evolved from the BLFQ PDF at the initial scale ${\mu_{0\pi}^2=0.240\pm0.024~\rm{GeV}^2}$ to the experimental scale of $16~\rm{GeV}^2$. The black solid, brown dot-dashed, and pink long-dashed lines are the accompanying valence quark, the sea quark, and the gluon distributions respectively all at $\mu^2=16$ GeV$^2$. Our valence PDF is compared with the original analysis of the FNAL-E-0615 experimental result~\cite{Conway:1989fs} as well as with the reanalysis of the FNAL-E-0615 experimental result~\cite{Chen:2016sno}. The red band corresponds to the LFHQCD prediction~\cite{deTeramond:2018ecg}.}
		\label{fpionE615}
	\end{center}
\end{figure*}

\begin{table*}
	\caption{Initial scales and the $\chi^2$/(d.o.f.) at the first three orders of the DGLAP equation. The $\chi^2$ are defined as the sum of square-difference of our results with respect to the center values of the FNAL-E-0615 experiment \cite{Conway:1989fs} and the CERN-NA-003 experiment \cite{Badier:1983mj}, both at the respective experimental scales.}\label{tab:fitting_scales}

\centering
		\begin{tabular}{ccccc}
\hline \hline
			~~Order~~ & ~~Initial scale of pion ~~&~~ Initial scale of kaon~~ &~~ E-0615 $\chi^2/$(d.o.f.) ~~&~~ NA-003 $\chi^2/$(d.o.f.)~~  \\ 
			\hline
			LO & $0.120\pm 0.012~\mathrm{GeV}^2$ & $0.133\pm 0.013~\mathrm{GeV}^2$ & $6.71$ & $0.88$ \\ 
			NLO & $0.205\pm 0.020~\mathrm{GeV}^2$ & $0.210\pm 0.021~\mathrm{GeV}^2$ & $4.67$ &$0.56$ \\ 
			NNLO & $0.240\pm 0.024~\mathrm{GeV}^2$ & $0.246\pm 0.024~\mathrm{GeV}^2$ & $3.64$ & $0.50$ \\ 
\hline			\hline 
		\end{tabular} 

\end{table*}

\begin{figure*}
	\begin{center}
		(a)\includegraphics[width=0.46\textwidth]{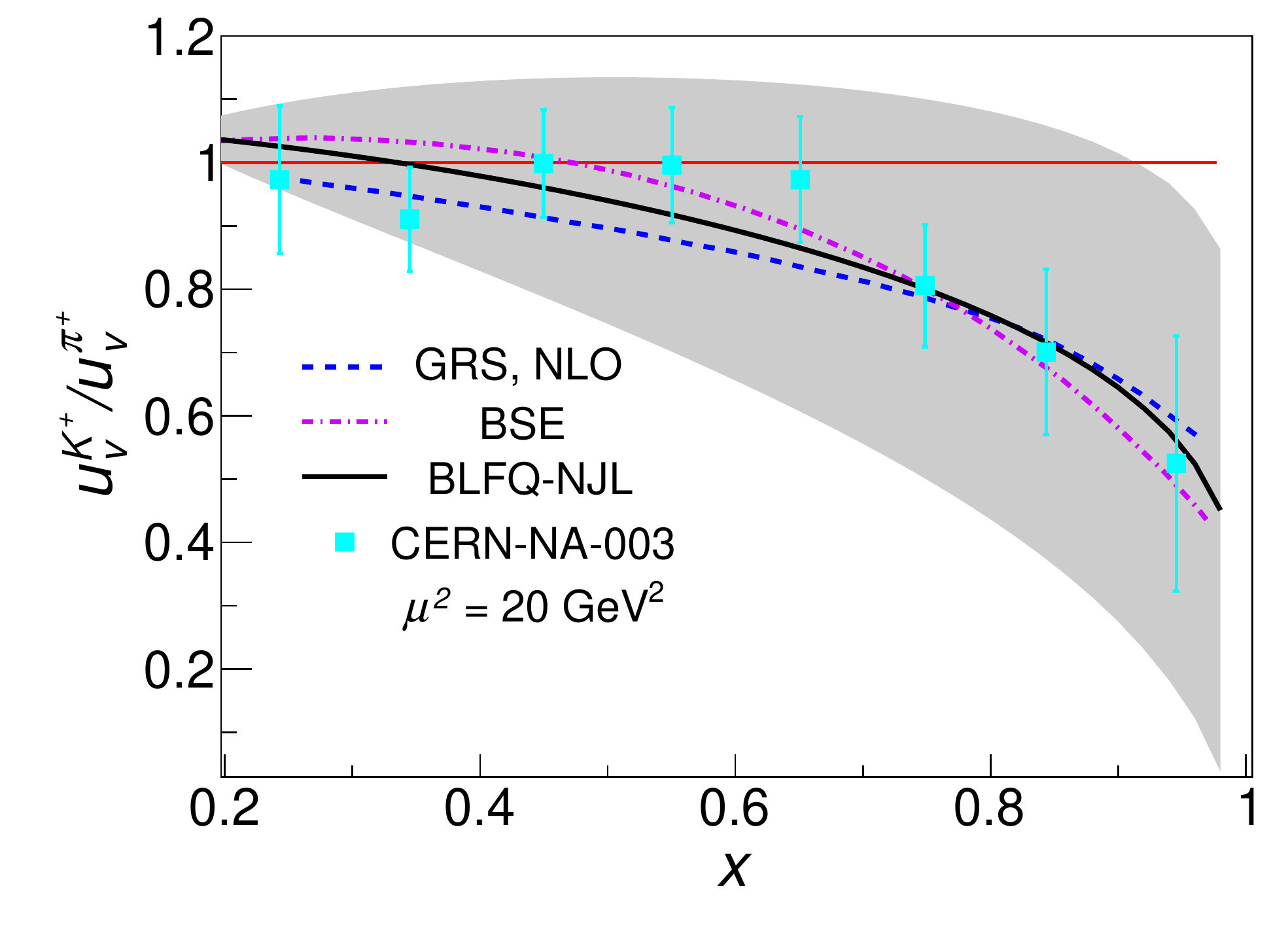}
		(b)\includegraphics[width=0.46\textwidth]{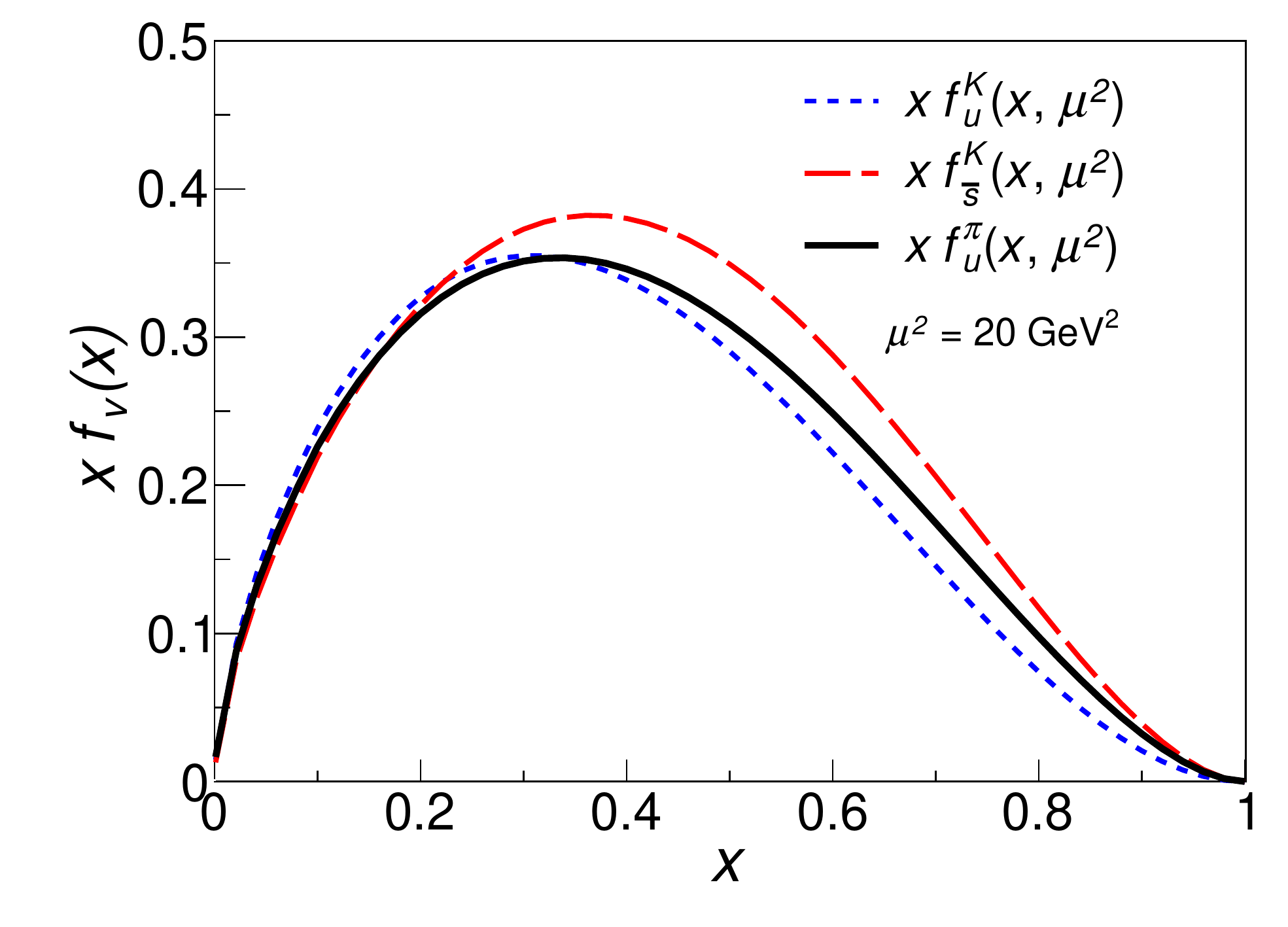}
		\caption{(Color online)~Illustrations of (a) the ratio of the u quark PDFs in the kaon to that in the pion, $u_{\rm{v}}^{K^+}/u_{\rm{v}}^{\pi^+}$ as a function of $x$ and; (b) comparison of the valence quark distributions in the kaon and the pion. In (b) the blue dashed and red long-dashed lines correspond to the u and the $\bar{\rm{s}}$ quark distributions in the kaon, respectively. The black solid line represents the valence quark distribution in the pion. The grey band in (a) corresponds to the QCD evolution from the initial scales ${\mu_{0\pi}^2=0.240\pm0.024~\rm{GeV}^2}$ for the pion and ${\mu_{0K}^2=0.246\pm0.024~\rm{GeV}^2}$ for the kaon. The discretized points with error bars in (a) are taken from the CERN-NA-003 Drell-Yan experiment~\cite{Badier:1983mj}. The blue dashed and magenta dashed dotted lines in (a) correspond the results obtained in the NLO Glück-Reya-Stratmann (GRS) model~\cite{Gluck:1997ww} and the prediction from the BSE~\cite{Nguyen:2011jy}, respectively.}
		\label{kaonpdf}
	\end{center}
\end{figure*}

\subsection{Valence quark PDFs in the pion and the kaon from BLFQ}\label{pdf_cal}
The LFWFs of the valence quarks in the $\pi^+$ meson and the $K^+$ meson have been solved in the BLFQ framework using the NJL interactions discussed in the previous subsection \cite{Jia:2018ary}. In the leading Fock sector, the LFWF for the mesons is written as  
\begin{align}
	\big\vert\Psi(P^+,\vec{P}^\perp)\big\rangle =&\sum_{r,s}\int_{0}^{1}\dfrac{dx}{4\pi x(1-x)}\int\dfrac{d\vec{\kappa}^\perp}{(2\pi)^2}\,\nonumber\\
	 &\times\psi_{rs}(x,\vec{\kappa}^\perp) b_r^\dagger(xP^+,\vec{\kappa}^\perp+x\vec{P}^\perp)\nonumber\\
	&\times d_s^\dagger((1-x)P^+,-\vec{\kappa}^\perp+(1-x)\vec{P}^\perp)|0\rangle,\label{eq:Psi_meson_qqbar}
\end{align}
where $P=k+p$ is the light front $3$-momentum of the meson, $x=k^+/P^+$ is the longitudinal momentum fraction carried by the valence quark, and $\vec{\kappa}^\perp=\vec{k}^\perp-x\vec{P}^\perp$ is the relative transverse momentum. The valence wave function is then expanded in the following orthonormal basis:
\begin{align} 
	&\quad \psi_{rs}(x,\vec{\kappa}^\perp)\nonumber\\
	& =\sum_{nml}\psi(n,m,l,r,s)\,\phi_{nm}\left(\dfrac{\vec{\kappa}^\perp}{\sqrt{x(1-x)}}\right)\chi_l(x),\label{eq:psi_rs_basis_expansions}
\end{align}
where $\phi_{nm}$ is the two-dimensional (2D) harmonic oscillator (HO) function, and $\chi_l$ is the longitudinal basis function. Here $n,\,m,$ and $l$ are basis quantum numbers corresponding to the radial excitation, the orbital angular momentum projection, and the longitudinal excitation, respectively. Explicitly, $\phi_{nm}$ is given by
	\begin{align}
	\phi_{nm}\left(\vec{q}^\perp;b \right)& =\dfrac{1}{b}\sqrt{\dfrac{4\pi n!}{(n+|m|)!}} \left(\dfrac{\vert\vec{q}^\perp\vert}{b}\right)^{|m|} \exp\left(-\dfrac{\vec{q}^{\perp 2}}{2b^2}\right)
	 \nonumber\\
	 &\quad\times L_n^{|m|} \left(\dfrac{\vec{q}^{\perp 2}}{b^2}\right)\,e^{im\varphi},\label{eq:def_phi_nm}
	\end{align}
	with $\tan(\varphi)=q^2/q^1$ and $L_n^{|m|}$ being the associated Laguerre function. Meanwhile, the longitudinal basis $\chi_l(x)$ is defined as
	\begin{align}
	&\quad \chi_l(x;\alpha,\beta)\nonumber\\
	&= \sqrt{4\pi(2l+\alpha+\beta+1)}\sqrt{\dfrac{\Gamma(l+1)\Gamma(l+\alpha+\beta+1)}{\Gamma(l+\alpha+1)\Gamma(l+\beta+1)}} \nonumber\\
	& \quad \times x^{\beta/2}(1-x)^{\alpha/2}\,P_l^{(\alpha,\beta)}(2x-1),\label{eq:def_chi_l}
	\end{align}
	with $P_{l}^{(\alpha,\beta)}(z)$ being the Jacobi polynomial and ${\alpha =2m_{\bar{q}}(m_q+m_{\bar{q}})/\kappa^2}$, ${\beta=2m_q(m_q+m_{\bar{q}})/\kappa^2}$. Here $m_{q}$ and $m_{\bar{q}}$ are the masses of the valence quark and the valence antiquark, respectively. 
	
	In order to numerically diagonalize $H_{\text{eff}}$, the infinite dimensional basis must be truncated. Because the NJL interactions do not couple to $\vert m\vert \geq 3$ basis states, we have a natural truncation for $m$~\cite{Jia:2018ary}. Specifically, we apply the following truncation to restrict the quantum numbers~\cite{Li:2015zda,Li:2017mlw,Jia:2018ary}:
	\begin{equation}
	0 \leq n \leq N_{\mathrm{max}}, \quad -2 \leq m \leq 2, \quad 0 \leq l \leq L_{\mathrm{max}},
	\label{eq:nmax}
	\end{equation}
	where $L_{\text{max}}$ determines the basis resolution in the longitudinal direction whereas $N_{\text{max}}$ controls the transverse momentum covered by 2D HO functions. Notice that our definition of $N_{\text{max}}$ in Eq.~(\ref{eq:nmax}) is different from that in Refs. \cite{Li:2017mlw,Tang:2018myz}.

	The probability of finding a quark inside the meson carrying the momentum fraction $x$ is then given by~\cite{Li:2017mlw}
	\begin{subequations}\label{eq:PDF_valence}
	\begin{align}
	& \quad f(x)\nonumber\\
	&=\dfrac{1}{4\pi\,x(1-x)}\sum_{rs}\int \dfrac{d\vec{\kappa}^\perp}{(2\pi)^2}\,\psi^*_{rs}(x,\vec{\kappa}^\perp)\,\psi_{rs}(x,\vec{\kappa}^\perp)\label{eq:PDF_valence_psi}\\
	& =\dfrac{1}{4\pi}\sum_{n,m,l',l,r,s}\psi^*(n,m,l',r,s)\psi(n,m,l,r,s)\, \chi_{l'}(x)\chi_{l}(x), \label{eq:PDF_valence_BLFQ}
	\end{align}
	\end{subequations}
	which is interpreted as the PDF for the valence quark. Correspondingly, the PDF for the valence antiquark is given by ${f(1-x)}$. In obtaining Eq.~\eqref{eq:PDF_valence_BLFQ} from Eq.~\eqref{eq:PDF_valence_psi}, the transverse integrals are evaluated exactly using the orthonormal property of the 2D HO functions. Equation~\eqref{eq:PDF_valence} implies the following momentum sum rule:
	\begin{equation}
	\int_{0}^{1}x\,f(x)\,dx+\int_{0}^{1}x\,f(1-x)\,dx=1,
	\end{equation}
	which indicates that, at our model scale, the valence quarks carry the entire momentum of the meson. Our normalization of the LFWF ensures that the normalizations of the PDFs for both valence quarks are $1$. 
	
	We then substitute the valence wave functions given by Eq.~\eqref{eq:psi_rs_basis_expansions} obtained from Ref.~\cite{Jia:2018ary} into Eq.~\eqref{eq:PDF_valence_BLFQ} to calculate the valence PDFs for the $\pi^+$ and the $K^{+}$. We show in the upper panels of Fig. 1 that with a fixed $L_{\mathrm{max}}$ the numerical PDFs oscillate about a single-peaked function, with the amplitude of the oscillation decreasing with increasing $L_{\mathrm{max}}$. Because the physical PDFs do not depend on the longitudinal cutoff, these oscillations are numerical artifacts. To remove such artifacts, we fit the resulting PDFs using the function
	\begin{equation}
	f(x)=x^a (1-x)^b/B(a+1,b+1)\label{eq:f_x_parametric},
	\end{equation}
	for each $L_{\mathrm{max}}\in \{8,\,12,\,16,\,20,\,24,\,28,\,32\}$. Here ${B(a+1,b+1)}$ is the Euler Beta function that ensures the normalization of Eq.~\eqref{eq:f_x_parametric}. Subsequently, we fit the $L_{\mathrm{max}}$ dependence of these fitting parameters by quadratic functions on $L^{-1}_{\mathrm{max}}$ and extrapolate to $L_{\mathrm{max}}\rightarrow +\infty$. The resulting fitting parameters and their extrapolations are given in Table~\ref{tab:L_max_dep} and the input PDFs of the pion and kaon corresponding to the extrapolations of the fitting parameters are shown in Fig.~\ref{fig:pikpdf}.
\section{PDFs, structure function, and cross sections}\label{sc:PDFs}
\subsection{PDFs and structure function}
\label{result}

By performing the QCD evolution, the valence-quark PDFs at high $\mu^2$ scale can be determined with the initial input using Eq.~(\ref{eq:f_x_parametric}) with parameters extrapolated to the infinite longitudinal basis cutoff as given in the last column of Table~\ref{tab:L_max_dep}. Specifically, we evolve our input PDFs to the relevant experimental scales ${\mu^2=16~\mathrm{GeV}^2}$ and ${\mu^2=20~\mathrm{GeV}^2}$ with independently adjustable initial scales of the pion and the kaon PDFs using the DGLAP equations~\cite{Dokshitzer:1977sg,Gribov:1972ri,Altarelli:1977zs}. Here, we use the higher order perturbative parton evolution toolkit (HOPPET) to numerically solve the DGLAP equations~\cite{Salam:2008qg}. We find that the initial scales increase when we progress from the leading order (LO) to NNLO. Meanwhile the evolved PDFs fit better to the experimental result demonstrated by smaller values of $\chi^2$ per degree of freedom (d.o.f.) at higher orders, as shown in Table \ref{tab:fitting_scales}. Since the results from the higher order DGLAP equation appear more reliable due to higher initial scales, only the results for the PDFs at NNLO are presented in this paper. 

We adopt ${\mu_{0\pi}^2=0.240\pm0.024~\rm{GeV}^2}$ for the initial scale of the pion PDF and ${\mu_{0K}^2=0.246\pm0.024~\rm{GeV}^2}$ for the initial scale of the kaon PDF which we determine by requiring the results after NNLO DGLAP evolution to fit both the pion PDF results from the FNAL-E-0615 experiment~\cite{Conway:1989fs} and the ratio $u^K_{\rm v}/u^\pi_{\rm v}$ result from the CERN-NA-003 experiment~\cite{Badier:1983mj}. 
 At our central value of the initial scales, the $\chi^2$ per d.o.f. for the fit of the pion PDF is 3.64, whereas for the ratio $u^K_{\rm v}/u^\pi_{\rm v}$, the corresponding value is 0.50.  The initial scales are the only adjustable parameters in this work and we assign them both a $10\%$ uncertainty. 
 We interpret the initial scales associated to our model as effective scales where the structures of the mesons are described by the motion of the valence quarks only. While applying the DGLAP equations, we impose the condition that the running coupling $\alpha_s(\mu^2)$ saturates in the infrared at a cutoff value of max $\{\alpha_{s}\}=1$.
 Note that the sea quark and the gluon distributions are absent in the initial scales of our model. The scale evolution allows quarks to emit and
absorb gluons, with the emitted gluons allowed to create
sea quarks as well as additional gluons.

\begin{figure*}
\begin{center}
(a)\includegraphics[width=0.465\textwidth]{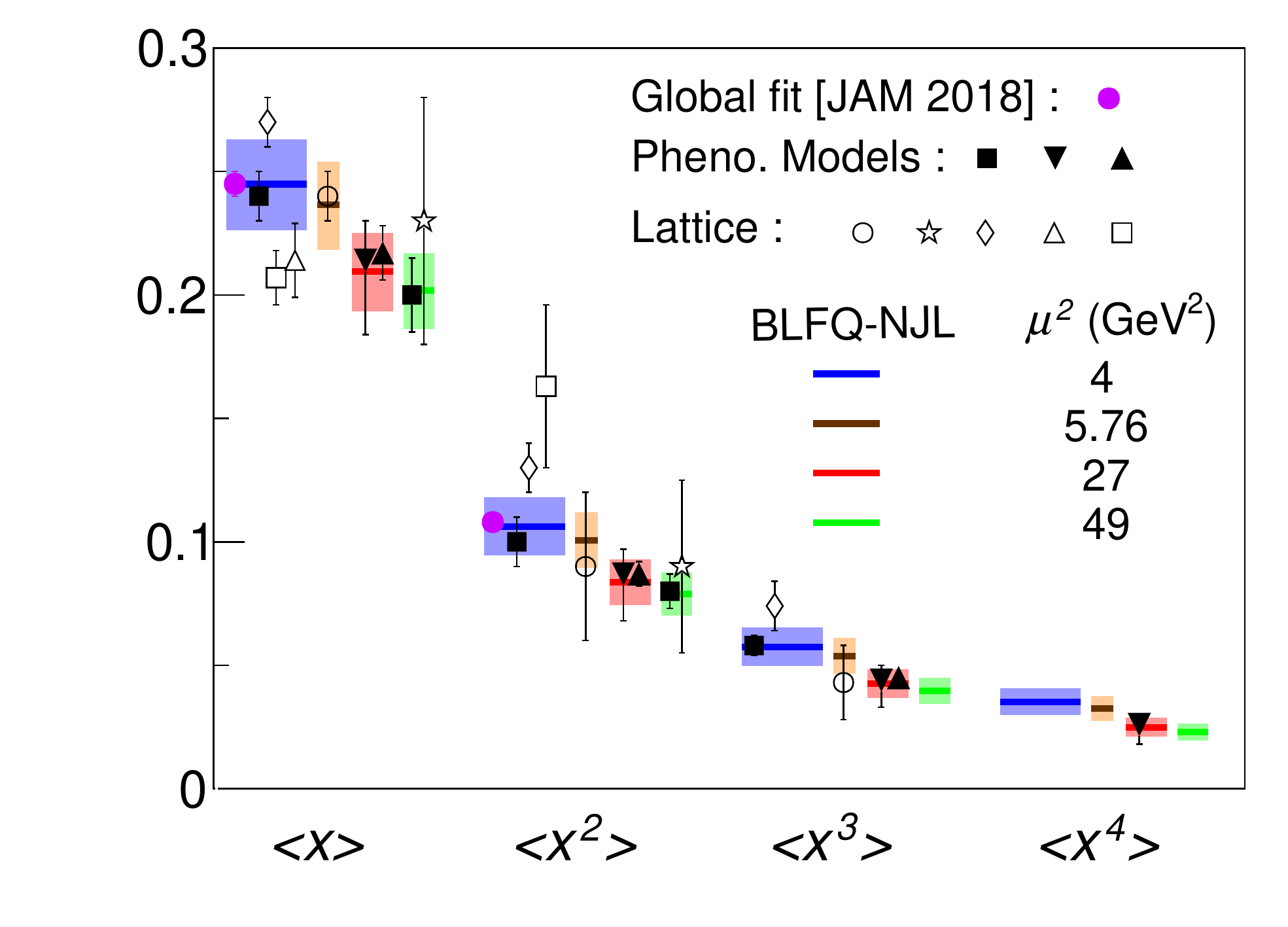}
(b)\includegraphics[width=0.465\textwidth]{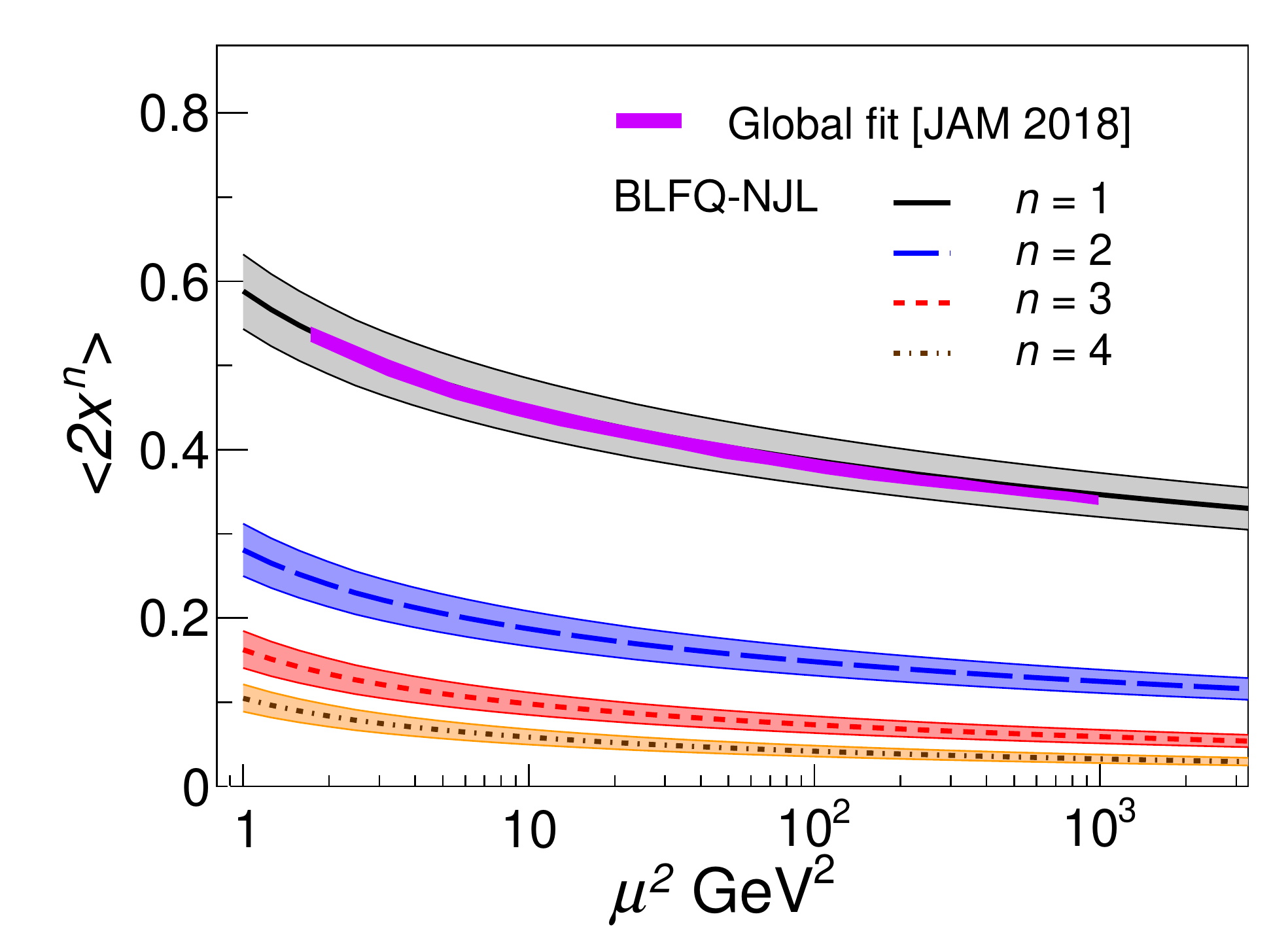}
\caption{(Color online)~(a) Comparison of the lowest four moments of valence quark distribution in the pion at different scales with the JAM global fit in Ref.~\cite{Barry:2018ort}, with lattice QCD results in Refs.~\cite{Brommel:2006zz,Martinelli:1987bh,Detmold:2003tm,Abdel-Rehim:2015owa,Oehm:2018jvm}, and with phenomenological models in Refs.~\cite{Nam:2012vm,Sutton:1991ay,Wijesooriya:2005ir}. Only results with uncertainties quoted are illustrated (see Table~\ref{moments_tab} for references and a more extensive listing). (b) The lowest four moments of the valence PDF as functions of the scale $\mu^2$. The colored horizontal bars in (a) and the lines in (b) with error bands are results of the present work taking into account the uncertainty in the initial scale ${\mu_{0\pi}^2=0.240\pm0.024~\rm{GeV}^2}$. The black solid line, the blue long-dashed line, the pink short-dashed line, and the yellow dot-dashed line correspond to ${n=1,\,2,\,3,}$ and $4$, respectively in Eq.~\eqref{pionmoment1}. The JAM global fit result for $\langle 2x \rangle$, shown as a purple band with $1\%$ uncertainty in the momentum fraction at the charm quark mass nearly coincides with our result and overlaps with our central result (black line) over a wide range of scales. }\label{moment}
\end{center}
\end{figure*}

In Fig.~\ref{fpionE615}, we show our result for the valence-quark PDF of the pion. We compare the valence-quark distribution after QCD evolution with the result from the FNAL-E-0615~\cite{Conway:1989fs} and with the reanalysis of the same result including soft gluon resummation~\cite{Chen:2016sno}. The error band in the valence-quark distributions is due to the spread in the initial scale $\mu_0^2=0.240\pm 0.024$ GeV$^2$ propagated by the QCD evolution. Our result favors the slower falloff in the large-$x$ region in the original analysis of the FNAL-E-0615 experiment. While in the intermediate region of $x$, our result is in agreement with the reanalysis of the FNAL-E-0615 result. The pion valence PDF from our model falls off at large $x$ as $(1-x)^{1.44}$, so there is a tension with the results obtained from the BSE~\cite{Chen:2016sno} and with the analysis in Ref.~\cite{Chen:2016sno} that incorporated the $(1-x)^2$ perturbative QCD falloff at large $x$ from the threshold resummation effects. We note, however, that there has been a recent fit to the FNAL-E-0615 result in LFHQCD which supports a linear falloff at high-$x$~\cite{deTeramond:2018ecg}.

\begin{table*}
	\caption{Comparison of the lowest four moments of the valence quark PDF in the pion based on the initial PDF from BLFQ-NJL model with the results from the global fit, lattice QCD, and phenomenological models at various scales. Results tabulated here at $\mu^2 \ge 4$ GeV$^2$ are also presented in Fig.~\ref{moment} (a).}\label{moments_tab}
\centering
\begin{tabular}{cccccc}
	\hline 	\hline &$\mu^2$ GeV$^2$&~~~~~~ $\langle x \rangle$~~~~~~ & ~~~~~~$\langle x^{2} \rangle$  ~~~~~~&~~~~~~ $\langle x^{3} \rangle$ ~~~~~~ & ~~~~~~$\langle x^{4} \rangle$ ~~~~~~\\
	\hline 	DSE-RL (2018) \cite{Bednar:2018mtf}&1.69&0.268&0.125&0.076&0.054\\
	WI-An (2018) \cite{Bednar:2018mtf}&&0.268&0.114&0.059&0.037\\
	    JAM global fit (2018)~\cite{Barry:2018ort}&&0.268&0.127&0.074&0.048\\
	BLFQ-NJL&&$0.271^{+0.020}_{-0.020}$&$0.124^{+0.014}_{-0.014}$&$0.069^{+0.009}_{-0.009}$&$0.044^{+0.007}_{-0.007}$\\
	\\
	Sutton (1992)~\cite{Sutton:1991ay}&  4  &$0.24\pm0.01$&$0.10\pm0.01$&$0.058\pm0.004$& \\
	Hecht (2001)~\cite{Hecht:2000xa}& &$0.24$&0.098&0.049& \\
	Chen (2016)~\cite{Chen:2016sno}&    &$0.26$&$0.11$&$0.052$& \\
	BSE (2018)~\cite{Shi:2018mcb}& &0.24& & &\\
	BSE (2019)~\cite{Ding:2019lwe} & &$0.24\pm0.02$ & &  &\\
	QCDSF/UKQCD (2007) [lattice QCD]~\cite{Brommel:2006zz}&  &$0.27\pm0.01$&$0.13\pm0.01$&$0.074\pm0.010$&\\
	DESY (2016) [lattice QCD]~\cite{Abdel-Rehim:2015owa} & &$0.214\pm0.015$ & &  &\\
	ETM (2018) [lattice QCD]~\cite{Oehm:2018jvm} & &$0.207\pm0.011$&$0.163\pm0.033$& &\\
	JAM global fit (2018)~\cite{Barry:2018ort}& &$0.245\pm 0.005$&$0.108\pm0.003$&&\\
	BLFQ-NJL& &$0.245^{+0.018}_{-0.018}$&$0.106^{+0.012}_{-0.012}$&$0.057^{+0.008}_{-0.008}$&$0.035^{+0.005}_{-0.005}$\\
\\
	Detmold (2003) [lattice QCD]~\cite{Detmold:2003tm}& 5.76 &$0.24\pm0.01$&$0.09\pm0.03$&$0.043\pm0.015$&\\
	BLFQ-NJL&  &$0.236^{+0.018}_{-0.018}$&$0.101^{+0.011}_{-0.011}$&$0.054^{+0.007}_{-0.007}$&$0.032^{+0.005}_{-0.005}$\\
\\
	Watanabe (2018)~\cite{Watanabe:2017pvl}&   27    &0.23&0.094&0.048&\\
	Nam (2012)~\cite{Nam:2012vm}&    &$0.214_{-0.030}^{+0.016}$&$0.087_{-0.019}^{+0.010}$&$0.044_{-0.011}^{+0.006}$&$0.026_{-0.008}^{+0.004}$ \\
	Wijesooriya (2005)~\cite{Wijesooriya:2005ir}&   &$0.217\pm0.011$&$0.087\pm0.005$&$0.045\pm0.003$& \\
	BLFQ-NJL&  &$0.210^{+0.016}_{-0.016}$&$0.084^{+0.009}_{-0.009}$&$0.043^{+0.006}_{-0.006}$&$0.025^{+0.004}_{-0.004}$\\
\\
	Sutton (1992) \cite{Sutton:1991ay} & 49 &$0.200\pm0.015$&$0.080\pm0.007$& & \\
	Martinell (1988) [lattice QCD]~\cite{Martinelli:1987bh}&    &$0.23\pm0.05$&$0.090\pm0.035$&    & \\
	BLFQ-NJL&   &$0.202^{+0.015}_{-0.015}$&$0.079^{+0.009}_{-0.009}$&$0.040^{+0.005}_{-0.005}$&$0.023^{+0.003}_{-0.003}$\\
	\hline 	\hline 
    \end{tabular}

\end{table*}
\begin{table*}
	\caption{Lowest four moments of valence quark distributions in the kaon based on the initial PDF from BLFQ-NJL model. Comparisons are made with results from Refs.~\cite{Chen:2016sno,Watanabe:2018qju}.}\label{moments_tab_kaon}

	\centering
	\begin{tabular}{ccccccc}
		\hline 		\hline  flavor&~~~&~$\mu^2$ GeV$^2$~&~~~~~~ $\langle x \rangle$~~~~~~ & ~~~~~~$\langle x^{2} \rangle$  ~~~~~~&~~~~~~ $\langle x^{3} \rangle$ ~~~~~~ & ~~~~~~$\langle x^{4} \rangle$ ~~~~~~ \\
		\hline 
		     $\rm{s}^{K}$&BLFQ-NJL&1&$0.320^{+0.024}_{-0.024}$&$0.158^{+0.018}_{-0.017}$&$0.093^{+0.013}_{-0.012}$&$0.061^{+0.010}_{-0.009}$\\
		$\rm u^{K}$&& &$0.282^{+0.021}_{-0.021}$&$0.128^{+0.014}_{-0.014}$&$0.071^{+0.010}_{-0.010}$&$0.044^{+0.007}_{-0.007}$\\
\\
		$\rm s^{K}$&BLFQ-NJL&4&$0.266^{+0.020}_{-0.020}$&$0.119^{+0.013}_{-0.013}$&$0.066^{+0.009}_{-0.009}$&$0.041^{+0.006}_{-0.006}$\\
		$\rm u^{K}$&&&$0.235^{+0.017}_{-0.018}$&$0.097^{+0.011}_{-0.011}$&$0.050^{+0.007}_{-0.007}$&$0.030^{+0.005}_{-0.005}$\\
\\
		$\rm s^{K}$&BLFQ-NJL&16&$0.237^{+0.018}_{-0.018}$&$0.100^{+0.011}_{-0.011}$&$0.052^{+0.007}_{-0.007}$&$0.031^{+0.005}_{-0.005}$\\
		$\rm u^{K}$&&&$0.209^{+0.015}_{-0.016}$&$0.081^{+0.009}_{-0.009}$&$0.040^{+0.006}_{-0.005}$&$0.023^{+0.004}_{-0.003}$\\
\\
		$\rm s^{K}$& Chen (2016)~\cite{Chen:2016sno}& 27 &0.36&0.17&0.092&\\
		& Watanabe (2018)~\cite{Watanabe:2018qju}& &0.24&0.096&0.049&\\
		&BLFQ-NJL&&$0.228^{+0.017}_{-0.017}	$&$0.094_{-0.010}^{+0.010}$ &	$0.049_{-0.007}^{+0.007}$&$0.029^{+0.005}_{-0.004}$\\
		$\rm u^{K}$& Chen (2016)~\cite{Chen:2016sno}&  &0.28&0.11&0.048&   \\
		& Watanabe (2018)~\cite{Watanabe:2018qju}& &0.23&0.091&0.045&\\
		&BLFQ-NJL&&$0.201^{+0.015}_{-0.015}	$&$0.077^{+0.009}_{-0.008}$ &	$0.037^{+0.005}_{-0.005}$&$0.021^{+0.003}_{-0.003}$\\
		\hline 		\hline 

    \end{tabular}
\end{table*}

\begin{figure*}
	\begin{center}
		\includegraphics[width=0.94\textwidth]{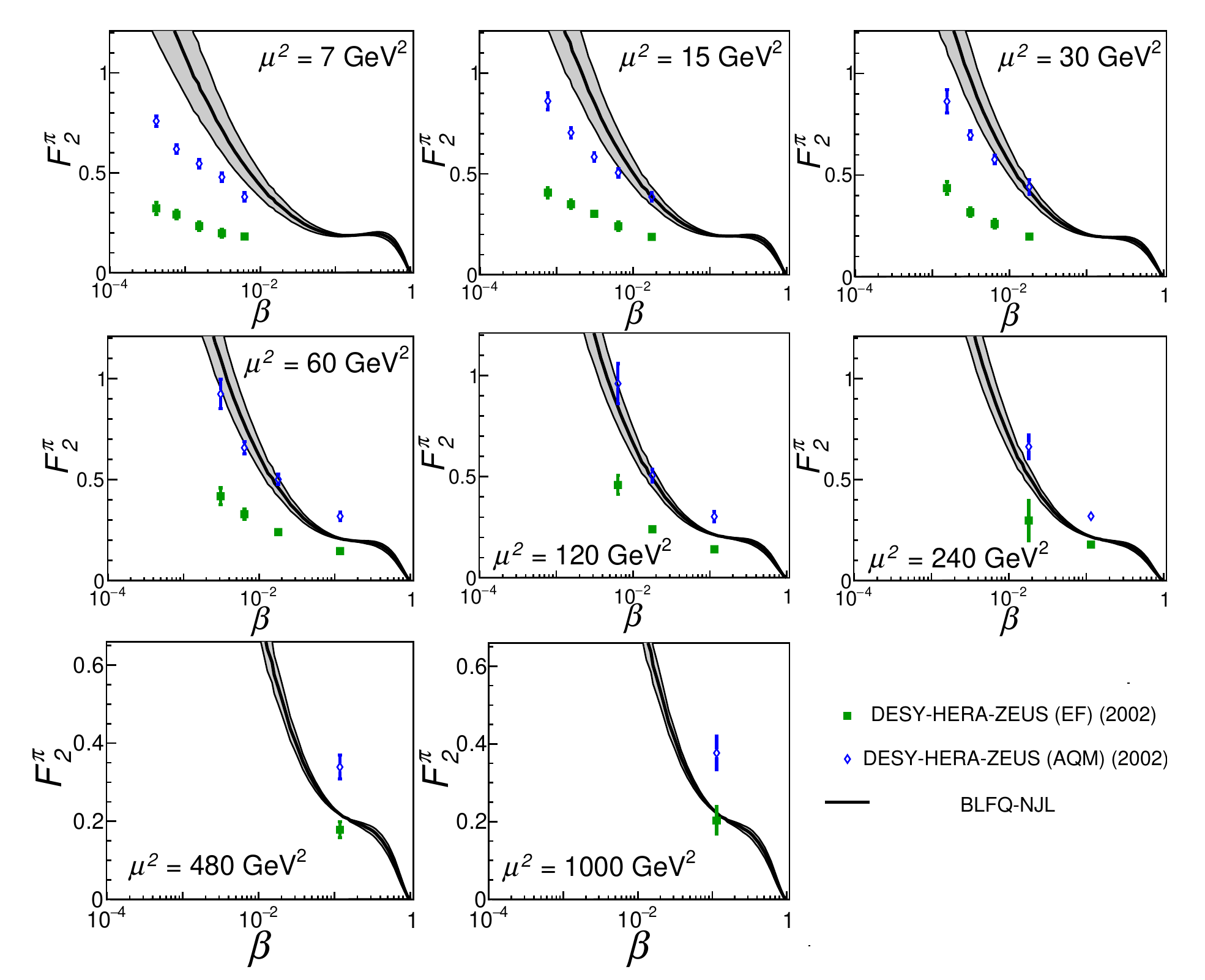}
		\caption{(Color online)~Structure function $F_2^{\pi}(\beta,\mu^2)$ for the pion as a function of $\beta$ at fixed experimental values of $\mu^2$. The data are taken from Ref.~\cite{Chekanov:2002pf} by the ZEUS Collaboration in DESY-HERA. The error bands are results of the present work taking the uncertainty in the initial scale ${\mu_{0\pi}^2=0.240\pm0.024~\rm{GeV}^2}$ into account.}
		\label{F21}
	\end{center}
\end{figure*}
\begin{figure*}
	\begin{center}
		\includegraphics[width=0.94\textwidth]{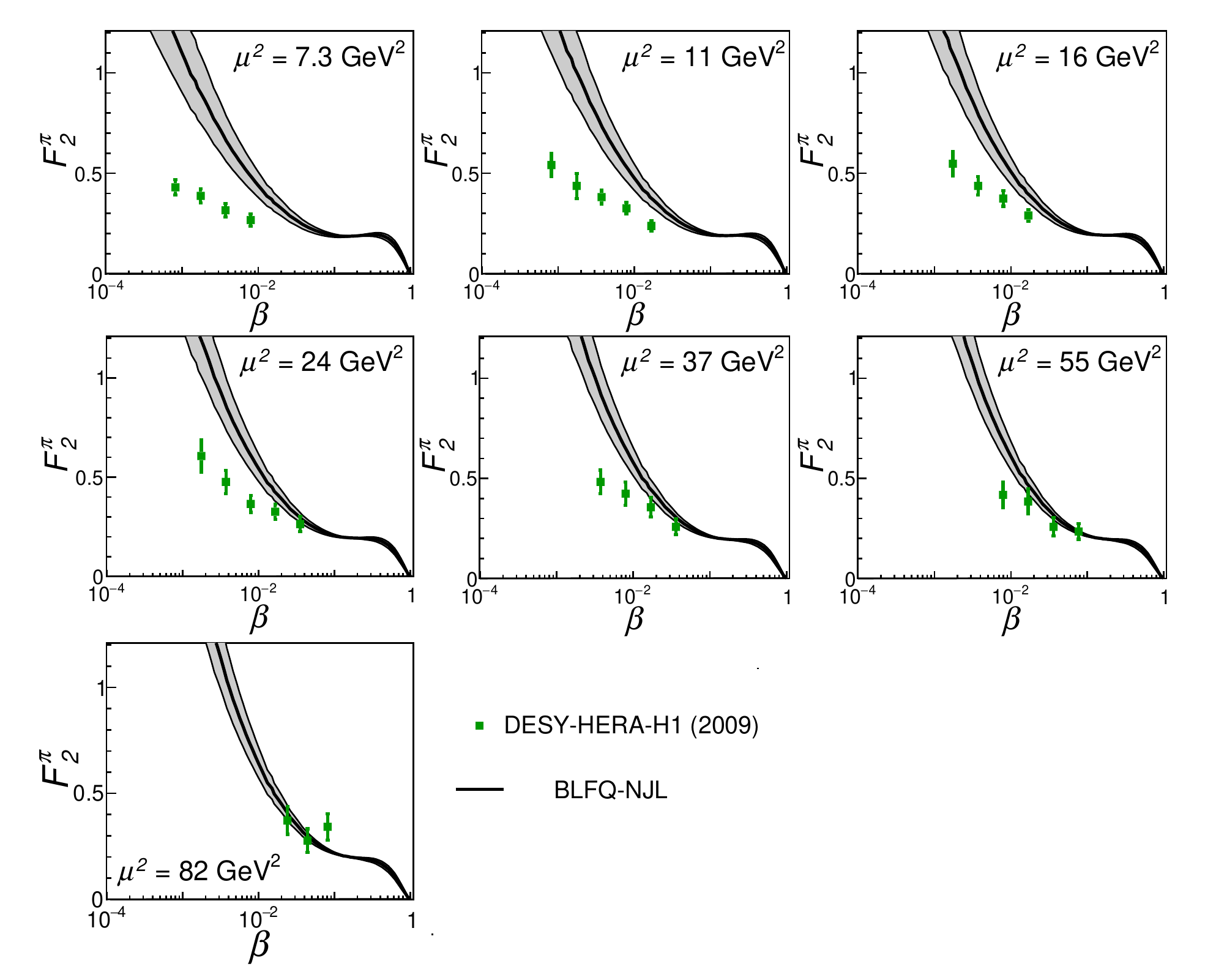}
		\caption{(Color online)~Structure function $F_2^{\pi}(\beta,\mu^2)$ for the pion as a function of $\beta$ at fixed experimental values of $\mu^2$. The data are taken from Ref.~\cite{Aaron:2010ab} by the H1 Collaboration in DESY-HERA. The error bands are results of the present work taking the uncertainty in the initial scale ${\mu_{0\pi}^2=0.240\pm0.024~\rm{GeV}^2}$ into account.}
		\label{F22}
	\end{center}
\end{figure*}

\begin{figure*}
	\begin{center}
		\includegraphics[width=0.47\textwidth]{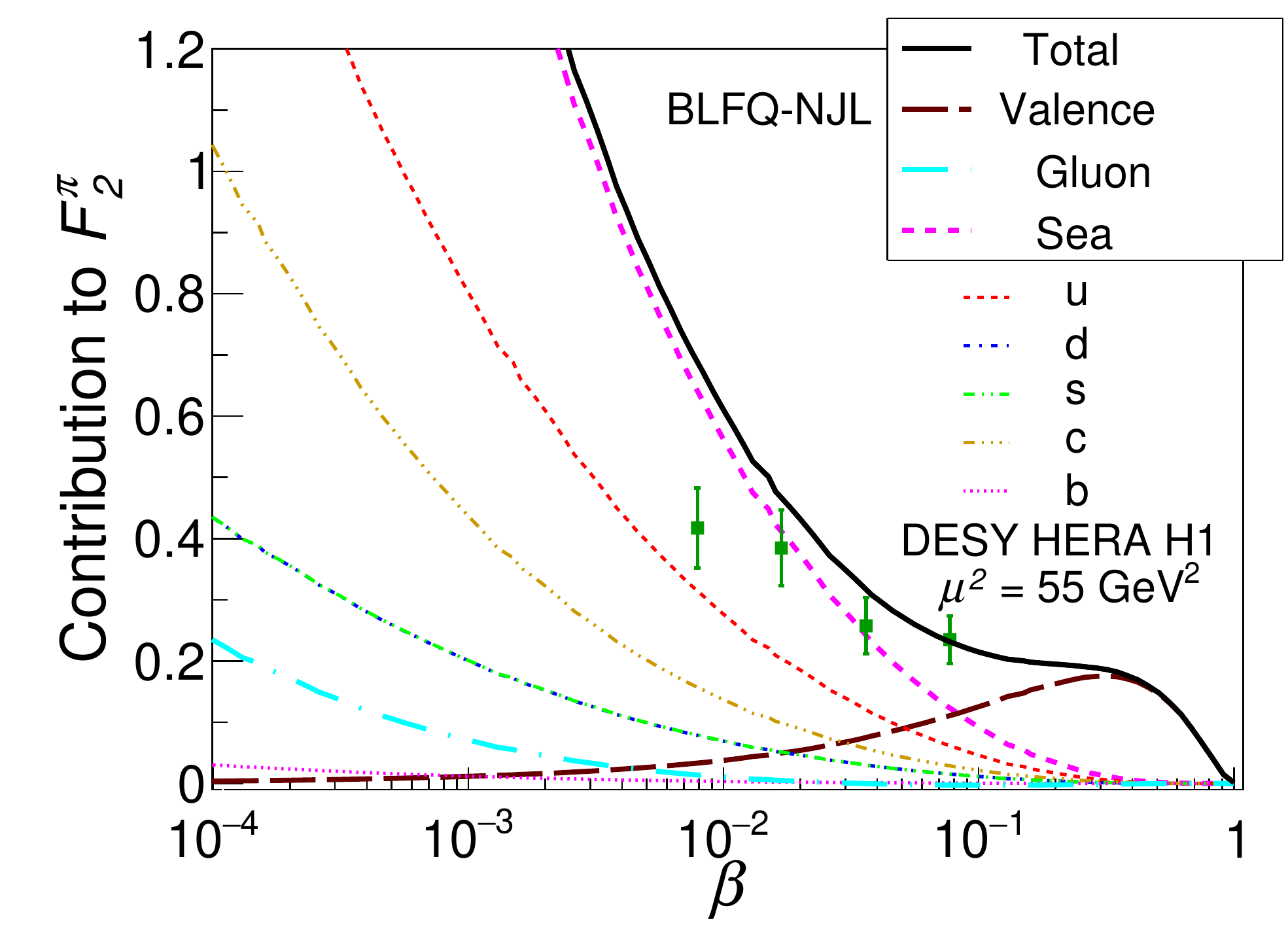}
		\caption{(Color online)~Contributions to the $F_2^{\pi}$ from the valence quark and sea quarks in different flavors. The data are taken from Ref.~\cite{Aaron:2010ab} by the H1 Collaboration in DESY-HERA. The error bands are results of the present work taking the uncertainty in the initial scale ${\mu_{0\pi}^2=0.240\pm0.024~\rm{GeV}^2}$ into account.}
		\label{F23}
	\end{center}
\end{figure*}

Another comparison can be made for the pion PDFs at the initial scale of Ref.~\cite{deTeramond:2018ecg} at $\mu_0^2=(1.12\pm0.32)~\mathrm{GeV}^2$. We find that at this scale the valence quarks carry 57\% of the pion's momentum from our model, close to the 54\% given by Ref.~\cite{deTeramond:2018ecg}. At the same scale, in contrast to the absence of gluon contributions in Ref.~\cite{deTeramond:2018ecg}, our model allocates 35\% of the pion's momentum to the gluons and 8\% to the sea quarks. 

In Fig.~\ref{kaonpdf} (a), we present the ratio of the u quark distributions in the kaon to that in the pion, with the valence-quark PDFs in the kaon shown in Fig.~\ref{kaonpdf} (b). We observe that at $\mu^2=20~\mathrm{GeV}^2$, our center value of $u_{\rm{v}}^{K}/u_{\rm{v}}^{\pi}$ is in good agreement with the result from CERN-NA-003 experiment~\cite{Badier:1983mj} as well as with a phenomenological quark model (GRS, NLO)~\cite{Gluck:1997ww} and the BSE approach~\cite{Nguyen:2011jy}. One notices that the ratio decreases as $x$ increases. This phenomena is rooted in the results shown in Fig.~\ref{kaonpdf} (b) where we compare the valence quark distributions of the kaon and the pion. We find that at the scale of $\mu^2=20~\mathrm{GeV}^2$  the distribution of the u quark PDF at high $x$ in the pion is above that in the kaon. This can be understood since $m_{\mathrm{s}}>m_{\mathrm{u}}$ the peak of the $\overline{\mathrm{s}}$ quark distribution in the kaon appears at higher $x$ compared to the u quark distribution. Therefore the $\overline{\mathrm{s}}$ quark carries larger momentum than the u quark does, reducing the probability of finding a u quark with high $x$ in the kaon. Specifically, the u quark PDF in the kaon falls off at large $x$ as $(1-x)^{1.60}$ whereas the same behavior in the pion is $(1-x)^{1.49}$. We also observe that in our model, the $\overline{\mathrm{s}}$ quark PDF in the kaon falls off as $(1-x)^{1.32}$. 

We further evaluate the lowest four nontrivial moments of the valence quark PDF defined as
\begin{eqnarray}
\langle x^{n} \rangle =\int_0^1 dx~ x^n f_v^{\pi/K}(x,\mu^2),~n=1,2,3,4.
\label{pionmoment1}
\end{eqnarray}
The corresponding moments of the pion PDF at different scales are shown in Fig.~\ref{moment} (a), together with the results from the global fit~\cite{Barry:2018ort}, lattice QCD~\cite{Brommel:2006zz,Martinelli:1987bh,Detmold:2003tm,Abdel-Rehim:2015owa,Oehm:2018jvm}, and several  phenomenological models~\cite{Nam:2012vm,Sutton:1991ay,Wijesooriya:2005ir}. Our predictions are in good agreement with Refs.~\cite{Martinelli:1987bh,Detmold:2003tm,Sutton:1991ay,Wijesooriya:2005ir,Nam:2012vm,Barry:2018ort}. The numerical values of the lowest four moments of the pion PDF at various scales are presented in Table~\ref{moments_tab}. 

The scale dependence of the lowest four moments of the pion valence quark PDF is presented in Fig.~\ref{moment} (b). These moments decrease uniformly as the scale $\mu^2$ increases, compensated by the increase in contributions from sea quarks and gluons. We find good agreement of our calculated $\langle x \rangle$ with the Jefferson Lab Angular Momentum Collaboration (JAM) global fit~\cite{Barry:2018ort} over nearly 3 decades of the $\mu^2$ scale within our uncertainty and close to the central value. Additionally, the numerical values of the lowest four moments of the valence quark PDFs in the kaon at various scales are presented in Table~\ref{moments_tab_kaon}.

\begin{figure*}
	\begin{center}
		(a)\includegraphics[width=0.47\linewidth]{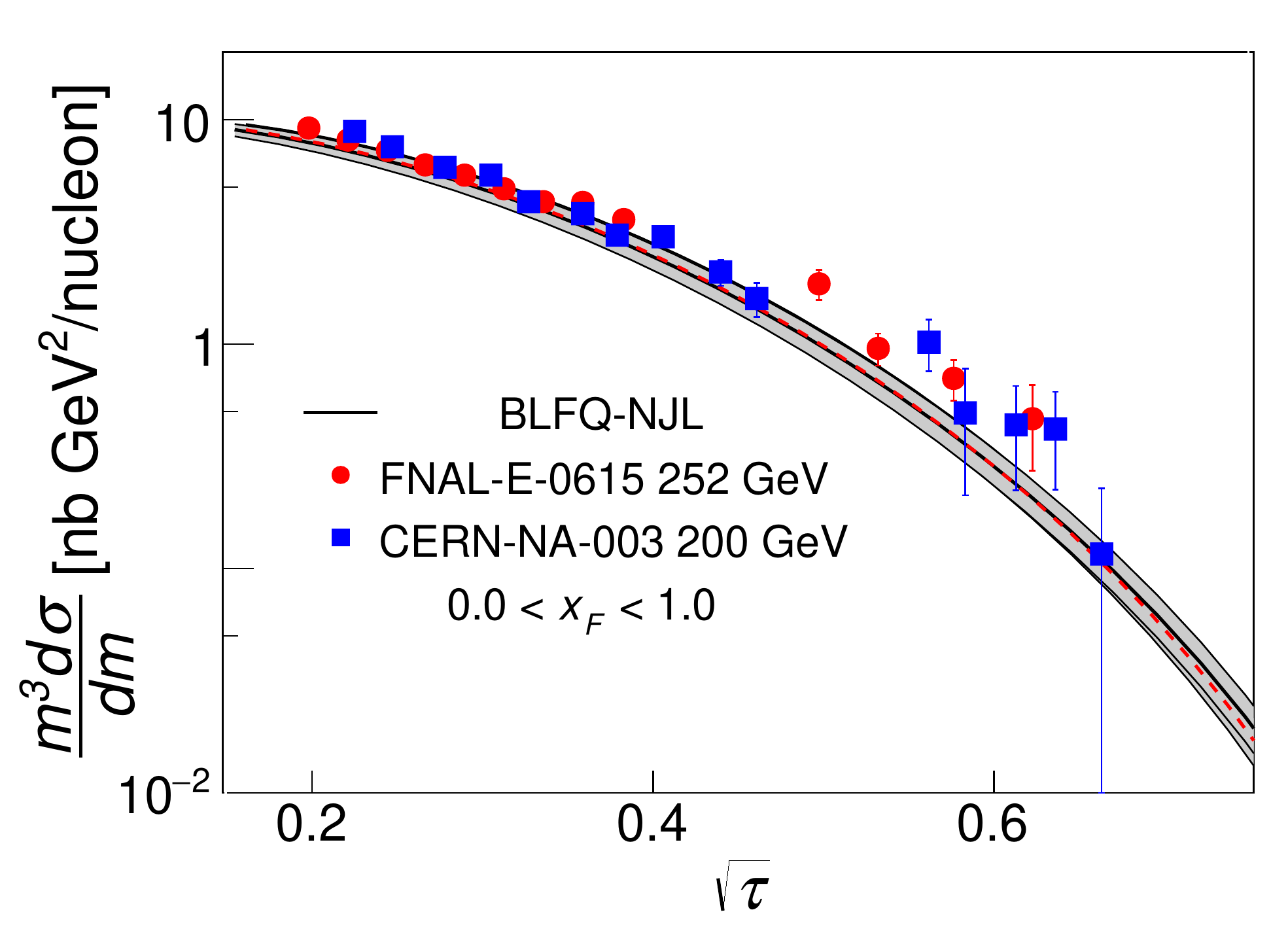}
		(b)\includegraphics[width=0.47\linewidth]{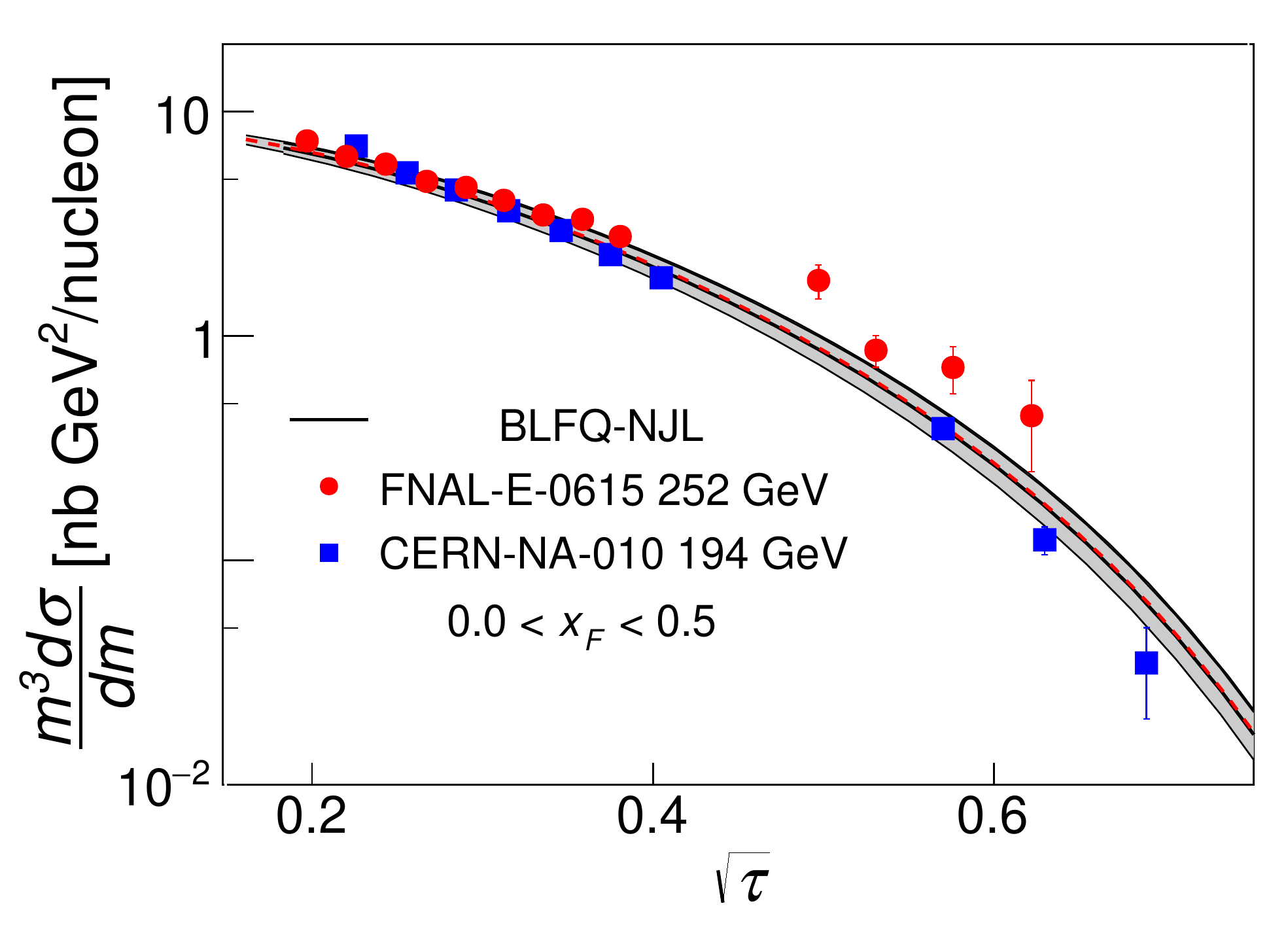}
		\caption{(Color online)~The cross section $m^3\,d{\bf \sigma}/dm$ for the $\pi^-$-nucleus Drell-Yan process as a function of $\sqrt{\tau}$ in the regions (a) $0<x_{\mathrm{F}}< 1$ and (b) $0<x_{\mathrm{F}}< 0.5$. The data of FNAL-E-0615 experiment with 252 GeV pions and CERN-NA003 with 200 GeV pions as well as CERN-NA-010 with 194 GeV pions are taken from Ref.~\cite{Conway:1989fs} and Refs.~\cite{Badier:1983mj,Betev:1985pf}, respectively. The error-bands are the cross section calculated from the BLFQ-NJL model taking the uncertainty in the initial scale ${\mu_{0\pi}^2=0.240\pm0.024~\rm{GeV}^2}$ into account. The FNAL-E-0615 and the CERN-NA-010 data both correspond to a tungsten target while the CERN-NA-003 data correspond to a platinum target. The black solid and the red dashed lines in (a) represent the cross sections evaluated using the tungsten and the platinum nuclear PDFs, respectively.}\label{cross1}
	\end{center}
\end{figure*}
\begin{figure*}
	\begin{center}
		\includegraphics[width=0.47\linewidth]{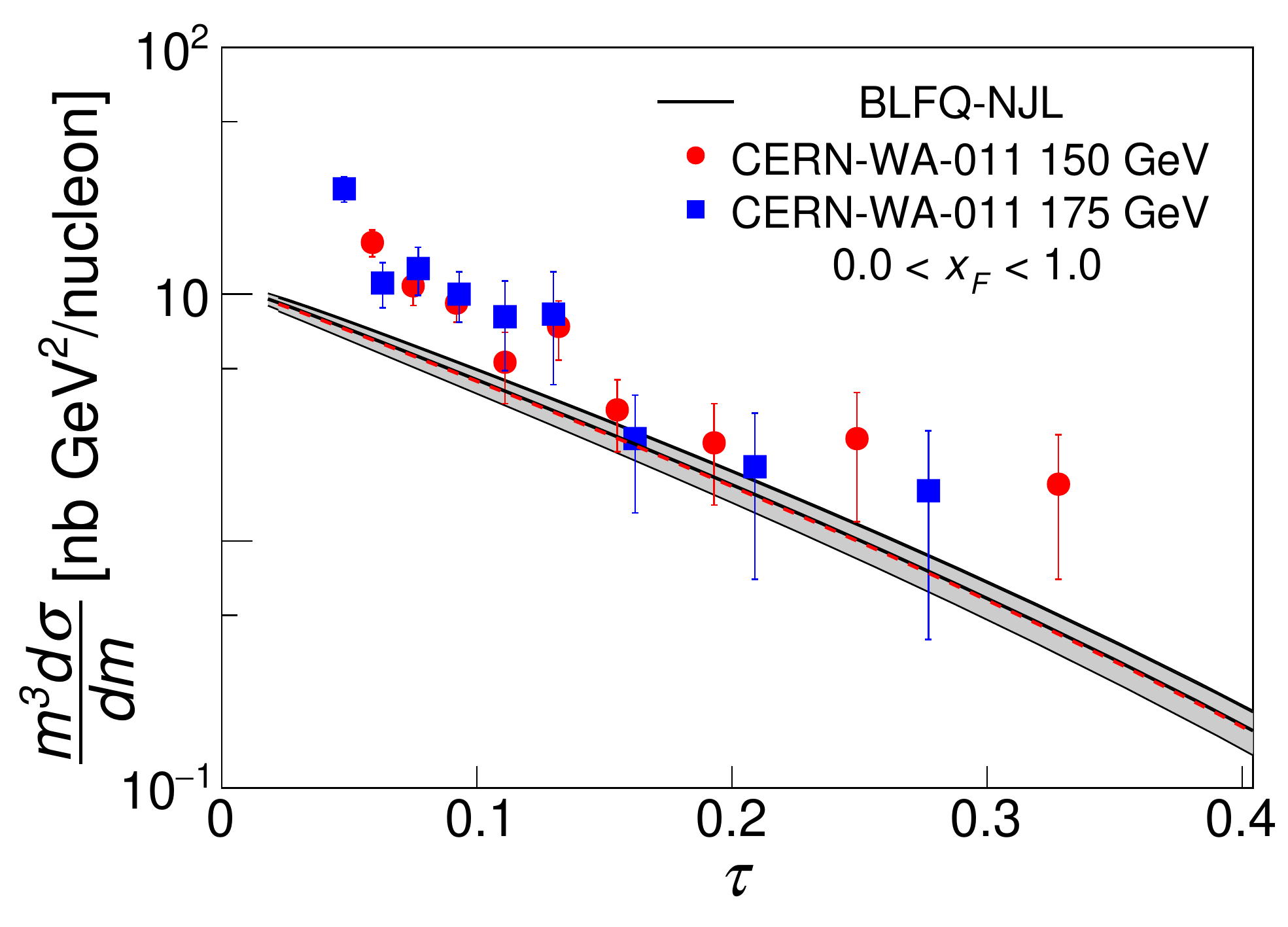}
		\caption{(Color online)~The cross section $m^3\,d{\bf \sigma}/dm$ for the $\pi^-$-nucleus Drell-Yan process as a function of $\tau$ in the region  $0<x_{\mathrm{F}}< 1$. The data of CERN-WA-011 experiment with 150 GeV pions and 175 GeV pions are taken from Ref.~\cite{Barate:1979da}. The error-band in the cross section represents results calculated from the BLFQ-NJL model taking the uncertainty in the initial scale ${\mu_{0\pi}^2=0.240\pm0.024~\rm{GeV}^2}$ into account. These CERN-WA-011 data correspond to a beryllium target.
		}
		\label{cross2}
	\end{center}
\end{figure*}
\begin{figure*}
	\begin{center}
		\includegraphics[width=0.47\linewidth]{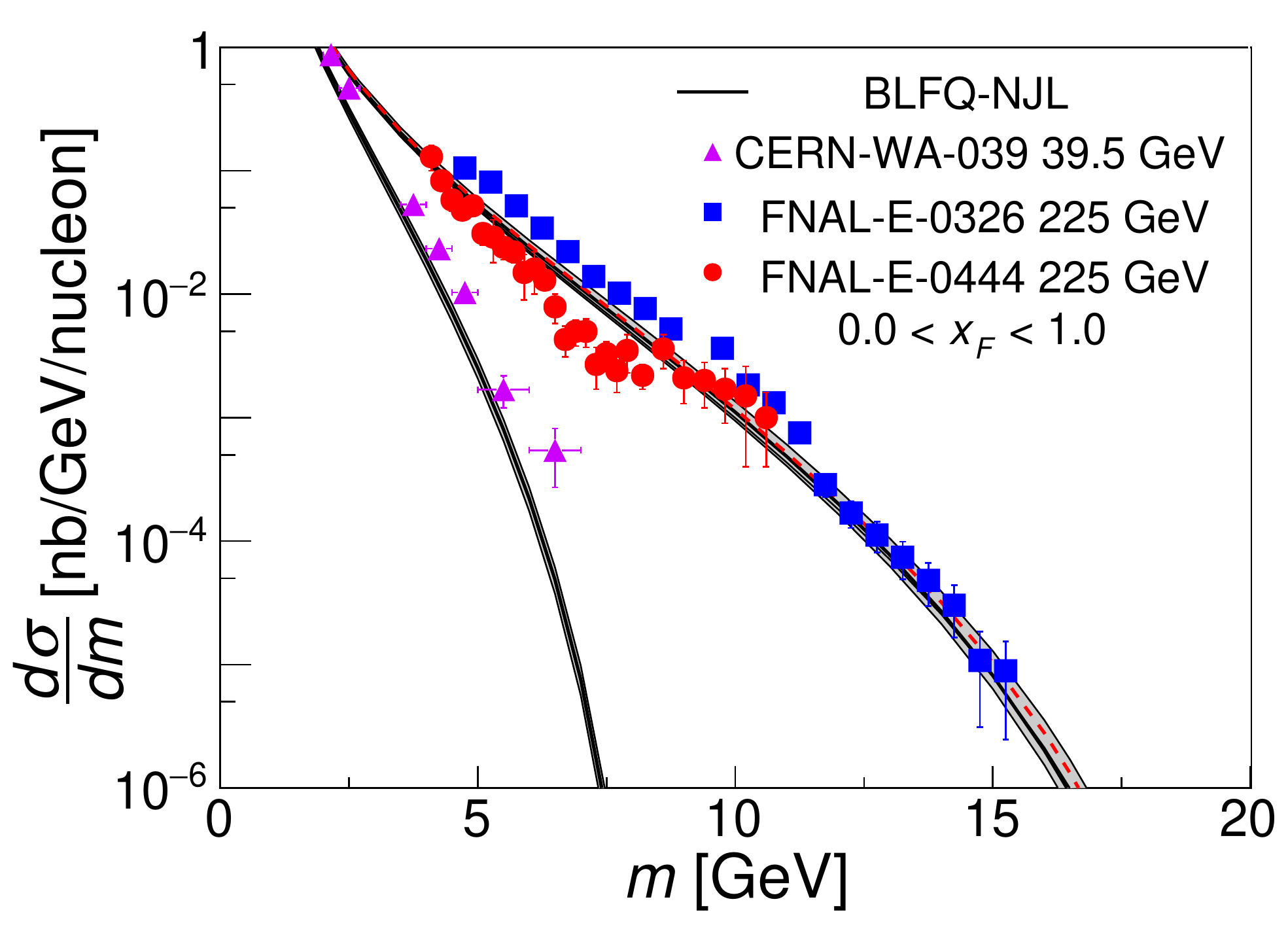}
		\caption{(Color online)~The cross section $d{\bf \sigma}/dm$ for the $\pi^-$-nucleus Drell-Yan process as a function of $m$ in the region  $0<x_{\mathrm{F}}< 1$. The data of the FNAL-E-0326 experiment with 225 GeV pions and FNAL-E-0444 experiment with 225 GeV pions are taken from Ref.~\cite{Greenlee:1985gd} and Ref.~\cite{Anderson:1979tt}, respectively. The same cross section is compared with the data of CERN-WA-039 experiment with 39.5 GeV pions~\cite{Corden:1980xf}. The error-bands in the cross section are results of the present work taking the uncertainty in the initial scale ${\mu_{0\pi}^2=0.240\pm0.024~\rm{GeV}^2}$ into account.	The FNAL-E-0444 data correspond to a carbon target whereas the FNAL-E-0326 and the CERN-WA-039 data correspond to a tungsten target. The black solid and the red dashed lines represent the cross sections evaluated using the tungsten and the carbon nuclear PDFs, respectively.}
		\label{cross3}
	\end{center}
\end{figure*}
With the pion PDFs known over a wide range of scales, we proceed to calculate the pion structure function $F^\pi_2(x_{\pi}=\beta,\mu^2)$ using the parton model. Specifically in the NLO in perturbative QCD, the structure function can be expressed in terms of the PDFs as \cite{Gluck:1991ng,Gluck:1994uf}
\begin{align}
F_{2}^{\pi}(\beta,\mu^{2})=&\sum_{q} e_{q}^{2}~\beta~ \Big\{f^{\pi}_{q}(\beta,\mu^{2})+f^{\pi}_{\bar{q}}(\beta,\mu^{2})+\frac{\alpha_s(\mu^2)}{2\pi}\nonumber\\
& \times \left[C_{q,2}\otimes (f_q^{\pi}+f_{\bar{q}}^{\pi})+2C_{g,2}\otimes f_g^{\pi}\right] \Big\},
\label{pionf2}
\end{align}
with
\begin{align}
&C_{q,2}[z]=\frac{4}{3}\left[ \frac{1+z^2}{1-z}\left(\rm{ln}\frac{1-z}{z}-\frac{3}{4}\right) \right]_+\nonumber,\\
&C_{g,2}[z]=\frac{1}{2}\left[ \left(z^2+(1-z)^2\right)\rm{ln}\frac{1-z}{z} -1 +8z(1-z)  \right]\nonumber,
\end{align}
and
\begin{eqnarray}
C\otimes f^{\pi}=\int_{\beta}^{1}\frac{dy}{y}C\left( \frac{\beta}{y} \right)f^{\pi}(y,\mu^{2})\nonumber
\label{pionf2nlo_int}.
\end{eqnarray}
Here $q$ is the flavor index and $e_{q}$ is the electric charge of the quark flavor $q$ in the units of the elementary charge while $g$ stands for the gluon. Here we have included heavy flavor contributions relevant to the scale of the structure functions. Our results for the structure function $F_2^{\pi}(x_{\pi}=\beta,\mu^{2})$ are shown in Fig.~\ref{F21} and Fig.~\ref{F22} in comparison with DESY-HERA-ZEUS~\cite{Chekanov:2002pf} and DESY-HERA-H1 data~\cite{Aaron:2010ab} at the respective experimental scales. Both the ZEUS and the H1 data were determined from the neutron production in $ep$ collisions, $ep\rightarrow e^\prime X n$ process. The $\beta$ on the horizontal axes in these figures is the parton momentum fraction relative to the pion which is defined as $\beta=x_p/(1-x_{\rm {L}})$, where $x_p$ is the parton momentum fraction relative to the proton. The momentum fraction carried by
the neutron relative to the proton is $x_{\rm {L}}=0.73$ \cite{Chekanov:2002pf,Aaron:2010ab}. The two different sets of ZEUS data in Fig.~\ref{F21} correspond to different pion fluxes used to determine $F_2^\pi$. One of them was obtained using the additive quark model (AQM) whereas the other is obtained using the effective one-pion-exchange flux (EF) in hadron-hadron charge-exchange reactions. The difference between these two results are attributed to the model dependence of the experimental analysis. Despite expecting that both AQM and EF are only valid when $x_{\rm {L}}\rightarrow 1$, our result appears to favor the AQM for $\mu^2\leq 240~\mathrm{GeV}^2$. 

We notice from Fig.~\ref{F21} and Fig.~\ref{F22} that our results deviate from the data at very low $x$. We expect that at low initial scale the DGLAP evolution with leading twist is not sufficient at low $x$~\cite{Zhu:2010qa,Boroun:2010zza}, and one needs to take into account of the higher twist corrections~\cite{Boroun:2013mgv,Boroun:2010zz,Devee:2014fna,Phukan:2017lzp,Rezaei:2010zz,Boroun:2009zzb,Lalung:2017omk}. On the other hand, our $F_2^{\pi}(x,\mu^{2})$ shows better agreement with data as $x$ increases. The component contributions from the valence quarks, sea quarks, and gluons to the total structure function $F_2^\pi$ of the pion at $55$ GeV$^2$ are shown in Fig.~\ref{F23}. We observe that at low $x$ the sea quark contribution dominates. However, at large $x$ the distribution is mostly accounted for by the valence quarks. 

\subsection{Cross section of the unpolarized Drell-Yan process}
\label{SecIII}
In this section we present the cross section of the Drell-Yan process using our BLFQ-NJL model for the pion PDF. The momenta of the incoming hadrons are denoted by $p_{1,2}$. We define $l$ and $l^\prime$ as the momenta of the two outgoing leptons.
The kinematics of the process are described by the invariant mass of the lepton pair $m$, center of mass energy square $s$, rapidity $Y$ or the Feynman
variable $x_{\mathrm{F}}$, and the variable $\tau$, $z$ and $y$. These variables are defined and related to each other by~\cite{Becher:2007ty}
\begin{eqnarray}
&&	s    =      (p_1+p_2)^2, 	\quad\quad\quad\quad  q=l+l^\prime,	\nonumber\\
&&	m^2  =      q^2 ,	\quad\quad\quad\quad\quad \nonumber Y =  \frac12\,\ln\frac{q_0+q_3}{q_0-q_3}\,,\;\;\;\nonumber\\
&& x_\mathrm{F} = x_1 - x_2	 , \quad\quad\quad\quad\quad  \tau = \frac{m^2}{s},	 \nonumber\\
&&z=\frac{m^2}{\hat{s}}=\frac{\tau}{x_1x_2}, \quad y=\frac{\frac{x_1}{x_2}e^{-2Y}-z}{(1-z)(1+\frac{x_1}{x_2}e^{-2Y})} \,,
\label{Eq:DY-kinematics}
\end{eqnarray}
where, $\hat{s}=x_1x_2s$. In the parton model, the $x_i$ denotes the fraction of the hadron momentum $p_i$ carried by the annihilating parton (or antiparton) and is given by
\begin{align}
\label{Eq:x1-x2}
 &     x_{1}= \sqrt{\frac{\tau}{z}\frac{1-(1-y)(1-z)}{1-y(1-z)}}e^{Y}, \nonumber\\
&      x_{2}= \sqrt{\frac{\tau}{z}\frac{1-y(1-z)}{1-(1-y)(1-z)}}e^{-Y}.
\end{align}

Explicitly in the NLO in perturbative QCD, the cross section in terms of the PDFs is given by \cite{Becher:2007ty,Anastasiou:2003yy,Anastasiou:2003ds,Barry:2018ort} 
\begin{align}
&\frac{m^3 d^2{\bf \sigma}}{dm\, dY}=\frac{8\pi\alpha^2}{9 }\frac{m^2}{s}\sum_{ij} \int dx_1 dx_2 \nonumber\\ &\times \widetilde{C}_{ij}(x_1,x_2,s,m,\mu^2) f_{i/\pi}(x_1,\mu^2) f_{j/N}(x_2,\mu^2),
\label{pionNnlo}
\end{align}
where $\widetilde{C}_{ij}$ are the hard-scattering kernels, which can be expanded in powers of the strong coupling $\alpha_s$. The sums extends over all possible partonic channels contributing at a given order in the expansion of $\widetilde{C}_{ij}$. At leading order, only the channels $(ij)=(q\bar q)$ and $(\bar q q)$ contribute, whereas at NLO ($\sim\alpha_s$), we must include $(ij)=(\bar q q), (q\bar q), (gq), (qg), (\bar q g), (g\bar q)$ in the sum. The expressions of the hard-scattering kernels at NLO are given in the appendix.
In order to evaluate Eq.~(\ref{pionNnlo}), we adopt the nuclear PDFs from the nCTEQ 2015~\cite{Kovarik:2015cma} at the experimental scale $\mu^2=16$ GeV$^2$ in conjunction with our pion PDFs at the same scale. While the PDFs for the tungsten and the beryllirum nuclei are readily available in Ref. \cite{Kovarik:2015cma}, we approximate the bound nucleon PDFs in the platinum nucleus by the corresponding bound nucleon PDFs in the gold nucleus.

After integrating out the $Y$ dependence of the differential cross section ${m^3 d^2{\bf \sigma}/dm \,dY}$, we obtain our results plotted as a function of either $\sqrt{\tau}$ in Fig.~\ref{cross1} or $\tau$ in Fig.~\ref{cross2} to compare with the experimental data. The FNAL-E-0615 and the CERN-NA-003 data in Fig.~\ref{cross1}(a) correspond to a tungsten and a platinum targets, respectively, whereas the data in Fig.~\ref{cross1}(b) correspond to a tungsten target. In Fig.~\ref{cross1}(a), we show the results evaluated using the tungsten and the platinum nuclear PDFs. We employ the tungsten  nuclear PDF to compute the cross shown section in Fig.~\ref{cross1}(b). We find that the cross sections per nucleon obtained by considering  the tungsten and the platinum nuclear PDFs are very close. In Fig.~\ref{cross2}, the CERN-WA-011 data represent a beryllium target and the same target nuclear PDF has been used by our approach to evaluate the cross section. In Fig.~\ref{cross3} we show the cross section $d{\bf \sigma}/dm $ as a function of $m$ and compare with the data of the FNAL-E-0326 experiment~\cite{Greenlee:1985gd} and the FNAL-E-0444 experiment~\cite{Anderson:1979tt} with 225 GeV pions, as well as with the data of CERN-WA-039 experiment with 39.5 GeV pions~\cite{Corden:1980xf}. 
Notice the FNAL-E-0444 data correspond to a carbon target whereas the FNAL-E-0326, and the CERN-WA-039 data represent a tungsten target. We use the corresponding target nuclear PDFs to calculate the cross section displayed in  Fig.~\ref{cross3}. Based on Figs.~\ref{cross1}-\ref{cross3}, we find that our results are in acceptable agreement with data from widely different experimental conditions~\cite{Conway:1989fs,Badier:1983mj,Betev:1985pf,Barate:1979da,Corden:1980xf,Greenlee:1985gd,Anderson:1979tt,Stirling:1993gc}. 

\section{Summary and conclusion}\label{SecVI}
We calculated the valence-quark PDFs of the pion and the kaon in the framework of the basis light front quantization from their light front wave functions. These wave functions were obtained as the eigenfunctions of the effective Hamiltonian, consisting of confinement potentials and the color-singlet Nambu--Jona-Lasinio interactions. The parameters in the BLFQ-NJL model were adjusted to reproduce the experimental mass spectrum and the charge radii of the light mesons~\cite{Jia:2018ary}. The initial scales of our PDFs, the only adjustable parameters in this work, have been obtained by consistently fitting both the evolved valence pion PDFs to the FNAL-E-0615 experiment~\cite{Conway:1989fs} and the evolved ratio of the up quark PDFs in the kaon to that in the pion to the CERN-NA-003 experimental result~\cite{Badier:1983mj}. The moments of the pion PDF have been found in agreement with the JAM global fit~\cite{Barry:2018ort}, with lattice QCD~\cite{Martinelli:1987bh,Detmold:2003tm}, as well as with phenomenological quark models~\cite{Sutton:1991ay,Wijesooriya:2005ir,Nam:2012vm,Bednar:2018mtf,Hecht:2000xa,Chen:2016sno,Shi:2018mcb,Ding:2019lwe,Nam:2012vm,Wijesooriya:2005ir} across various scales.

We have subsequently calculated the structure function $F_2^{\pi}(x,Q^2)$ for the pion, the large $x$ behavior of which is consistent with the DESY-HERA experiment~\cite{Chekanov:2002pf,Aaron:2010ab}. However, the discrepancies at small $x$ for the structure function suggest the need to include the higher-twist corrections and a non-vanishing initial gluon distribution to the DGLAP evolution. We have also studied the cross sections of the pion-nucleus induced Drell-Yan process in comparison with Refs.~\cite{Conway:1989fs,Badier:1983mj,Stirling:1993gc,Barate:1979da,Greenlee:1985gd,Anderson:1979tt,Corden:1980xf}, finding reasonable agreement with these various experimental data. 
These comparisons affirm the robustness of the BLFQ-NJL model with QCD evolution as a theoretical method to describe the structures of the pion and the kaon in the language of parton distribution functions.
\\
\\

\begin{acknowledgments}
We thank Y. Li, S. Xu, J. Huston, W. Zhu, J. Gao, X. Chen, T. Liu, and R. Wang for many insightful discussions. C. M. is supported by the National Natural Science Foundation of China (NSFC) under the grants No. 11850410436 and No. 11950410753. X. Z. is supported by  Key Research Program of Frontier Sciences, CAS, Grant No. ZDBS-LY-7020.  S. J. and J. P. V. are supported  by the Department of Energy under grants No.~DE-FG02-87ER40371 and No.~DE-SC0018223 (SciDAC4/NUCLEI).
\end{acknowledgments}
\appendix
\section{ }
The expressions of $\widetilde{C}_{ij}(x_1,x_2,s,m^2,\mu^2)$ are given by~\cite{Anastasiou:2003yy} 
\begin{widetext} 
\begin{eqnarray} 
\widetilde{C}_{ij}(x_1,x_2,s,m^2,\mu^2)=\left\vert\frac{dzdy}{dx_1dx_2}\right\vert\frac{C_{ij}(z,y,m^2,\mu^2)}{[1-y(1-z)][1-(1-y)(1-z)]},
\end{eqnarray} 
\begin{align}
\frac{C_{q\bar{q}}}{e_q^2}=&\delta(1-z)\frac{\delta(y)+\delta(1-y)}{2}\Big[1+\frac{4\alpha_s(\mu^2)}{3\pi}\Big(\frac{3}{2}{\rm ln}\frac{m^2}{\mu^2}+\frac{2\pi^2}{3}-6 \Big) \Big]\nonumber\\
&+\frac{4\alpha_s(\mu^2)}{3\pi}\Big\{ \frac{\delta(y)+\delta(1-y)}{2}\Big[(1+z^2)\Big[\frac{1}{1-z}{\rm ln}\frac{m^2(1-z)^2}{\mu^2z}  \Big]_+ +1-z  \Big]  \nonumber\\
&\quad +\frac{1}{2}\Big[1+\frac{(1-z)^2}{z}y(1-y)\Big]\Big[\frac{1+z^2}{1-z}\left(\Big[\frac{1}{y}\Big]_+ +\Big[\frac{1}{1-y}\Big]_+ \right)-2(1-z) \Big]\Big\},\nonumber\\
\frac{C_{qg}}{e_q^2}=&\frac{\alpha_s(\mu^2)}{4\pi}\Big\{ \delta(y)\Big[\left(z^2+(1-z)^2\right){\rm ln}\frac{m^2(1-z)^2}{\mu^2z}  +2z(1-z)  \Big]  \nonumber\\
& +\Big[1+\frac{(1-z)^2}{z}y(1-y)\Big]\Big[(z^2+(1-z)^2)\Big[\frac{1}{y}\Big]_+ +2z(1-z)+(1-z)^2 y\Big]\Big\},
\end{align}
\begin{align}
C_{\bar{q}q}=C_{q\bar{q}},\quad C_{\bar{q}g}=C_{qg},\quad C_{gq}=C_{g\bar{q}}=C_{qg}\big\vert_{y\rightarrow1-y},
\end{align}
respectively, where the plus prescription is defined as
\begin{align}
\int dt f(t)\Big[\frac{1}{x-t}\Big]_+ =\int dt \Big(f(t)-f(x)\Big)\Big[\frac{1}{x-t}\Big].
\end{align}
\end{widetext} 
\bibliography{BLFQ_NJL_fulltex_bib.bib}

\begin{thebibliography}{106}%
\makeatletter
\providecommand \@ifxundefined [1]{%
 \@ifx{#1\undefined}
}%
\providecommand \@ifnum [1]{%
 \ifnum #1\expandafter \@firstoftwo
 \else \expandafter \@secondoftwo
 \fi
}%
\providecommand \@ifx [1]{%
 \ifx #1\expandafter \@firstoftwo
 \else \expandafter \@secondoftwo
 \fi
}%
\providecommand \natexlab [1]{#1}%
\providecommand \enquote  [1]{``#1''}%
\providecommand \bibnamefont  [1]{#1}%
\providecommand \bibfnamefont [1]{#1}%
\providecommand \citenamefont [1]{#1}%
\providecommand \href@noop [0]{\@secondoftwo}%
\providecommand \href [0]{\begingroup \@sanitize@url \@href}%
\providecommand \@href[1]{\@@startlink{#1}\@@href}%
\providecommand \@@href[1]{\endgroup#1\@@endlink}%
\providecommand \@sanitize@url [0]{\catcode `\\12\catcode `\$12\catcode
  `\&12\catcode `\#12\catcode `\^12\catcode `\_12\catcode `\%12\relax}%
\providecommand \@@startlink[1]{}%
\providecommand \@@endlink[0]{}%
\providecommand \url  [0]{\begingroup\@sanitize@url \@url }%
\providecommand \@url [1]{\endgroup\@href {#1}{\urlprefix }}%
\providecommand \urlprefix  [0]{URL }%
\providecommand \Eprint [0]{\href }%
\providecommand \doibase [0]{http://dx.doi.org/}%
\providecommand \selectlanguage [0]{\@gobble}%
\providecommand \bibinfo  [0]{\@secondoftwo}%
\providecommand \bibfield  [0]{\@secondoftwo}%
\providecommand \translation [1]{[#1]}%
\providecommand \BibitemOpen [0]{}%
\providecommand \bibitemStop [0]{}%
\providecommand \bibitemNoStop [0]{.\EOS\space}%
\providecommand \EOS [0]{\spacefactor3000\relax}%
\providecommand \BibitemShut  [1]{\csname bibitem#1\endcsname}%
\let\auto@bib@innerbib\@empty
\bibitem [{\citenamefont {{NA10 Collaboration}}\ \emph
  {et~al.}(1987)\citenamefont {{NA10 Collaboration}}, \citenamefont {Bordalo},
  \citenamefont {Busson}, \citenamefont {Kluberg}, \citenamefont {Romana},
  \citenamefont {Salmeron}, \citenamefont {Vall\'{e}e}, \citenamefont
  {Blaising}, \citenamefont {Degr\'{e}}, \citenamefont {Juillot}, \citenamefont
  {Morand} \emph {et~al.}}]{Bordalo:1987cs}%
  \BibitemOpen
  \bibfield  {author} {\bibinfo {author} {\bibnamefont {{NA10 Collaboration}}},
  \bibinfo {author} {\bibfnamefont {P.}~\bibnamefont {Bordalo}}, \bibinfo
  {author} {\bibfnamefont {P.}~\bibnamefont {Busson}}, \bibinfo {author}
  {\bibfnamefont {L.}~\bibnamefont {Kluberg}}, \bibinfo {author} {\bibfnamefont
  {A.}~\bibnamefont {Romana}}, \bibinfo {author} {\bibfnamefont
  {R.}~\bibnamefont {Salmeron}}, \bibinfo {author} {\bibfnamefont
  {C.}~\bibnamefont {Vall\'{e}e}}, \bibinfo {author} {\bibfnamefont
  {J.}~\bibnamefont {Blaising}}, \bibinfo {author} {\bibfnamefont
  {A.}~\bibnamefont {Degr\'{e}}}, \bibinfo {author} {\bibfnamefont
  {P.}~\bibnamefont {Juillot}}, \bibinfo {author} {\bibfnamefont
  {R.}~\bibnamefont {Morand}},  \emph {et~al.},\ }\href {\doibase
  https://doi.org/10.1016/0370-2693(87)91253-6} {\bibfield  {journal} {\bibinfo
   {journal} {Phys. Lett. B}\ }\textbf {\bibinfo {volume} {193}},\ \bibinfo
  {pages} {368 } (\bibinfo {year} {1987})}\BibitemShut {NoStop}%
\bibitem [{\citenamefont {Freudenreich}(1990)}]{Freudenreich:1990mu}%
  \BibitemOpen
  \bibfield  {author} {\bibinfo {author} {\bibfnamefont {K.}~\bibnamefont
  {Freudenreich}},\ }\href {\doibase 10.1142/S0217751X90001586} {\bibfield
  {journal} {\bibinfo  {journal} {Int. J. Mod. Phys. A}\ }\textbf {\bibinfo
  {volume} {05}},\ \bibinfo {pages} {3643} (\bibinfo {year}
  {1990})}\BibitemShut {NoStop}%
\bibitem [{\citenamefont {Sutton}\ \emph {et~al.}(1992)\citenamefont {Sutton},
  \citenamefont {Martin}, \citenamefont {Roberts},\ and\ \citenamefont
  {Stirling}}]{Sutton:1991ay}%
  \BibitemOpen
  \bibfield  {author} {\bibinfo {author} {\bibfnamefont {P.~J.}\ \bibnamefont
  {Sutton}}, \bibinfo {author} {\bibfnamefont {A.~D.}\ \bibnamefont {Martin}},
  \bibinfo {author} {\bibfnamefont {R.~G.}\ \bibnamefont {Roberts}}, \ and\
  \bibinfo {author} {\bibfnamefont {W.~J.}\ \bibnamefont {Stirling}},\ }\href
  {\doibase 10.1103/PhysRevD.45.2349} {\bibfield  {journal} {\bibinfo
  {journal} {Phys. Rev. D}\ }\textbf {\bibinfo {volume} {45}},\ \bibinfo
  {pages} {2349} (\bibinfo {year} {1992})}\BibitemShut {NoStop}%
\bibitem [{\citenamefont {Gl{\"u}ck}\ \emph {et~al.}(1999)\citenamefont
  {Gl{\"u}ck}, \citenamefont {Reya},\ and\ \citenamefont
  {Schienbein}}]{Gluck:1999xe}%
  \BibitemOpen
  \bibfield  {author} {\bibinfo {author} {\bibfnamefont {M.}~\bibnamefont
  {Gl{\"u}ck}}, \bibinfo {author} {\bibfnamefont {E.}~\bibnamefont {Reya}}, \
  and\ \bibinfo {author} {\bibfnamefont {I.}~\bibnamefont {Schienbein}},\
  }\href {\doibase 10.1007/s100529900124} {\bibfield  {journal} {\bibinfo
  {journal} {Eur. Phys. J. C}\ }\textbf {\bibinfo {volume} {10}},\ \bibinfo
  {pages} {313} (\bibinfo {year} {1999})}\BibitemShut {NoStop}%
\bibitem [{\citenamefont {Wijesooriya}\ \emph {et~al.}(2005)\citenamefont
  {Wijesooriya}, \citenamefont {Reimer},\ and\ \citenamefont
  {Holt}}]{Wijesooriya:2005ir}%
  \BibitemOpen
  \bibfield  {author} {\bibinfo {author} {\bibfnamefont {K.}~\bibnamefont
  {Wijesooriya}}, \bibinfo {author} {\bibfnamefont {P.~E.}\ \bibnamefont
  {Reimer}}, \ and\ \bibinfo {author} {\bibfnamefont {R.~J.}\ \bibnamefont
  {Holt}},\ }\href {\doibase 10.1103/PhysRevC.72.065203} {\bibfield  {journal}
  {\bibinfo  {journal} {Phys. Rev. C}\ }\textbf {\bibinfo {volume} {72}},\
  \bibinfo {pages} {065203} (\bibinfo {year} {2005})}\BibitemShut {NoStop}%
\bibitem [{\citenamefont {Conway}\ \emph {et~al.}(1989)\citenamefont {Conway},
  \citenamefont {Adolphsen}, \citenamefont {Alexander}, \citenamefont
  {Anderson}, \citenamefont {Heinrich}, \citenamefont {Pilcher}, \citenamefont
  {Possoz}, \citenamefont {Rosenberg}, \citenamefont {Biino}, \citenamefont
  {Greenhalgh} \emph {et~al.}}]{Conway:1989fs}%
  \BibitemOpen
  \bibfield  {author} {\bibinfo {author} {\bibfnamefont {J.~S.}\ \bibnamefont
  {Conway}}, \bibinfo {author} {\bibfnamefont {C.~E.}\ \bibnamefont
  {Adolphsen}}, \bibinfo {author} {\bibfnamefont {J.~P.}\ \bibnamefont
  {Alexander}}, \bibinfo {author} {\bibfnamefont {K.~J.}\ \bibnamefont
  {Anderson}}, \bibinfo {author} {\bibfnamefont {J.~G.}\ \bibnamefont
  {Heinrich}}, \bibinfo {author} {\bibfnamefont {J.~E.}\ \bibnamefont
  {Pilcher}}, \bibinfo {author} {\bibfnamefont {A.}~\bibnamefont {Possoz}},
  \bibinfo {author} {\bibfnamefont {E.~I.}\ \bibnamefont {Rosenberg}}, \bibinfo
  {author} {\bibfnamefont {C.}~\bibnamefont {Biino}}, \bibinfo {author}
  {\bibfnamefont {J.~F.}\ \bibnamefont {Greenhalgh}},  \emph {et~al.},\ }\href
  {\doibase 10.1103/PhysRevD.39.92} {\bibfield  {journal} {\bibinfo  {journal}
  {Phys. Rev. D}\ }\textbf {\bibinfo {volume} {39}},\ \bibinfo {pages} {92}
  (\bibinfo {year} {1989})}\BibitemShut {NoStop}%
\bibitem [{\citenamefont {{NA3 Collaboration}}\ \emph
  {et~al.}(1983)\citenamefont {{NA3 Collaboration}}, \citenamefont {Badier},
  \citenamefont {Boucrot}, \citenamefont {Bourotte}, \citenamefont {Burgun},
  \citenamefont {Callot}, \citenamefont {Charpentier}, \citenamefont {Crozon},
  \citenamefont {Decamp}, \citenamefont {Delpierre}, \citenamefont {Gandois}
  \emph {et~al.}}]{Badier:1983mj}%
  \BibitemOpen
  \bibfield  {author} {\bibinfo {author} {\bibnamefont {{NA3 Collaboration}}},
  \bibinfo {author} {\bibfnamefont {J.}~\bibnamefont {Badier}}, \bibinfo
  {author} {\bibfnamefont {J.}~\bibnamefont {Boucrot}}, \bibinfo {author}
  {\bibfnamefont {J.}~\bibnamefont {Bourotte}}, \bibinfo {author}
  {\bibfnamefont {G.}~\bibnamefont {Burgun}}, \bibinfo {author} {\bibfnamefont
  {O.}~\bibnamefont {Callot}}, \bibinfo {author} {\bibfnamefont
  {P.}~\bibnamefont {Charpentier}}, \bibinfo {author} {\bibfnamefont
  {M.}~\bibnamefont {Crozon}}, \bibinfo {author} {\bibfnamefont
  {D.}~\bibnamefont {Decamp}}, \bibinfo {author} {\bibfnamefont
  {P.}~\bibnamefont {Delpierre}}, \bibinfo {author} {\bibfnamefont
  {B.}~\bibnamefont {Gandois}},  \emph {et~al.},\ }\href {\doibase
  10.1007/BF01573728} {\bibfield  {journal} {\bibinfo  {journal} {Zeitschrift
  f{\"u}r Physik C Particles and Fields}\ }\textbf {\bibinfo {volume} {18}},\
  \bibinfo {pages} {281} (\bibinfo {year} {1983})}\BibitemShut {NoStop}%
\bibitem [{\citenamefont {Aicher}\ \emph {et~al.}(2010)\citenamefont {Aicher},
  \citenamefont {Sch\"afer},\ and\ \citenamefont {Vogelsang}}]{Aicher:2010cb}%
  \BibitemOpen
  \bibfield  {author} {\bibinfo {author} {\bibfnamefont {M.}~\bibnamefont
  {Aicher}}, \bibinfo {author} {\bibfnamefont {A.}~\bibnamefont {Sch\"afer}}, \
  and\ \bibinfo {author} {\bibfnamefont {W.}~\bibnamefont {Vogelsang}},\ }\href
  {\doibase 10.1103/PhysRevLett.105.252003} {\bibfield  {journal} {\bibinfo
  {journal} {Phys. Rev. Lett.}\ }\textbf {\bibinfo {volume} {105}},\ \bibinfo
  {pages} {252003} (\bibinfo {year} {2010})}\BibitemShut {NoStop}%
\bibitem [{\citenamefont {Watanabe}\ \emph
  {et~al.}(2018{\natexlab{a}})\citenamefont {Watanabe}, \citenamefont
  {Sawada},\ and\ \citenamefont {Kao}}]{Watanabe:2017pvl}%
  \BibitemOpen
  \bibfield  {author} {\bibinfo {author} {\bibfnamefont {A.}~\bibnamefont
  {Watanabe}}, \bibinfo {author} {\bibfnamefont {T.}~\bibnamefont {Sawada}}, \
  and\ \bibinfo {author} {\bibfnamefont {C.~W.}\ \bibnamefont {Kao}},\ }\href
  {\doibase 10.1103/PhysRevD.97.074015} {\bibfield  {journal} {\bibinfo
  {journal} {Phys. Rev. D}\ }\textbf {\bibinfo {volume} {97}},\ \bibinfo
  {pages} {074015} (\bibinfo {year} {2018}{\natexlab{a}})}\BibitemShut
  {NoStop}%
\bibitem [{\citenamefont {Hecht}\ \emph {et~al.}(2001)\citenamefont {Hecht},
  \citenamefont {Roberts},\ and\ \citenamefont {Schmidt}}]{Hecht:2000xa}%
  \BibitemOpen
  \bibfield  {author} {\bibinfo {author} {\bibfnamefont {M.~B.}\ \bibnamefont
  {Hecht}}, \bibinfo {author} {\bibfnamefont {C.~D.}\ \bibnamefont {Roberts}},
  \ and\ \bibinfo {author} {\bibfnamefont {S.~M.}\ \bibnamefont {Schmidt}},\
  }\href {\doibase 10.1103/PhysRevC.63.025213} {\bibfield  {journal} {\bibinfo
  {journal} {Phys. Rev. C}\ }\textbf {\bibinfo {volume} {63}},\ \bibinfo
  {pages} {025213} (\bibinfo {year} {2001})}\BibitemShut {NoStop}%
\bibitem [{\citenamefont {Nam}(2012)}]{Nam:2012vm}%
  \BibitemOpen
  \bibfield  {author} {\bibinfo {author} {\bibfnamefont {S.-i.}\ \bibnamefont
  {Nam}},\ }\href {\doibase 10.1103/PhysRevD.86.074005} {\bibfield  {journal}
  {\bibinfo  {journal} {Phys. Rev. D}\ }\textbf {\bibinfo {volume} {86}},\
  \bibinfo {pages} {074005} (\bibinfo {year} {2012})}\BibitemShut {NoStop}%
\bibitem [{\citenamefont {Detmold}\ \emph {et~al.}(2003)\citenamefont
  {Detmold}, \citenamefont {Melnitchouk},\ and\ \citenamefont
  {Thomas}}]{Detmold:2003tm}%
  \BibitemOpen
  \bibfield  {author} {\bibinfo {author} {\bibfnamefont {W.}~\bibnamefont
  {Detmold}}, \bibinfo {author} {\bibfnamefont {W.}~\bibnamefont
  {Melnitchouk}}, \ and\ \bibinfo {author} {\bibfnamefont {A.~W.}\ \bibnamefont
  {Thomas}},\ }\href {\doibase 10.1103/PhysRevD.68.034025} {\bibfield
  {journal} {\bibinfo  {journal} {Phys. Rev. D}\ }\textbf {\bibinfo {volume}
  {68}},\ \bibinfo {pages} {034025} (\bibinfo {year} {2003})}\BibitemShut
  {NoStop}%
\bibitem [{\citenamefont {Holt}\ and\ \citenamefont
  {Roberts}(2010)}]{Holt:2010vj}%
  \BibitemOpen
  \bibfield  {author} {\bibinfo {author} {\bibfnamefont {R.~J.}\ \bibnamefont
  {Holt}}\ and\ \bibinfo {author} {\bibfnamefont {C.~D.}\ \bibnamefont
  {Roberts}},\ }\href {\doibase 10.1103/RevModPhys.82.2991} {\bibfield
  {journal} {\bibinfo  {journal} {Rev. Mod. Phys.}\ }\textbf {\bibinfo {volume}
  {82}},\ \bibinfo {pages} {2991} (\bibinfo {year} {2010})}\BibitemShut
  {NoStop}%
\bibitem [{\citenamefont {Pumplin}\ \emph {et~al.}(2002)\citenamefont
  {Pumplin}, \citenamefont {Stump}, \citenamefont {Huston}, \citenamefont
  {Lai}, \citenamefont {Nadolsky},\ and\ \citenamefont
  {Tung}}]{Pumplin:2002vw}%
  \BibitemOpen
  \bibfield  {author} {\bibinfo {author} {\bibfnamefont {J.}~\bibnamefont
  {Pumplin}}, \bibinfo {author} {\bibfnamefont {D.~R.}\ \bibnamefont {Stump}},
  \bibinfo {author} {\bibfnamefont {J.}~\bibnamefont {Huston}}, \bibinfo
  {author} {\bibfnamefont {H.-L.}\ \bibnamefont {Lai}}, \bibinfo {author}
  {\bibfnamefont {P.}~\bibnamefont {Nadolsky}}, \ and\ \bibinfo {author}
  {\bibfnamefont {W.-K.}\ \bibnamefont {Tung}},\ }\href {\doibase
  10.1088/1126-6708/2002/07/012} {\bibfield  {journal} {\bibinfo  {journal} {J.
  High Energy Phys.}\ }\textbf {\bibinfo {volume} {2002}},\ \bibinfo {pages}
  {012} (\bibinfo {year} {2002})}\BibitemShut {NoStop}%
\bibitem [{\citenamefont {Ball}\ \emph {et~al.}(2017)\citenamefont {Ball},
  \citenamefont {Bertone}, \citenamefont {Carrazza}, \citenamefont {Debbio},
  \citenamefont {Forte}, \citenamefont {Groth-Merrild}, \citenamefont
  {Guffanti}, \citenamefont {Hartland}, \citenamefont {Kassabov}, \citenamefont
  {Latorre} \emph {et~al.}}]{Ball:2017nwa}%
  \BibitemOpen
  \bibfield  {author} {\bibinfo {author} {\bibfnamefont {R.~D.}\ \bibnamefont
  {Ball}}, \bibinfo {author} {\bibfnamefont {V.}~\bibnamefont {Bertone}},
  \bibinfo {author} {\bibfnamefont {S.}~\bibnamefont {Carrazza}}, \bibinfo
  {author} {\bibfnamefont {L.~D.}\ \bibnamefont {Debbio}}, \bibinfo {author}
  {\bibfnamefont {S.}~\bibnamefont {Forte}}, \bibinfo {author} {\bibfnamefont
  {P.}~\bibnamefont {Groth-Merrild}}, \bibinfo {author} {\bibfnamefont
  {A.}~\bibnamefont {Guffanti}}, \bibinfo {author} {\bibfnamefont {N.~P.}\
  \bibnamefont {Hartland}}, \bibinfo {author} {\bibfnamefont {Z.}~\bibnamefont
  {Kassabov}}, \bibinfo {author} {\bibfnamefont {J.~I.}\ \bibnamefont
  {Latorre}},  \emph {et~al.},\ }\href {\doibase
  10.1140/epjc/s10052-017-5199-5} {\bibfield  {journal} {\bibinfo  {journal}
  {Eur. Phys. J. C}\ }\textbf {\bibinfo {volume} {77}},\ \bibinfo {pages} {663}
  (\bibinfo {year} {2017})}\BibitemShut {NoStop}%
\bibitem [{\citenamefont {Alekhin}\ \emph {et~al.}(2017)\citenamefont
  {Alekhin}, \citenamefont {Bl\"umlein}, \citenamefont {Moch},\ and\
  \citenamefont {Pla\ifmmode \check{c}\else \v{c}\fi{}akyt\ifmmode~\dot{e}\else
  \.{e}\fi{}}}]{Alekhin:2017kpj}%
  \BibitemOpen
  \bibfield  {author} {\bibinfo {author} {\bibfnamefont {S.}~\bibnamefont
  {Alekhin}}, \bibinfo {author} {\bibfnamefont {J.}~\bibnamefont {Bl\"umlein}},
  \bibinfo {author} {\bibfnamefont {S.}~\bibnamefont {Moch}}, \ and\ \bibinfo
  {author} {\bibfnamefont {R.}~\bibnamefont {Pla\ifmmode \check{c}\else
  \v{c}\fi{}akyt\ifmmode~\dot{e}\else \.{e}\fi{}}},\ }\href {\doibase
  10.1103/PhysRevD.96.014011} {\bibfield  {journal} {\bibinfo  {journal} {Phys.
  Rev. D}\ }\textbf {\bibinfo {volume} {96}},\ \bibinfo {pages} {014011}
  (\bibinfo {year} {2017})}\BibitemShut {NoStop}%
\bibitem [{\citenamefont {Dulat}\ \emph {et~al.}(2016)\citenamefont {Dulat},
  \citenamefont {Hou}, \citenamefont {Gao}, \citenamefont {Guzzi},
  \citenamefont {Huston}, \citenamefont {Nadolsky}, \citenamefont {Pumplin},
  \citenamefont {Schmidt}, \citenamefont {Stump},\ and\ \citenamefont
  {Yuan}}]{Dulat:2015mca}%
  \BibitemOpen
  \bibfield  {author} {\bibinfo {author} {\bibfnamefont {S.}~\bibnamefont
  {Dulat}}, \bibinfo {author} {\bibfnamefont {T.-J.}\ \bibnamefont {Hou}},
  \bibinfo {author} {\bibfnamefont {J.}~\bibnamefont {Gao}}, \bibinfo {author}
  {\bibfnamefont {M.}~\bibnamefont {Guzzi}}, \bibinfo {author} {\bibfnamefont
  {J.}~\bibnamefont {Huston}}, \bibinfo {author} {\bibfnamefont
  {P.}~\bibnamefont {Nadolsky}}, \bibinfo {author} {\bibfnamefont
  {J.}~\bibnamefont {Pumplin}}, \bibinfo {author} {\bibfnamefont
  {C.}~\bibnamefont {Schmidt}}, \bibinfo {author} {\bibfnamefont
  {D.}~\bibnamefont {Stump}}, \ and\ \bibinfo {author} {\bibfnamefont {C.-P.}\
  \bibnamefont {Yuan}},\ }\href {\doibase 10.1103/PhysRevD.93.033006}
  {\bibfield  {journal} {\bibinfo  {journal} {Phys. Rev. D}\ }\textbf {\bibinfo
  {volume} {93}},\ \bibinfo {pages} {033006} (\bibinfo {year}
  {2016})}\BibitemShut {NoStop}%
\bibitem [{\citenamefont {Harland-Lang}\ \emph {et~al.}(2015)\citenamefont
  {Harland-Lang}, \citenamefont {Martin}, \citenamefont {Motylinski},\ and\
  \citenamefont {Thorne}}]{Harland-Lang:2014zoa}%
  \BibitemOpen
  \bibfield  {author} {\bibinfo {author} {\bibfnamefont {L.~A.}\ \bibnamefont
  {Harland-Lang}}, \bibinfo {author} {\bibfnamefont {A.~D.}\ \bibnamefont
  {Martin}}, \bibinfo {author} {\bibfnamefont {P.}~\bibnamefont {Motylinski}},
  \ and\ \bibinfo {author} {\bibfnamefont {R.~S.}\ \bibnamefont {Thorne}},\
  }\href {\doibase 10.1140/epjc/s10052-015-3397-6} {\bibfield  {journal}
  {\bibinfo  {journal} {Eur. Phys. J. C}\ }\textbf {\bibinfo {volume} {75}},\
  \bibinfo {pages} {204} (\bibinfo {year} {2015})}\BibitemShut {NoStop}%
\bibitem [{\citenamefont {Bednar}\ \emph {et~al.}(2018)\citenamefont {Bednar},
  \citenamefont {Cloët},\ and\ \citenamefont {Tandy}}]{Bednar:2018mtf}%
  \BibitemOpen
  \bibfield  {author} {\bibinfo {author} {\bibfnamefont {K.~D.}\ \bibnamefont
  {Bednar}}, \bibinfo {author} {\bibfnamefont {I.~C.}\ \bibnamefont {Cloët}},
  \ and\ \bibinfo {author} {\bibfnamefont {P.~C.}\ \bibnamefont {Tandy}},\
  }\href@noop {} {\  (\bibinfo {year} {2018})},\ \Eprint
  {http://arxiv.org/abs/1811.12310} {arXiv:1811.12310 [nucl-th]} \BibitemShut
  {NoStop}%
\bibitem [{\citenamefont {Thomas}(2007)}]{Thomas:2007bc}%
  \BibitemOpen
  \bibfield  {author} {\bibinfo {author} {\bibfnamefont {A.~W.}\ \bibnamefont
  {Thomas}},\ }\bibfield  {booktitle} {\emph {\bibinfo {booktitle}
  {{Proceedings, Yukawa International Seminar on New Frontiers in QCD - Exotic
  Hadrons and Hadronic Matter (YKIS2006): Kyoto, Japan, November 20-December 8,
  2006}}},\ }\href {\doibase 10.1143/PTPS.168.614} {\bibfield  {journal}
  {\bibinfo  {journal} {Prog. Theor. Phys.}\ }\textbf {\bibinfo {volume}
  {168}},\ \bibinfo {pages} {614} (\bibinfo {year} {2007})},\ \Eprint
  {http://arxiv.org/abs/0711.2259} {arXiv:0711.2259 [nucl-th]} \BibitemShut
  {NoStop}%
\bibitem [{\citenamefont {Theberge}\ \emph {et~al.}(1980)\citenamefont
  {Theberge}, \citenamefont {Thomas},\ and\ \citenamefont
  {Miller}}]{Theberge:1980ye}%
  \BibitemOpen
  \bibfield  {author} {\bibinfo {author} {\bibfnamefont {S.}~\bibnamefont
  {Theberge}}, \bibinfo {author} {\bibfnamefont {A.~W.}\ \bibnamefont
  {Thomas}}, \ and\ \bibinfo {author} {\bibfnamefont {G.~A.}\ \bibnamefont
  {Miller}},\ }\href {\doibase 10.1103/physrevd.23.2106.2,
  10.1103/PhysRevD.22.2838} {\bibfield  {journal} {\bibinfo  {journal} {Phys.
  Rev.}\ }\textbf {\bibinfo {volume} {D22}},\ \bibinfo {pages} {2838} (\bibinfo
  {year} {1980})},\ \bibinfo {note} {[Erratum: Phys.
  Rev.D23,2106(1981)]}\BibitemShut {NoStop}%
\bibitem [{\citenamefont {Thomas}(1984)}]{Thomas:1982kv}%
  \BibitemOpen
  \bibfield  {author} {\bibinfo {author} {\bibfnamefont {A.~W.}\ \bibnamefont
  {Thomas}},\ }\href {\doibase 10.1007/978-1-4613-9892-9_1} {\bibfield
  {journal} {\bibinfo  {journal} {Advances in Nuclear Physics}\ }\textbf
  {\bibinfo {volume} {13}},\ \bibinfo {pages} {1} (\bibinfo {year}
  {1984})}\BibitemShut {NoStop}%
\bibitem [{\citenamefont {Woods}\ \emph {et~al.}(1988)\citenamefont {Woods},
  \citenamefont {Nishikawa}, \citenamefont {Patterson}, \citenamefont {Wah},
  \citenamefont {Winstein}, \citenamefont {Winston}, \citenamefont {Yamamoto},
  \citenamefont {Swallow}, \citenamefont {Bock}, \citenamefont {Coleman} \emph
  {et~al.}}]{Woods:1988za}%
  \BibitemOpen
  \bibfield  {author} {\bibinfo {author} {\bibfnamefont {M.}~\bibnamefont
  {Woods}}, \bibinfo {author} {\bibfnamefont {K.}~\bibnamefont {Nishikawa}},
  \bibinfo {author} {\bibfnamefont {J.}~\bibnamefont {Patterson}}, \bibinfo
  {author} {\bibfnamefont {Y.}~\bibnamefont {Wah}}, \bibinfo {author}
  {\bibfnamefont {B.}~\bibnamefont {Winstein}}, \bibinfo {author}
  {\bibfnamefont {R.}~\bibnamefont {Winston}}, \bibinfo {author} {\bibfnamefont
  {H.}~\bibnamefont {Yamamoto}}, \bibinfo {author} {\bibfnamefont
  {E.}~\bibnamefont {Swallow}}, \bibinfo {author} {\bibfnamefont
  {G.}~\bibnamefont {Bock}}, \bibinfo {author} {\bibfnamefont {R.}~\bibnamefont
  {Coleman}},  \emph {et~al.},\ }\href {\doibase 10.1103/PhysRevLett.60.1695}
  {\bibfield  {journal} {\bibinfo  {journal} {Phys. Rev. Lett.}\ }\textbf
  {\bibinfo {volume} {60}},\ \bibinfo {pages} {1695} (\bibinfo {year}
  {1988})}\BibitemShut {NoStop}%
\bibitem [{\citenamefont {{NA31 Collaboration}}\ \emph
  {et~al.}(1993)\citenamefont {{NA31 Collaboration}}, \citenamefont {Barr},
  \citenamefont {Barr}, \citenamefont {Buchholz}, \citenamefont {Carosi},
  \citenamefont {Coward}, \citenamefont {Cundy}, \citenamefont {Doble},
  \citenamefont {Gatignon}, \citenamefont {Gibson}, \citenamefont {Grafstrom}
  \emph {et~al.}}]{Barr:1993rx}%
  \BibitemOpen
  \bibfield  {author} {\bibinfo {author} {\bibnamefont {{NA31 Collaboration}}},
  \bibinfo {author} {\bibfnamefont {G.~D.}\ \bibnamefont {Barr}}, \bibinfo
  {author} {\bibfnamefont {G.}~\bibnamefont {Barr}}, \bibinfo {author}
  {\bibfnamefont {P.}~\bibnamefont {Buchholz}}, \bibinfo {author}
  {\bibfnamefont {R.}~\bibnamefont {Carosi}}, \bibinfo {author} {\bibfnamefont
  {D.}~\bibnamefont {Coward}}, \bibinfo {author} {\bibfnamefont
  {D.}~\bibnamefont {Cundy}}, \bibinfo {author} {\bibfnamefont
  {N.}~\bibnamefont {Doble}}, \bibinfo {author} {\bibfnamefont
  {L.}~\bibnamefont {Gatignon}}, \bibinfo {author} {\bibfnamefont
  {V.}~\bibnamefont {Gibson}}, \bibinfo {author} {\bibfnamefont
  {P.}~\bibnamefont {Grafstrom}},  \emph {et~al.},\ }\href {\doibase
  10.1016/0370-2693(93)91599-I} {\bibfield  {journal} {\bibinfo  {journal}
  {Phys. Lett.}\ }\textbf {\bibinfo {volume} {B317}},\ \bibinfo {pages} {233}
  (\bibinfo {year} {1993})}\BibitemShut {NoStop}%
\bibitem [{\citenamefont {Gibbons}\ \emph {et~al.}(1993)\citenamefont
  {Gibbons}, \citenamefont {Barker}, \citenamefont {Briere}, \citenamefont
  {Makoff}, \citenamefont {Papadimitriou}, \citenamefont {Patterson},
  \citenamefont {Schwingenheuer}, \citenamefont {Somalwar}, \citenamefont
  {Wah}, \citenamefont {Winstein} \emph {et~al.}}]{Gibbons:1993zq}%
  \BibitemOpen
  \bibfield  {author} {\bibinfo {author} {\bibfnamefont {L.~K.}\ \bibnamefont
  {Gibbons}}, \bibinfo {author} {\bibfnamefont {A.}~\bibnamefont {Barker}},
  \bibinfo {author} {\bibfnamefont {R.~A.}\ \bibnamefont {Briere}}, \bibinfo
  {author} {\bibfnamefont {G.}~\bibnamefont {Makoff}}, \bibinfo {author}
  {\bibfnamefont {V.}~\bibnamefont {Papadimitriou}}, \bibinfo {author}
  {\bibfnamefont {J.}~\bibnamefont {Patterson}}, \bibinfo {author}
  {\bibfnamefont {B.}~\bibnamefont {Schwingenheuer}}, \bibinfo {author}
  {\bibfnamefont {S.}~\bibnamefont {Somalwar}}, \bibinfo {author}
  {\bibfnamefont {Y.}~\bibnamefont {Wah}}, \bibinfo {author} {\bibfnamefont
  {B.}~\bibnamefont {Winstein}},  \emph {et~al.},\ }\href {\doibase
  10.1103/PhysRevLett.70.1203} {\bibfield  {journal} {\bibinfo  {journal}
  {Phys. Rev. Lett.}\ }\textbf {\bibinfo {volume} {70}},\ \bibinfo {pages}
  {1203} (\bibinfo {year} {1993})}\BibitemShut {NoStop}%
\bibitem [{\citenamefont {{JAM Collaboration}}\ \emph
  {et~al.}(2018)\citenamefont {{JAM Collaboration}}, \citenamefont {Barry},
  \citenamefont {Sato}, \citenamefont {Melnitchouk},\ and\ \citenamefont
  {Ji}}]{Barry:2018ort}%
  \BibitemOpen
  \bibfield  {author} {\bibinfo {author} {\bibnamefont {{JAM Collaboration}}},
  \bibinfo {author} {\bibfnamefont {P.~C.}\ \bibnamefont {Barry}}, \bibinfo
  {author} {\bibfnamefont {N.}~\bibnamefont {Sato}}, \bibinfo {author}
  {\bibfnamefont {W.}~\bibnamefont {Melnitchouk}}, \ and\ \bibinfo {author}
  {\bibfnamefont {C.-R.}\ \bibnamefont {Ji}},\ }\href {\doibase
  10.1103/PhysRevLett.121.152001} {\bibfield  {journal} {\bibinfo  {journal}
  {Phys. Rev. Lett.}\ }\textbf {\bibinfo {volume} {121}},\ \bibinfo {pages}
  {152001} (\bibinfo {year} {2018})}\BibitemShut {NoStop}%
\bibitem [{\citenamefont {Frederico}\ and\ \citenamefont
  {Miller}(1994)}]{Frederico:1994dx}%
  \BibitemOpen
  \bibfield  {author} {\bibinfo {author} {\bibfnamefont {T.}~\bibnamefont
  {Frederico}}\ and\ \bibinfo {author} {\bibfnamefont {G.~A.}\ \bibnamefont
  {Miller}},\ }\href {\doibase 10.1103/PhysRevD.50.210} {\bibfield  {journal}
  {\bibinfo  {journal} {Phys. Rev. D}\ }\textbf {\bibinfo {volume} {50}},\
  \bibinfo {pages} {210} (\bibinfo {year} {1994})}\BibitemShut {NoStop}%
\bibitem [{\citenamefont {Shigetani}\ \emph {et~al.}(1993)\citenamefont
  {Shigetani}, \citenamefont {Suzuki},\ and\ \citenamefont
  {Toki}}]{Shigetani:1993dx}%
  \BibitemOpen
  \bibfield  {author} {\bibinfo {author} {\bibfnamefont {T.}~\bibnamefont
  {Shigetani}}, \bibinfo {author} {\bibfnamefont {K.}~\bibnamefont {Suzuki}}, \
  and\ \bibinfo {author} {\bibfnamefont {H.}~\bibnamefont {Toki}},\ }\href
  {\doibase https://doi.org/10.1016/0370-2693(93)91302-4} {\bibfield  {journal}
  {\bibinfo  {journal} {Phys. Lett. B}\ }\textbf {\bibinfo {volume} {308}},\
  \bibinfo {pages} {383 } (\bibinfo {year} {1993})}\BibitemShut {NoStop}%
\bibitem [{\citenamefont {Broniowski}\ \emph {et~al.}(2008)\citenamefont
  {Broniowski}, \citenamefont {Arriola},\ and\ \citenamefont
  {Golec-Biernat}}]{Broniowski:2007si}%
  \BibitemOpen
  \bibfield  {author} {\bibinfo {author} {\bibfnamefont {W.}~\bibnamefont
  {Broniowski}}, \bibinfo {author} {\bibfnamefont {E.~R.}\ \bibnamefont
  {Arriola}}, \ and\ \bibinfo {author} {\bibfnamefont {K.}~\bibnamefont
  {Golec-Biernat}},\ }\href {\doibase 10.1103/PhysRevD.77.034023} {\bibfield
  {journal} {\bibinfo  {journal} {Phys. Rev. D}\ }\textbf {\bibinfo {volume}
  {77}},\ \bibinfo {pages} {034023} (\bibinfo {year} {2008})}\BibitemShut
  {NoStop}%
\bibitem [{\citenamefont {Gutsche}\ \emph
  {et~al.}(2015{\natexlab{a}})\citenamefont {Gutsche}, \citenamefont
  {Lyubovitskij}, \citenamefont {Schmidt},\ and\ \citenamefont
  {Vega}}]{Gutsche:2014zua}%
  \BibitemOpen
  \bibfield  {author} {\bibinfo {author} {\bibfnamefont {T.}~\bibnamefont
  {Gutsche}}, \bibinfo {author} {\bibfnamefont {V.~E.}\ \bibnamefont
  {Lyubovitskij}}, \bibinfo {author} {\bibfnamefont {I.}~\bibnamefont
  {Schmidt}}, \ and\ \bibinfo {author} {\bibfnamefont {A.}~\bibnamefont
  {Vega}},\ }\href {\doibase 10.1088/0954-3899/42/9/095005} {\bibfield
  {journal} {\bibinfo  {journal} {J. Phys.}\ }\textbf {\bibinfo {volume}
  {G42}},\ \bibinfo {pages} {095005} (\bibinfo {year} {2015}{\natexlab{a}})},\
  \Eprint {http://arxiv.org/abs/1410.6424} {arXiv:1410.6424 [hep-ph]}
  \BibitemShut {NoStop}%
\bibitem [{\citenamefont {Gutsche}\ \emph
  {et~al.}(2015{\natexlab{b}})\citenamefont {Gutsche}, \citenamefont
  {Lyubovitskij}, \citenamefont {Schmidt},\ and\ \citenamefont
  {Vega}}]{Gutsche:2013zia}%
  \BibitemOpen
  \bibfield  {author} {\bibinfo {author} {\bibfnamefont {T.}~\bibnamefont
  {Gutsche}}, \bibinfo {author} {\bibfnamefont {V.~E.}\ \bibnamefont
  {Lyubovitskij}}, \bibinfo {author} {\bibfnamefont {I.}~\bibnamefont
  {Schmidt}}, \ and\ \bibinfo {author} {\bibfnamefont {A.}~\bibnamefont
  {Vega}},\ }\href {\doibase 10.1103/PhysRevD.92.019902} {\bibfield  {journal}
  {\bibinfo  {journal} {Phys. Rev. D}\ }\textbf {\bibinfo {volume} {92}},\
  \bibinfo {pages} {019902} (\bibinfo {year} {2015}{\natexlab{b}})},\ \bibinfo
  {note} {[Erratum: Phys. Rev.D92,no.1,019902(2015)]}\BibitemShut {NoStop}%
\bibitem [{\citenamefont {Ahmady}\ \emph {et~al.}(2018)\citenamefont {Ahmady},
  \citenamefont {Mondal},\ and\ \citenamefont {Sandapen}}]{Ahmady:2018muv}%
  \BibitemOpen
  \bibfield  {author} {\bibinfo {author} {\bibfnamefont {M.}~\bibnamefont
  {Ahmady}}, \bibinfo {author} {\bibfnamefont {C.}~\bibnamefont {Mondal}}, \
  and\ \bibinfo {author} {\bibfnamefont {R.}~\bibnamefont {Sandapen}},\ }\href
  {\doibase 10.1103/PhysRevD.98.034010} {\bibfield  {journal} {\bibinfo
  {journal} {Phys. Rev. D}\ }\textbf {\bibinfo {volume} {98}},\ \bibinfo
  {pages} {034010} (\bibinfo {year} {2018})}\BibitemShut {NoStop}%
\bibitem [{\citenamefont {{HLFHS Collaboration}}\ \emph
  {et~al.}(2018)\citenamefont {{HLFHS Collaboration}}, \citenamefont
  {de~T\'eramond}, \citenamefont {Liu}, \citenamefont {Sufian}, \citenamefont
  {Dosch}, \citenamefont {Brodsky},\ and\ \citenamefont
  {Deur}}]{deTeramond:2018ecg}%
  \BibitemOpen
  \bibfield  {author} {\bibinfo {author} {\bibnamefont {{HLFHS
  Collaboration}}}, \bibinfo {author} {\bibfnamefont {G.~F.}\ \bibnamefont
  {de~T\'eramond}}, \bibinfo {author} {\bibfnamefont {T.}~\bibnamefont {Liu}},
  \bibinfo {author} {\bibfnamefont {R.~S.}\ \bibnamefont {Sufian}}, \bibinfo
  {author} {\bibfnamefont {H.~G.}\ \bibnamefont {Dosch}}, \bibinfo {author}
  {\bibfnamefont {S.~J.}\ \bibnamefont {Brodsky}}, \ and\ \bibinfo {author}
  {\bibfnamefont {A.}~\bibnamefont {Deur}},\ }\href {\doibase
  10.1103/PhysRevLett.120.182001} {\bibfield  {journal} {\bibinfo  {journal}
  {Phys. Rev. Lett.}\ }\textbf {\bibinfo {volume} {120}},\ \bibinfo {pages}
  {182001} (\bibinfo {year} {2018})}\BibitemShut {NoStop}%
\bibitem [{\citenamefont {{QCDSF-UKQCD Collaboration}}\ \emph
  {et~al.}(2007)\citenamefont {{QCDSF-UKQCD Collaboration}}, \citenamefont
  {Brommel}, \citenamefont {Diehl}, \citenamefont {Gockeler}, \citenamefont
  {Hagler}, \citenamefont {Horsley}, \citenamefont {Nakamura}, \citenamefont
  {Pleiter}, \citenamefont {Rakow}, \citenamefont {Schafer}, \citenamefont
  {Schierholz}, \citenamefont {Stuben}, \citenamefont {Brommel},\ and\
  \citenamefont {Zanotti}}]{Brommel:2006zz}%
  \BibitemOpen
  \bibfield  {author} {\bibinfo {author} {\bibnamefont {{QCDSF-UKQCD
  Collaboration}}}, \bibinfo {author} {\bibfnamefont {D.}~\bibnamefont
  {Brommel}}, \bibinfo {author} {\bibfnamefont {M.}~\bibnamefont {Diehl}},
  \bibinfo {author} {\bibfnamefont {M.}~\bibnamefont {Gockeler}}, \bibinfo
  {author} {\bibfnamefont {P.}~\bibnamefont {Hagler}}, \bibinfo {author}
  {\bibfnamefont {R.}~\bibnamefont {Horsley}}, \bibinfo {author} {\bibfnamefont
  {Y.}~\bibnamefont {Nakamura}}, \bibinfo {author} {\bibfnamefont
  {D.}~\bibnamefont {Pleiter}}, \bibinfo {author} {\bibfnamefont {P.~E.}\
  \bibnamefont {Rakow}}, \bibinfo {author} {\bibfnamefont {A.}~\bibnamefont
  {Schafer}}, \bibinfo {author} {\bibfnamefont {G.}~\bibnamefont {Schierholz}},
  \bibinfo {author} {\bibfnamefont {H.}~\bibnamefont {Stuben}}, \bibinfo
  {author} {\bibfnamefont {D.}~\bibnamefont {Brommel}}, \ and\ \bibinfo
  {author} {\bibfnamefont {J.~M.}\ \bibnamefont {Zanotti}},\ }\bibfield
  {booktitle} {\emph {\bibinfo {booktitle} {{Proceedings, 25th International
  Symposium on Lattice field theory (Lattice 2007): Regensburg, Germany, July
  30-August 4, 2007}}},\ }\href {\doibase 10.22323/1.042.0140} {\bibfield
  {journal} {\bibinfo  {journal} {PoS}\ }\textbf {\bibinfo {volume}
  {LATTICE2007}},\ \bibinfo {pages} {140} (\bibinfo {year} {2007})}\BibitemShut
  {NoStop}%
\bibitem [{\citenamefont {Martinelli}\ and\ \citenamefont
  {Sachrajda}(1988)}]{Martinelli:1987bh}%
  \BibitemOpen
  \bibfield  {author} {\bibinfo {author} {\bibfnamefont {G.}~\bibnamefont
  {Martinelli}}\ and\ \bibinfo {author} {\bibfnamefont {C.}~\bibnamefont
  {Sachrajda}},\ }\href {\doibase https://doi.org/10.1016/0550-3213(88)90445-2}
  {\bibfield  {journal} {\bibinfo  {journal} {Nucl. Phys. B}\ }\textbf
  {\bibinfo {volume} {306}},\ \bibinfo {pages} {865 } (\bibinfo {year}
  {1988})}\BibitemShut {NoStop}%
\bibitem [{\citenamefont {Abdel-Rehim}\ \emph {et~al.}(2015)\citenamefont
  {Abdel-Rehim}, \citenamefont {Alexandrou}, \citenamefont {Constantinou},
  \citenamefont {Dimopoulos}, \citenamefont {Frezzotti}, \citenamefont
  {Hadjiyiannakou}, \citenamefont {Jansen}, \citenamefont {Kallidonis},
  \citenamefont {Kostrzewa}, \citenamefont {Koutsou} \emph
  {et~al.}}]{Abdel-Rehim:2015owa}%
  \BibitemOpen
  \bibfield  {author} {\bibinfo {author} {\bibfnamefont {A.}~\bibnamefont
  {Abdel-Rehim}}, \bibinfo {author} {\bibfnamefont {C.}~\bibnamefont
  {Alexandrou}}, \bibinfo {author} {\bibfnamefont {M.}~\bibnamefont
  {Constantinou}}, \bibinfo {author} {\bibfnamefont {P.}~\bibnamefont
  {Dimopoulos}}, \bibinfo {author} {\bibfnamefont {R.}~\bibnamefont
  {Frezzotti}}, \bibinfo {author} {\bibfnamefont {K.}~\bibnamefont
  {Hadjiyiannakou}}, \bibinfo {author} {\bibfnamefont {K.}~\bibnamefont
  {Jansen}}, \bibinfo {author} {\bibfnamefont {C.}~\bibnamefont {Kallidonis}},
  \bibinfo {author} {\bibfnamefont {B.}~\bibnamefont {Kostrzewa}}, \bibinfo
  {author} {\bibfnamefont {G.}~\bibnamefont {Koutsou}},  \emph {et~al.},\
  }\href {\doibase 10.1103/PhysRevD.92.114513} {\bibfield  {journal} {\bibinfo
  {journal} {Phys. Rev. D}\ }\textbf {\bibinfo {volume} {92}},\ \bibinfo
  {pages} {114513} (\bibinfo {year} {2015})},\ \bibinfo {note} {[Erratum: Phys.
  Rev.D93,no.3,039904(2016)]}\BibitemShut {NoStop}%
\bibitem [{\citenamefont {{ETM Collaboration}}\ \emph
  {et~al.}(2019)\citenamefont {{ETM Collaboration}}, \citenamefont {Oehm},
  \citenamefont {Alexandrou}, \citenamefont {Constantinou}, \citenamefont
  {Jansen}, \citenamefont {Koutsou}, \citenamefont {Kostrzewa}, \citenamefont
  {Steffens}, \citenamefont {Urbach},\ and\ \citenamefont
  {Zafeiropoulos}}]{Oehm:2018jvm}%
  \BibitemOpen
  \bibfield  {author} {\bibinfo {author} {\bibnamefont {{ETM Collaboration}}},
  \bibinfo {author} {\bibfnamefont {M.}~\bibnamefont {Oehm}}, \bibinfo {author}
  {\bibfnamefont {C.}~\bibnamefont {Alexandrou}}, \bibinfo {author}
  {\bibfnamefont {M.}~\bibnamefont {Constantinou}}, \bibinfo {author}
  {\bibfnamefont {K.}~\bibnamefont {Jansen}}, \bibinfo {author} {\bibfnamefont
  {G.}~\bibnamefont {Koutsou}}, \bibinfo {author} {\bibfnamefont
  {B.}~\bibnamefont {Kostrzewa}}, \bibinfo {author} {\bibfnamefont
  {F.}~\bibnamefont {Steffens}}, \bibinfo {author} {\bibfnamefont
  {C.}~\bibnamefont {Urbach}}, \ and\ \bibinfo {author} {\bibfnamefont
  {S.}~\bibnamefont {Zafeiropoulos}},\ }\href {\doibase
  10.1103/PhysRevD.99.014508} {\bibfield  {journal} {\bibinfo  {journal} {Phys.
  Rev. D}\ }\textbf {\bibinfo {volume} {99}},\ \bibinfo {pages} {014508}
  (\bibinfo {year} {2019})}\BibitemShut {NoStop}%
\bibitem [{\citenamefont {Sufian}\ \emph {et~al.}(2019)\citenamefont {Sufian},
  \citenamefont {Karpie}, \citenamefont {Egerer}, \citenamefont {Orginos},
  \citenamefont {Qiu},\ and\ \citenamefont {Richards}}]{Sufian:2019bol}%
  \BibitemOpen
  \bibfield  {author} {\bibinfo {author} {\bibfnamefont {R.~S.}\ \bibnamefont
  {Sufian}}, \bibinfo {author} {\bibfnamefont {J.}~\bibnamefont {Karpie}},
  \bibinfo {author} {\bibfnamefont {C.}~\bibnamefont {Egerer}}, \bibinfo
  {author} {\bibfnamefont {K.}~\bibnamefont {Orginos}}, \bibinfo {author}
  {\bibfnamefont {J.-W.}\ \bibnamefont {Qiu}}, \ and\ \bibinfo {author}
  {\bibfnamefont {D.~G.}\ \bibnamefont {Richards}},\ }\href {\doibase
  10.1103/PhysRevD.99.074507} {\bibfield  {journal} {\bibinfo  {journal} {Phys.
  Rev.}\ }\textbf {\bibinfo {volume} {D99}},\ \bibinfo {pages} {074507}
  (\bibinfo {year} {2019})},\ \Eprint {http://arxiv.org/abs/1901.03921}
  {arXiv:1901.03921 [hep-lat]} \BibitemShut {NoStop}%
\bibitem [{\citenamefont {Lin}\ \emph {et~al.}(2018)\citenamefont {Lin},
  \citenamefont {Nocera}, \citenamefont {Olness}, \citenamefont {Orginos},
  \citenamefont {Rojo}, \citenamefont {Accardi}, \citenamefont {Alexandrou},
  \citenamefont {Bacchetta}, \citenamefont {Bozzi}, \citenamefont {Chen} \emph
  {et~al.}}]{Lin:2017snn}%
  \BibitemOpen
  \bibfield  {author} {\bibinfo {author} {\bibfnamefont {H.-W.}\ \bibnamefont
  {Lin}}, \bibinfo {author} {\bibfnamefont {E.~R.}\ \bibnamefont {Nocera}},
  \bibinfo {author} {\bibfnamefont {F.}~\bibnamefont {Olness}}, \bibinfo
  {author} {\bibfnamefont {K.}~\bibnamefont {Orginos}}, \bibinfo {author}
  {\bibfnamefont {J.}~\bibnamefont {Rojo}}, \bibinfo {author} {\bibfnamefont
  {A.}~\bibnamefont {Accardi}}, \bibinfo {author} {\bibfnamefont
  {C.}~\bibnamefont {Alexandrou}}, \bibinfo {author} {\bibfnamefont
  {A.}~\bibnamefont {Bacchetta}}, \bibinfo {author} {\bibfnamefont
  {G.}~\bibnamefont {Bozzi}}, \bibinfo {author} {\bibfnamefont {J.-W.}\
  \bibnamefont {Chen}},  \emph {et~al.},\ }\href {\doibase
  10.1016/j.ppnp.2018.01.007} {\bibfield  {journal} {\bibinfo  {journal} {Prog.
  Part. Nucl. Phys.}\ }\textbf {\bibinfo {volume} {100}},\ \bibinfo {pages}
  {107} (\bibinfo {year} {2018})},\ \Eprint {http://arxiv.org/abs/1711.07916}
  {arXiv:1711.07916 [hep-ph]} \BibitemShut {NoStop}%
\bibitem [{\citenamefont {Melnitchouk}(2003)}]{Melnitchouk:2002gh}%
  \BibitemOpen
  \bibfield  {author} {\bibinfo {author} {\bibfnamefont {W.}~\bibnamefont
  {Melnitchouk}},\ }\href {\doibase 10.1140/epja/i2003-10006-6} {\bibfield
  {journal} {\bibinfo  {journal} {Eur. Phys. J. A}\ }\textbf {\bibinfo {volume}
  {17}},\ \bibinfo {pages} {223} (\bibinfo {year} {2003})}\BibitemShut
  {NoStop}%
\bibitem [{\citenamefont {Farrar}\ and\ \citenamefont
  {Jackson}(1979)}]{Farrar:1979aw}%
  \BibitemOpen
  \bibfield  {author} {\bibinfo {author} {\bibfnamefont {G.~R.}\ \bibnamefont
  {Farrar}}\ and\ \bibinfo {author} {\bibfnamefont {D.~R.}\ \bibnamefont
  {Jackson}},\ }\href {\doibase 10.1103/PhysRevLett.43.246} {\bibfield
  {journal} {\bibinfo  {journal} {Phys. Rev. Lett.}\ }\textbf {\bibinfo
  {volume} {43}},\ \bibinfo {pages} {246} (\bibinfo {year} {1979})}\BibitemShut
  {NoStop}%
\bibitem [{\citenamefont {Berger}\ and\ \citenamefont
  {Brodsky}(1979)}]{Berger:1979du}%
  \BibitemOpen
  \bibfield  {author} {\bibinfo {author} {\bibfnamefont {E.~L.}\ \bibnamefont
  {Berger}}\ and\ \bibinfo {author} {\bibfnamefont {S.~J.}\ \bibnamefont
  {Brodsky}},\ }\href {\doibase 10.1103/PhysRevLett.42.940} {\bibfield
  {journal} {\bibinfo  {journal} {Phys. Rev. Lett.}\ }\textbf {\bibinfo
  {volume} {42}},\ \bibinfo {pages} {940} (\bibinfo {year} {1979})}\BibitemShut
  {NoStop}%
\bibitem [{\citenamefont {Brodsky}\ and\ \citenamefont
  {Yuan}(2006)}]{Brodsky:2006hj}%
  \BibitemOpen
  \bibfield  {author} {\bibinfo {author} {\bibfnamefont {S.~J.}\ \bibnamefont
  {Brodsky}}\ and\ \bibinfo {author} {\bibfnamefont {F.}~\bibnamefont {Yuan}},\
  }\href {\doibase 10.1103/PhysRevD.74.094018} {\bibfield  {journal} {\bibinfo
  {journal} {Phys. Rev. D}\ }\textbf {\bibinfo {volume} {74}},\ \bibinfo
  {pages} {094018} (\bibinfo {year} {2006})}\BibitemShut {NoStop}%
\bibitem [{\citenamefont {Yuan}(2004)}]{Yuan:2003fs}%
  \BibitemOpen
  \bibfield  {author} {\bibinfo {author} {\bibfnamefont {F.}~\bibnamefont
  {Yuan}},\ }\href {\doibase 10.1103/PhysRevD.69.051501} {\bibfield  {journal}
  {\bibinfo  {journal} {Phys. Rev. D}\ }\textbf {\bibinfo {volume} {69}},\
  \bibinfo {pages} {051501} (\bibinfo {year} {2004})}\BibitemShut {NoStop}%
\bibitem [{\citenamefont {Ding}\ \emph {et~al.}(2019)\citenamefont {Ding},
  \citenamefont {Raya}, \citenamefont {Binosi}, \citenamefont {Chang},
  \citenamefont {Roberts},\ and\ \citenamefont {Schmidt}}]{Ding:2019lwe}%
  \BibitemOpen
  \bibfield  {author} {\bibinfo {author} {\bibfnamefont {M.}~\bibnamefont
  {Ding}}, \bibinfo {author} {\bibfnamefont {K.}~\bibnamefont {Raya}}, \bibinfo
  {author} {\bibfnamefont {D.}~\bibnamefont {Binosi}}, \bibinfo {author}
  {\bibfnamefont {L.}~\bibnamefont {Chang}}, \bibinfo {author} {\bibfnamefont
  {C.~D.}\ \bibnamefont {Roberts}}, \ and\ \bibinfo {author} {\bibfnamefont
  {S.~M.}\ \bibnamefont {Schmidt}},\ }\href@noop {} {\  (\bibinfo {year}
  {2019})},\ \Eprint {http://arxiv.org/abs/1905.05208} {arXiv:1905.05208
  [nucl-th]} \BibitemShut {NoStop}%
\bibitem [{\citenamefont {Nguyen}\ \emph {et~al.}(2011)\citenamefont {Nguyen},
  \citenamefont {Bashir}, \citenamefont {Roberts},\ and\ \citenamefont
  {Tandy}}]{Nguyen:2011jy}%
  \BibitemOpen
  \bibfield  {author} {\bibinfo {author} {\bibfnamefont {T.}~\bibnamefont
  {Nguyen}}, \bibinfo {author} {\bibfnamefont {A.}~\bibnamefont {Bashir}},
  \bibinfo {author} {\bibfnamefont {C.~D.}\ \bibnamefont {Roberts}}, \ and\
  \bibinfo {author} {\bibfnamefont {P.~C.}\ \bibnamefont {Tandy}},\ }\href
  {\doibase 10.1103/PhysRevC.83.062201} {\bibfield  {journal} {\bibinfo
  {journal} {Phys. Rev. C}\ }\textbf {\bibinfo {volume} {83}},\ \bibinfo
  {pages} {062201} (\bibinfo {year} {2011})}\BibitemShut {NoStop}%
\bibitem [{\citenamefont {Shi}\ \emph {et~al.}(2018)\citenamefont {Shi},
  \citenamefont {Mezrag},\ and\ \citenamefont {Zong}}]{Shi:2018mcb}%
  \BibitemOpen
  \bibfield  {author} {\bibinfo {author} {\bibfnamefont {C.}~\bibnamefont
  {Shi}}, \bibinfo {author} {\bibfnamefont {C.}~\bibnamefont {Mezrag}}, \ and\
  \bibinfo {author} {\bibfnamefont {H.-s.}\ \bibnamefont {Zong}},\ }\href
  {\doibase 10.1103/PhysRevD.98.054029} {\bibfield  {journal} {\bibinfo
  {journal} {Phys. Rev. D}\ }\textbf {\bibinfo {volume} {98}},\ \bibinfo
  {pages} {054029} (\bibinfo {year} {2018})}\BibitemShut {NoStop}%
\bibitem [{\citenamefont {Chen}\ \emph {et~al.}(2016)\citenamefont {Chen},
  \citenamefont {Chang}, \citenamefont {Roberts}, \citenamefont {Wan},\ and\
  \citenamefont {Zong}}]{Chen:2016sno}%
  \BibitemOpen
  \bibfield  {author} {\bibinfo {author} {\bibfnamefont {C.}~\bibnamefont
  {Chen}}, \bibinfo {author} {\bibfnamefont {L.}~\bibnamefont {Chang}},
  \bibinfo {author} {\bibfnamefont {C.~D.}\ \bibnamefont {Roberts}}, \bibinfo
  {author} {\bibfnamefont {S.}~\bibnamefont {Wan}}, \ and\ \bibinfo {author}
  {\bibfnamefont {H.-S.}\ \bibnamefont {Zong}},\ }\href {\doibase
  10.1103/PhysRevD.93.074021} {\bibfield  {journal} {\bibinfo  {journal} {Phys.
  Rev. D}\ }\textbf {\bibinfo {volume} {93}},\ \bibinfo {pages} {074021}
  (\bibinfo {year} {2016})}\BibitemShut {NoStop}%
\bibitem [{\citenamefont {Hutauruk}\ \emph {et~al.}(2016)\citenamefont
  {Hutauruk}, \citenamefont {Clo\"et},\ and\ \citenamefont
  {Thomas}}]{Hutauruk:2016sug}%
  \BibitemOpen
  \bibfield  {author} {\bibinfo {author} {\bibfnamefont {P.~T.~P.}\
  \bibnamefont {Hutauruk}}, \bibinfo {author} {\bibfnamefont {I.~C.}\
  \bibnamefont {Clo\"et}}, \ and\ \bibinfo {author} {\bibfnamefont {A.~W.}\
  \bibnamefont {Thomas}},\ }\href {\doibase 10.1103/PhysRevC.94.035201}
  {\bibfield  {journal} {\bibinfo  {journal} {Phys. Rev. C}\ }\textbf {\bibinfo
  {volume} {94}},\ \bibinfo {pages} {035201} (\bibinfo {year}
  {2016})}\BibitemShut {NoStop}%
\bibitem [{\citenamefont {Davidson}\ and\ \citenamefont
  {Ruiz~Arriola}(2002)}]{Davidson:2001cc}%
  \BibitemOpen
  \bibfield  {author} {\bibinfo {author} {\bibfnamefont {R.~M.}\ \bibnamefont
  {Davidson}}\ and\ \bibinfo {author} {\bibfnamefont {E.}~\bibnamefont
  {Ruiz~Arriola}},\ }\href@noop {} {\bibfield  {journal} {\bibinfo  {journal}
  {Acta Phys. Polon.}\ }\textbf {\bibinfo {volume} {B33}},\ \bibinfo {pages}
  {1791} (\bibinfo {year} {2002})},\ \Eprint
  {http://arxiv.org/abs/hep-ph/0110291} {arXiv:hep-ph/0110291 [hep-ph]}
  \BibitemShut {NoStop}%
\bibitem [{\citenamefont {Xu}\ \emph {et~al.}(2018)\citenamefont {Xu},
  \citenamefont {Chang}, \citenamefont {Roberts},\ and\ \citenamefont
  {Zong}}]{Xu:2018eii}%
  \BibitemOpen
  \bibfield  {author} {\bibinfo {author} {\bibfnamefont {S.-S.}\ \bibnamefont
  {Xu}}, \bibinfo {author} {\bibfnamefont {L.}~\bibnamefont {Chang}}, \bibinfo
  {author} {\bibfnamefont {C.~D.}\ \bibnamefont {Roberts}}, \ and\ \bibinfo
  {author} {\bibfnamefont {H.-S.}\ \bibnamefont {Zong}},\ }\href {\doibase
  10.1103/PhysRevD.97.094014} {\bibfield  {journal} {\bibinfo  {journal} {Phys.
  Rev. D}\ }\textbf {\bibinfo {volume} {97}},\ \bibinfo {pages} {094014}
  (\bibinfo {year} {2018})}\BibitemShut {NoStop}%
\bibitem [{\citenamefont {Broniowski}\ and\ \citenamefont
  {Arriola}(2017)}]{Broniowski:2017wbr}%
  \BibitemOpen
  \bibfield  {author} {\bibinfo {author} {\bibfnamefont {W.}~\bibnamefont
  {Broniowski}}\ and\ \bibinfo {author} {\bibfnamefont {E.~R.}\ \bibnamefont
  {Arriola}},\ }\href {\doibase https://doi.org/10.1016/j.physletb.2017.08.055}
  {\bibfield  {journal} {\bibinfo  {journal} {Phys. Lett. B}\ }\textbf
  {\bibinfo {volume} {773}},\ \bibinfo {pages} {385 } (\bibinfo {year}
  {2017})}\BibitemShut {NoStop}%
\bibitem [{\citenamefont {Radyushkin}(2017)}]{Radyushkin:2017gjd}%
  \BibitemOpen
  \bibfield  {author} {\bibinfo {author} {\bibfnamefont {A.~V.}\ \bibnamefont
  {Radyushkin}},\ }\href {\doibase 10.1103/PhysRevD.95.056020} {\bibfield
  {journal} {\bibinfo  {journal} {Phys. Rev. D}\ }\textbf {\bibinfo {volume}
  {95}},\ \bibinfo {pages} {056020} (\bibinfo {year} {2017})}\BibitemShut
  {NoStop}%
\bibitem [{\citenamefont {Vary}\ \emph {et~al.}(2010)\citenamefont {Vary},
  \citenamefont {Honkanen}, \citenamefont {Li}, \citenamefont {Maris},
  \citenamefont {Brodsky}, \citenamefont {Harindranath}, \citenamefont
  {de~Teramond}, \citenamefont {Sternberg}, \citenamefont {Ng},\ and\
  \citenamefont {Yang}}]{Vary:2009gt}%
  \BibitemOpen
  \bibfield  {author} {\bibinfo {author} {\bibfnamefont {J.~P.}\ \bibnamefont
  {Vary}}, \bibinfo {author} {\bibfnamefont {H.}~\bibnamefont {Honkanen}},
  \bibinfo {author} {\bibfnamefont {J.}~\bibnamefont {Li}}, \bibinfo {author}
  {\bibfnamefont {P.}~\bibnamefont {Maris}}, \bibinfo {author} {\bibfnamefont
  {S.~J.}\ \bibnamefont {Brodsky}}, \bibinfo {author} {\bibfnamefont
  {A.}~\bibnamefont {Harindranath}}, \bibinfo {author} {\bibfnamefont {G.~F.}\
  \bibnamefont {de~Teramond}}, \bibinfo {author} {\bibfnamefont
  {P.}~\bibnamefont {Sternberg}}, \bibinfo {author} {\bibfnamefont {E.~G.}\
  \bibnamefont {Ng}}, \ and\ \bibinfo {author} {\bibfnamefont {C.}~\bibnamefont
  {Yang}},\ }\href {\doibase 10.1103/PhysRevC.81.035205} {\bibfield  {journal}
  {\bibinfo  {journal} {Phys. Rev. C}\ }\textbf {\bibinfo {volume} {81}},\
  \bibinfo {pages} {035205} (\bibinfo {year} {2010})}\BibitemShut {NoStop}%
\bibitem [{\citenamefont {Wiecki}\ \emph {et~al.}(2015)\citenamefont {Wiecki},
  \citenamefont {Li}, \citenamefont {Zhao}, \citenamefont {Maris},\ and\
  \citenamefont {Vary}}]{Wiecki:2014ola}%
  \BibitemOpen
  \bibfield  {author} {\bibinfo {author} {\bibfnamefont {P.}~\bibnamefont
  {Wiecki}}, \bibinfo {author} {\bibfnamefont {Y.}~\bibnamefont {Li}}, \bibinfo
  {author} {\bibfnamefont {X.}~\bibnamefont {Zhao}}, \bibinfo {author}
  {\bibfnamefont {P.}~\bibnamefont {Maris}}, \ and\ \bibinfo {author}
  {\bibfnamefont {J.~P.}\ \bibnamefont {Vary}},\ }\href {\doibase
  10.1103/PhysRevD.91.105009} {\bibfield  {journal} {\bibinfo  {journal} {Phys.
  Rev. D}\ }\textbf {\bibinfo {volume} {91}},\ \bibinfo {pages} {105009}
  (\bibinfo {year} {2015})}\BibitemShut {NoStop}%
\bibitem [{\citenamefont {Li}\ \emph {et~al.}(2016)\citenamefont {Li},
  \citenamefont {Maris}, \citenamefont {Zhao},\ and\ \citenamefont
  {Vary}}]{Li:2015zda}%
  \BibitemOpen
  \bibfield  {author} {\bibinfo {author} {\bibfnamefont {Y.}~\bibnamefont
  {Li}}, \bibinfo {author} {\bibfnamefont {P.}~\bibnamefont {Maris}}, \bibinfo
  {author} {\bibfnamefont {X.}~\bibnamefont {Zhao}}, \ and\ \bibinfo {author}
  {\bibfnamefont {J.~P.}\ \bibnamefont {Vary}},\ }\href {\doibase
  https://doi.org/10.1016/j.physletb.2016.04.065} {\bibfield  {journal}
  {\bibinfo  {journal} {Phys. Lett. B}\ }\textbf {\bibinfo {volume} {758}},\
  \bibinfo {pages} {118 } (\bibinfo {year} {2016})}\BibitemShut {NoStop}%
\bibitem [{\citenamefont {Brodsky}\ \emph {et~al.}(1998)\citenamefont
  {Brodsky}, \citenamefont {Pauli},\ and\ \citenamefont
  {Pinsky}}]{Brodsky:1997de}%
  \BibitemOpen
  \bibfield  {author} {\bibinfo {author} {\bibfnamefont {S.~J.}\ \bibnamefont
  {Brodsky}}, \bibinfo {author} {\bibfnamefont {H.-C.}\ \bibnamefont {Pauli}},
  \ and\ \bibinfo {author} {\bibfnamefont {S.~S.}\ \bibnamefont {Pinsky}},\
  }\href {\doibase 10.1016/S0370-1573(97)00089-6} {\bibfield  {journal}
  {\bibinfo  {journal} {Phys. Rept.}\ }\textbf {\bibinfo {volume} {301}},\
  \bibinfo {pages} {299} (\bibinfo {year} {1998})},\ \Eprint
  {http://arxiv.org/abs/hep-ph/9705477} {arXiv:hep-ph/9705477 [hep-ph]}
  \BibitemShut {NoStop}%
\bibitem [{\citenamefont {Zhao}\ \emph {et~al.}(2014)\citenamefont {Zhao},
  \citenamefont {Honkanen}, \citenamefont {Maris}, \citenamefont {Vary},\ and\
  \citenamefont {Brodsky}}]{Zhao:2014xaa}%
  \BibitemOpen
  \bibfield  {author} {\bibinfo {author} {\bibfnamefont {X.}~\bibnamefont
  {Zhao}}, \bibinfo {author} {\bibfnamefont {H.}~\bibnamefont {Honkanen}},
  \bibinfo {author} {\bibfnamefont {P.}~\bibnamefont {Maris}}, \bibinfo
  {author} {\bibfnamefont {J.~P.}\ \bibnamefont {Vary}}, \ and\ \bibinfo
  {author} {\bibfnamefont {S.~J.}\ \bibnamefont {Brodsky}},\ }\href {\doibase
  10.1016/j.physletb.2014.08.020} {\bibfield  {journal} {\bibinfo  {journal}
  {Phys. Lett.}\ }\textbf {\bibinfo {volume} {B737}},\ \bibinfo {pages} {65}
  (\bibinfo {year} {2014})},\ \Eprint {http://arxiv.org/abs/1402.4195}
  {arXiv:1402.4195 [nucl-th]} \BibitemShut {NoStop}%
\bibitem [{\citenamefont {Li}\ \emph {et~al.}(2017)\citenamefont {Li},
  \citenamefont {Maris},\ and\ \citenamefont {Vary}}]{Li:2017mlw}%
  \BibitemOpen
  \bibfield  {author} {\bibinfo {author} {\bibfnamefont {Y.}~\bibnamefont
  {Li}}, \bibinfo {author} {\bibfnamefont {P.}~\bibnamefont {Maris}}, \ and\
  \bibinfo {author} {\bibfnamefont {J.~P.}\ \bibnamefont {Vary}},\ }\href
  {\doibase 10.1103/PhysRevD.96.016022} {\bibfield  {journal} {\bibinfo
  {journal} {Phys. Rev. D}\ }\textbf {\bibinfo {volume} {96}},\ \bibinfo
  {pages} {016022} (\bibinfo {year} {2017})}\BibitemShut {NoStop}%
\bibitem [{\citenamefont {Tang}\ \emph {et~al.}(2018)\citenamefont {Tang},
  \citenamefont {Li}, \citenamefont {Maris},\ and\ \citenamefont
  {Vary}}]{Tang:2018myz}%
  \BibitemOpen
  \bibfield  {author} {\bibinfo {author} {\bibfnamefont {S.}~\bibnamefont
  {Tang}}, \bibinfo {author} {\bibfnamefont {Y.}~\bibnamefont {Li}}, \bibinfo
  {author} {\bibfnamefont {P.}~\bibnamefont {Maris}}, \ and\ \bibinfo {author}
  {\bibfnamefont {J.~P.}\ \bibnamefont {Vary}},\ }\href {\doibase
  10.1103/PhysRevD.98.114038} {\bibfield  {journal} {\bibinfo  {journal} {Phys.
  Rev.}\ }\textbf {\bibinfo {volume} {D98}},\ \bibinfo {pages} {114038}
  (\bibinfo {year} {2018})},\ \Eprint {http://arxiv.org/abs/1810.05971}
  {arXiv:1810.05971 [nucl-th]} \BibitemShut {NoStop}%
\bibitem [{\citenamefont {Jia}\ and\ \citenamefont {Vary}(2019)}]{Jia:2018ary}%
  \BibitemOpen
  \bibfield  {author} {\bibinfo {author} {\bibfnamefont {S.}~\bibnamefont
  {Jia}}\ and\ \bibinfo {author} {\bibfnamefont {J.~P.}\ \bibnamefont {Vary}},\
  }\href {\doibase 10.1103/PhysRevC.99.035206} {\bibfield  {journal} {\bibinfo
  {journal} {Phys. Rev. C}\ }\textbf {\bibinfo {volume} {99}},\ \bibinfo
  {pages} {035206} (\bibinfo {year} {2019})}\BibitemShut {NoStop}%
\bibitem [{\citenamefont {Zhao}\ \emph
  {et~al.}(2013{\natexlab{a}})\citenamefont {Zhao}, \citenamefont {Ilderton},
  \citenamefont {Maris},\ and\ \citenamefont {Vary}}]{Zhao:2013jia}%
  \BibitemOpen
  \bibfield  {author} {\bibinfo {author} {\bibfnamefont {X.}~\bibnamefont
  {Zhao}}, \bibinfo {author} {\bibfnamefont {A.}~\bibnamefont {Ilderton}},
  \bibinfo {author} {\bibfnamefont {P.}~\bibnamefont {Maris}}, \ and\ \bibinfo
  {author} {\bibfnamefont {J.~P.}\ \bibnamefont {Vary}},\ }\href {\doibase
  10.1016/j.physletb.2013.09.030} {\bibfield  {journal} {\bibinfo  {journal}
  {Phys. Lett.}\ }\textbf {\bibinfo {volume} {B726}},\ \bibinfo {pages} {856}
  (\bibinfo {year} {2013}{\natexlab{a}})},\ \Eprint
  {http://arxiv.org/abs/1309.5338} {arXiv:1309.5338 [nucl-th]} \BibitemShut
  {NoStop}%
\bibitem [{\citenamefont {Leitão}\ \emph {et~al.}(2017)\citenamefont
  {Leitão}, \citenamefont {Li}, \citenamefont {Maris}, \citenamefont {Peña},
  \citenamefont {Stadler}, \citenamefont {Vary},\ and\ \citenamefont
  {Biernat}}]{Leitao:2017esb}%
  \BibitemOpen
  \bibfield  {author} {\bibinfo {author} {\bibfnamefont {S.}~\bibnamefont
  {Leitão}}, \bibinfo {author} {\bibfnamefont {Y.}~\bibnamefont {Li}},
  \bibinfo {author} {\bibfnamefont {P.}~\bibnamefont {Maris}}, \bibinfo
  {author} {\bibfnamefont {M.~T.}\ \bibnamefont {Peña}}, \bibinfo {author}
  {\bibfnamefont {A.}~\bibnamefont {Stadler}}, \bibinfo {author} {\bibfnamefont
  {J.~P.}\ \bibnamefont {Vary}}, \ and\ \bibinfo {author} {\bibfnamefont
  {E.~P.}\ \bibnamefont {Biernat}},\ }\href {\doibase
  10.1140/epjc/s10052-017-5248-0} {\bibfield  {journal} {\bibinfo  {journal}
  {Eur. Phys. J.}\ }\textbf {\bibinfo {volume} {C77}},\ \bibinfo {pages} {696}
  (\bibinfo {year} {2017})},\ \Eprint {http://arxiv.org/abs/1705.06178}
  {arXiv:1705.06178 [hep-ph]} \BibitemShut {NoStop}%
\bibitem [{\citenamefont {Adhikari}\ \emph {et~al.}(2016)\citenamefont
  {Adhikari}, \citenamefont {Li}, \citenamefont {Zhao}, \citenamefont {Maris},
  \citenamefont {Vary},\ and\ \citenamefont {Abd El-Hady}}]{Adhikari:2016idg}%
  \BibitemOpen
  \bibfield  {author} {\bibinfo {author} {\bibfnamefont {L.}~\bibnamefont
  {Adhikari}}, \bibinfo {author} {\bibfnamefont {Y.}~\bibnamefont {Li}},
  \bibinfo {author} {\bibfnamefont {X.}~\bibnamefont {Zhao}}, \bibinfo {author}
  {\bibfnamefont {P.}~\bibnamefont {Maris}}, \bibinfo {author} {\bibfnamefont
  {J.~P.}\ \bibnamefont {Vary}}, \ and\ \bibinfo {author} {\bibfnamefont
  {A.}~\bibnamefont {Abd El-Hady}},\ }\href {\doibase
  10.1103/PhysRevC.93.055202} {\bibfield  {journal} {\bibinfo  {journal} {Phys.
  Rev.}\ }\textbf {\bibinfo {volume} {C93}},\ \bibinfo {pages} {055202}
  (\bibinfo {year} {2016})},\ \Eprint {http://arxiv.org/abs/1602.06027}
  {arXiv:1602.06027 [nucl-th]} \BibitemShut {NoStop}%
\bibitem [{\citenamefont {Zhao}\ \emph
  {et~al.}(2013{\natexlab{b}})\citenamefont {Zhao}, \citenamefont {Ilderton},
  \citenamefont {Maris},\ and\ \citenamefont {Vary}}]{Zhao:2013cma}%
  \BibitemOpen
  \bibfield  {author} {\bibinfo {author} {\bibfnamefont {X.}~\bibnamefont
  {Zhao}}, \bibinfo {author} {\bibfnamefont {A.}~\bibnamefont {Ilderton}},
  \bibinfo {author} {\bibfnamefont {P.}~\bibnamefont {Maris}}, \ and\ \bibinfo
  {author} {\bibfnamefont {J.~P.}\ \bibnamefont {Vary}},\ }\href {\doibase
  10.1103/PhysRevD.88.065014} {\bibfield  {journal} {\bibinfo  {journal} {Phys.
  Rev.}\ }\textbf {\bibinfo {volume} {D88}},\ \bibinfo {pages} {065014}
  (\bibinfo {year} {2013}{\natexlab{b}})},\ \Eprint
  {http://arxiv.org/abs/1303.3273} {arXiv:1303.3273 [nucl-th]} \BibitemShut
  {NoStop}%
\bibitem [{\citenamefont {Chen}\ \emph {et~al.}(2017)\citenamefont {Chen},
  \citenamefont {Li}, \citenamefont {Maris}, \citenamefont {Tuchin},\ and\
  \citenamefont {Vary}}]{Chen:2016dlk}%
  \BibitemOpen
  \bibfield  {author} {\bibinfo {author} {\bibfnamefont {G.}~\bibnamefont
  {Chen}}, \bibinfo {author} {\bibfnamefont {Y.}~\bibnamefont {Li}}, \bibinfo
  {author} {\bibfnamefont {P.}~\bibnamefont {Maris}}, \bibinfo {author}
  {\bibfnamefont {K.}~\bibnamefont {Tuchin}}, \ and\ \bibinfo {author}
  {\bibfnamefont {J.~P.}\ \bibnamefont {Vary}},\ }\href {\doibase
  10.1016/j.physletb.2017.04.024} {\bibfield  {journal} {\bibinfo  {journal}
  {Phys. Lett.}\ }\textbf {\bibinfo {volume} {B769}},\ \bibinfo {pages} {477}
  (\bibinfo {year} {2017})},\ \Eprint {http://arxiv.org/abs/1610.04945}
  {arXiv:1610.04945 [nucl-th]} \BibitemShut {NoStop}%
\bibitem [{\citenamefont {Li}\ \emph {et~al.}(2018{\natexlab{a}})\citenamefont
  {Li}, \citenamefont {Li}, \citenamefont {Maris},\ and\ \citenamefont
  {Vary}}]{Li:2018uif}%
  \BibitemOpen
  \bibfield  {author} {\bibinfo {author} {\bibfnamefont {M.}~\bibnamefont
  {Li}}, \bibinfo {author} {\bibfnamefont {Y.}~\bibnamefont {Li}}, \bibinfo
  {author} {\bibfnamefont {P.}~\bibnamefont {Maris}}, \ and\ \bibinfo {author}
  {\bibfnamefont {J.~P.}\ \bibnamefont {Vary}},\ }\href {\doibase
  10.1103/PhysRevD.98.034024} {\bibfield  {journal} {\bibinfo  {journal} {Phys.
  Rev.}\ }\textbf {\bibinfo {volume} {D98}},\ \bibinfo {pages} {034024}
  (\bibinfo {year} {2018}{\natexlab{a}})},\ \Eprint
  {http://arxiv.org/abs/1803.11519} {arXiv:1803.11519 [hep-ph]} \BibitemShut
  {NoStop}%
\bibitem [{\citenamefont {Li}\ \emph {et~al.}(2018{\natexlab{b}})\citenamefont
  {Li}, \citenamefont {Maris},\ and\ \citenamefont {Vary}}]{Li:2017uug}%
  \BibitemOpen
  \bibfield  {author} {\bibinfo {author} {\bibfnamefont {Y.}~\bibnamefont
  {Li}}, \bibinfo {author} {\bibfnamefont {P.}~\bibnamefont {Maris}}, \ and\
  \bibinfo {author} {\bibfnamefont {J.}~\bibnamefont {Vary}},\ }\href {\doibase
  10.1103/PhysRevD.97.054034} {\bibfield  {journal} {\bibinfo  {journal} {Phys.
  Rev.}\ }\textbf {\bibinfo {volume} {D97}},\ \bibinfo {pages} {054034}
  (\bibinfo {year} {2018}{\natexlab{b}})},\ \Eprint
  {http://arxiv.org/abs/1712.03467} {arXiv:1712.03467 [hep-ph]} \BibitemShut
  {NoStop}%
\bibitem [{\citenamefont {Chakrabarti}\ \emph {et~al.}(2014)\citenamefont
  {Chakrabarti}, \citenamefont {Zhao}, \citenamefont {Honkanen}, \citenamefont
  {Manohar}, \citenamefont {Maris},\ and\ \citenamefont
  {Vary}}]{Chakrabarti:2014cwa}%
  \BibitemOpen
  \bibfield  {author} {\bibinfo {author} {\bibfnamefont {D.}~\bibnamefont
  {Chakrabarti}}, \bibinfo {author} {\bibfnamefont {X.}~\bibnamefont {Zhao}},
  \bibinfo {author} {\bibfnamefont {H.}~\bibnamefont {Honkanen}}, \bibinfo
  {author} {\bibfnamefont {R.}~\bibnamefont {Manohar}}, \bibinfo {author}
  {\bibfnamefont {P.}~\bibnamefont {Maris}}, \ and\ \bibinfo {author}
  {\bibfnamefont {J.~P.}\ \bibnamefont {Vary}},\ }\href {\doibase
  10.1103/PhysRevD.89.116004} {\bibfield  {journal} {\bibinfo  {journal} {Phys.
  Rev.}\ }\textbf {\bibinfo {volume} {D89}},\ \bibinfo {pages} {116004}
  (\bibinfo {year} {2014})},\ \Eprint {http://arxiv.org/abs/1403.0704}
  {arXiv:1403.0704 [hep-ph]} \BibitemShut {NoStop}%
\bibitem [{\citenamefont {Adhikari}\ \emph {et~al.}(2019)\citenamefont
  {Adhikari}, \citenamefont {Li}, \citenamefont {Li},\ and\ \citenamefont
  {Vary}}]{Adhikari:2018umb}%
  \BibitemOpen
  \bibfield  {author} {\bibinfo {author} {\bibfnamefont {L.}~\bibnamefont
  {Adhikari}}, \bibinfo {author} {\bibfnamefont {Y.}~\bibnamefont {Li}},
  \bibinfo {author} {\bibfnamefont {M.}~\bibnamefont {Li}}, \ and\ \bibinfo
  {author} {\bibfnamefont {J.~P.}\ \bibnamefont {Vary}},\ }\href {\doibase
  10.1103/PhysRevC.99.035208} {\bibfield  {journal} {\bibinfo  {journal} {Phys.
  Rev.}\ }\textbf {\bibinfo {volume} {C99}},\ \bibinfo {pages} {035208}
  (\bibinfo {year} {2019})},\ \Eprint {http://arxiv.org/abs/1809.06475}
  {arXiv:1809.06475 [hep-ph]} \BibitemShut {NoStop}%
\bibitem [{\citenamefont {Jia}\ and\ \citenamefont {Vary}(2018)}]{Jia:2018hxd}%
  \BibitemOpen
  \bibfield  {author} {\bibinfo {author} {\bibfnamefont {S.}~\bibnamefont
  {Jia}}\ and\ \bibinfo {author} {\bibfnamefont {J.~P.}\ \bibnamefont {Vary}},\
  }\href@noop {} {\  (\bibinfo {year} {2018})},\ \Eprint
  {http://arxiv.org/abs/1812.09340} {arXiv:1812.09340 [nucl-th]} \BibitemShut
  {NoStop}%
\bibitem [{\citenamefont {Lan}\ \emph {et~al.}(2019)\citenamefont {Lan},
  \citenamefont {Mondal}, \citenamefont {Jia}, \citenamefont {Zhao},\ and\
  \citenamefont {Vary}}]{Lan:2019vui}%
  \BibitemOpen
  \bibfield  {author} {\bibinfo {author} {\bibfnamefont {J.}~\bibnamefont
  {Lan}}, \bibinfo {author} {\bibfnamefont {C.}~\bibnamefont {Mondal}},
  \bibinfo {author} {\bibfnamefont {S.}~\bibnamefont {Jia}}, \bibinfo {author}
  {\bibfnamefont {X.}~\bibnamefont {Zhao}}, \ and\ \bibinfo {author}
  {\bibfnamefont {J.~P.}\ \bibnamefont {Vary}},\ }\href {\doibase
  10.1103/PhysRevLett.122.172001} {\bibfield  {journal} {\bibinfo  {journal}
  {Phys. Rev. Lett.}\ }\textbf {\bibinfo {volume} {122}},\ \bibinfo {pages}
  {172001} (\bibinfo {year} {2019})},\ \Eprint
  {http://arxiv.org/abs/1901.11430} {arXiv:1901.11430 [nucl-th]} \BibitemShut
  {NoStop}%
\bibitem [{\citenamefont {Brodsky}\ \emph {et~al.}(2015)\citenamefont
  {Brodsky}, \citenamefont {de~T\'{e}ramond}, \citenamefont {Dosch},\ and\
  \citenamefont {Erlich}}]{Brodsky:2014yha}%
  \BibitemOpen
  \bibfield  {author} {\bibinfo {author} {\bibfnamefont {S.~J.}\ \bibnamefont
  {Brodsky}}, \bibinfo {author} {\bibfnamefont {G.~F.}\ \bibnamefont
  {de~T\'{e}ramond}}, \bibinfo {author} {\bibfnamefont {H.~G.}\ \bibnamefont
  {Dosch}}, \ and\ \bibinfo {author} {\bibfnamefont {J.}~\bibnamefont
  {Erlich}},\ }\href {\doibase https://doi.org/10.1016/j.physrep.2015.05.001}
  {\bibfield  {journal} {\bibinfo  {journal} {Phys. Rept.}\ }\textbf {\bibinfo
  {volume} {584}},\ \bibinfo {pages} {1} (\bibinfo {year} {2015})}\BibitemShut
  {NoStop}%
\bibitem [{\citenamefont {Klimt}\ \emph {et~al.}(1990)\citenamefont {Klimt},
  \citenamefont {Lutz}, \citenamefont {Vogl},\ and\ \citenamefont
  {Weise}}]{Klimt:1989pm}%
  \BibitemOpen
  \bibfield  {author} {\bibinfo {author} {\bibfnamefont {S.}~\bibnamefont
  {Klimt}}, \bibinfo {author} {\bibfnamefont {M.}~\bibnamefont {Lutz}},
  \bibinfo {author} {\bibfnamefont {U.}~\bibnamefont {Vogl}}, \ and\ \bibinfo
  {author} {\bibfnamefont {W.}~\bibnamefont {Weise}},\ }\href {\doibase
  https://doi.org/10.1016/0375-9474(90)90123-4} {\bibfield  {journal} {\bibinfo
   {journal} {Nucl. Phys. A}\ }\textbf {\bibinfo {volume} {516}},\ \bibinfo
  {pages} {429 } (\bibinfo {year} {1990})}\BibitemShut {NoStop}%
\bibitem [{\citenamefont {Dokshitzer}(1977)}]{Dokshitzer:1977sg}%
  \BibitemOpen
  \bibfield  {author} {\bibinfo {author} {\bibfnamefont {Y.~L.}\ \bibnamefont
  {Dokshitzer}},\ }\href@noop {} {\bibfield  {journal} {\bibinfo  {journal}
  {Sov. Phys. JETP}\ }\textbf {\bibinfo {volume} {46}},\ \bibinfo {pages} {641}
  (\bibinfo {year} {1977})},\ \bibinfo {note} {[Zh. Eksp. Teor.
  Fiz.73,1216(1977)]}\BibitemShut {NoStop}%
\bibitem [{\citenamefont {Gribov}\ and\ \citenamefont
  {Lipatov}(1972)}]{Gribov:1972ri}%
  \BibitemOpen
  \bibfield  {author} {\bibinfo {author} {\bibfnamefont {V.~N.}\ \bibnamefont
  {Gribov}}\ and\ \bibinfo {author} {\bibfnamefont {L.~N.}\ \bibnamefont
  {Lipatov}},\ }\href@noop {} {\bibfield  {journal} {\bibinfo  {journal} {Sov.
  J. Nucl. Phys.}\ }\textbf {\bibinfo {volume} {15}},\ \bibinfo {pages} {438}
  (\bibinfo {year} {1972})},\ \bibinfo {note} {[Yad.
  Fiz.15,781(1972)]}\BibitemShut {NoStop}%
\bibitem [{\citenamefont {Altarelli}\ and\ \citenamefont
  {Parisi}(1977)}]{Altarelli:1977zs}%
  \BibitemOpen
  \bibfield  {author} {\bibinfo {author} {\bibfnamefont {G.}~\bibnamefont
  {Altarelli}}\ and\ \bibinfo {author} {\bibfnamefont {G.}~\bibnamefont
  {Parisi}},\ }\href {\doibase https://doi.org/10.1016/0550-3213(77)90384-4}
  {\bibfield  {journal} {\bibinfo  {journal} {Nucl. Phys. B}\ }\textbf
  {\bibinfo {volume} {126}},\ \bibinfo {pages} {298 } (\bibinfo {year}
  {1977})}\BibitemShut {NoStop}%
\bibitem [{\citenamefont {Kovarik}\ \emph {et~al.}(2016)\citenamefont {Kovarik}
  \emph {et~al.}}]{Kovarik:2015cma}%
  \BibitemOpen
  \bibfield  {author} {\bibinfo {author} {\bibfnamefont {K.}~\bibnamefont
  {Kovarik}} \emph {et~al.},\ }\href {\doibase 10.1103/PhysRevD.93.085037}
  {\bibfield  {journal} {\bibinfo  {journal} {Phys. Rev.}\ }\textbf {\bibinfo
  {volume} {D93}},\ \bibinfo {pages} {085037} (\bibinfo {year} {2016})},\
  \Eprint {http://arxiv.org/abs/1509.00792} {arXiv:1509.00792 [hep-ph]}
  \BibitemShut {NoStop}%
\bibitem [{\citenamefont {Vogl}\ \emph {et~al.}(1990)\citenamefont {Vogl},
  \citenamefont {Lutz}, \citenamefont {Klimt},\ and\ \citenamefont
  {Weise}}]{Vogl:1989ea}%
  \BibitemOpen
  \bibfield  {author} {\bibinfo {author} {\bibfnamefont {U.}~\bibnamefont
  {Vogl}}, \bibinfo {author} {\bibfnamefont {M.~F.~M.}\ \bibnamefont {Lutz}},
  \bibinfo {author} {\bibfnamefont {S.}~\bibnamefont {Klimt}}, \ and\ \bibinfo
  {author} {\bibfnamefont {W.}~\bibnamefont {Weise}},\ }\href {\doibase
  10.1016/0375-9474(90)90124-5} {\bibfield  {journal} {\bibinfo  {journal}
  {Nucl. Phys.}\ }\textbf {\bibinfo {volume} {A516}},\ \bibinfo {pages} {469}
  (\bibinfo {year} {1990})}\BibitemShut {NoStop}%
\bibitem [{\citenamefont {Vogl}\ and\ \citenamefont
  {Weise}(1991)}]{Vogl:1991qt}%
  \BibitemOpen
  \bibfield  {author} {\bibinfo {author} {\bibfnamefont {U.}~\bibnamefont
  {Vogl}}\ and\ \bibinfo {author} {\bibfnamefont {W.}~\bibnamefont {Weise}},\
  }\href {\doibase 10.1016/0146-6410(91)90005-9} {\bibfield  {journal}
  {\bibinfo  {journal} {Prog. Part. Nucl. Phys.}\ }\textbf {\bibinfo {volume}
  {27}},\ \bibinfo {pages} {195} (\bibinfo {year} {1991})}\BibitemShut
  {NoStop}%
\bibitem [{\citenamefont {Klevansky}(1992)}]{Klevansky:1992qe}%
  \BibitemOpen
  \bibfield  {author} {\bibinfo {author} {\bibfnamefont {S.~P.}\ \bibnamefont
  {Klevansky}},\ }\href {\doibase 10.1103/RevModPhys.64.649} {\bibfield
  {journal} {\bibinfo  {journal} {Rev. Mod. Phys.}\ }\textbf {\bibinfo {volume}
  {64}},\ \bibinfo {pages} {649} (\bibinfo {year} {1992})}\BibitemShut
  {NoStop}%
\bibitem [{\citenamefont {Gl{\"u}ck}\ \emph {et~al.}(1998)\citenamefont
  {Gl{\"u}ck}, \citenamefont {Reya},\ and\ \citenamefont
  {Stratmann}}]{Gluck:1997ww}%
  \BibitemOpen
  \bibfield  {author} {\bibinfo {author} {\bibfnamefont {M.}~\bibnamefont
  {Gl{\"u}ck}}, \bibinfo {author} {\bibfnamefont {E.}~\bibnamefont {Reya}}, \
  and\ \bibinfo {author} {\bibfnamefont {M.}~\bibnamefont {Stratmann}},\ }\href
  {\doibase 10.1007/s100520050130} {\bibfield  {journal} {\bibinfo  {journal}
  {Eur. Phys. J. C}\ }\textbf {\bibinfo {volume} {2}},\ \bibinfo {pages} {159}
  (\bibinfo {year} {1998})}\BibitemShut {NoStop}%
\bibitem [{\citenamefont {Salam}\ and\ \citenamefont
  {Rojo}(2009)}]{Salam:2008qg}%
  \BibitemOpen
  \bibfield  {author} {\bibinfo {author} {\bibfnamefont {G.}~\bibnamefont
  {Salam}}\ and\ \bibinfo {author} {\bibfnamefont {J.}~\bibnamefont {Rojo}},\
  }\href {\doibase https://doi.org/10.1016/j.cpc.2008.08.010} {\bibfield
  {journal} {\bibinfo  {journal} {Comput. Phys. Commun.}\ }\textbf {\bibinfo
  {volume} {180}},\ \bibinfo {pages} {120 } (\bibinfo {year}
  {2009})}\BibitemShut {NoStop}%
\bibitem [{\citenamefont {Watanabe}\ \emph
  {et~al.}(2018{\natexlab{b}})\citenamefont {Watanabe}, \citenamefont
  {Sawada},\ and\ \citenamefont {Kao}}]{Watanabe:2018qju}%
  \BibitemOpen
  \bibfield  {author} {\bibinfo {author} {\bibfnamefont {A.}~\bibnamefont
  {Watanabe}}, \bibinfo {author} {\bibfnamefont {T.}~\bibnamefont {Sawada}}, \
  and\ \bibinfo {author} {\bibfnamefont {C.~W.}\ \bibnamefont {Kao}},\
  }\bibfield  {booktitle} {\emph {\bibinfo {booktitle} {{Proceedings, 21st
  High-Energy Physics International Conference in Quantum Chromodynamics (QCD
  18): Montpellier, France, July 2-6, 2018}}},\ }\href {\doibase
  10.1016/j.nuclphysbps.2018.12.021} {\bibfield  {journal} {\bibinfo  {journal}
  {Nucl. Part. Phys. Proc.}\ }\textbf {\bibinfo {volume} {300-302}},\ \bibinfo
  {pages} {121} (\bibinfo {year} {2018}{\natexlab{b}})},\ \Eprint
  {http://arxiv.org/abs/1810.04032} {arXiv:1810.04032 [hep-ph]} \BibitemShut
  {NoStop}%
\bibitem [{\citenamefont {{ZEUS Collaboration}}\ \emph
  {et~al.}(2002)\citenamefont {{ZEUS Collaboration}}, \citenamefont {Chekanov},
  \citenamefont {Krakauer}, \citenamefont {Magill}, \citenamefont {Musgrave},
  \citenamefont {Pellegrino}, \citenamefont {Repond}, \citenamefont {Yoshida},
  \citenamefont {Mattingly}, \citenamefont {Antonioli}, \citenamefont {Bari}
  \emph {et~al.}}]{Chekanov:2002pf}%
  \BibitemOpen
  \bibfield  {author} {\bibinfo {author} {\bibnamefont {{ZEUS Collaboration}}},
  \bibinfo {author} {\bibfnamefont {S.}~\bibnamefont {Chekanov}}, \bibinfo
  {author} {\bibfnamefont {D.}~\bibnamefont {Krakauer}}, \bibinfo {author}
  {\bibfnamefont {S.}~\bibnamefont {Magill}}, \bibinfo {author} {\bibfnamefont
  {B.}~\bibnamefont {Musgrave}}, \bibinfo {author} {\bibfnamefont
  {A.}~\bibnamefont {Pellegrino}}, \bibinfo {author} {\bibfnamefont
  {J.}~\bibnamefont {Repond}}, \bibinfo {author} {\bibfnamefont
  {R.}~\bibnamefont {Yoshida}}, \bibinfo {author} {\bibfnamefont
  {M.}~\bibnamefont {Mattingly}}, \bibinfo {author} {\bibfnamefont
  {P.}~\bibnamefont {Antonioli}}, \bibinfo {author} {\bibfnamefont
  {G.}~\bibnamefont {Bari}},  \emph {et~al.},\ }\href {\doibase
  10.1016/S0550-3213(02)00439-X} {\bibfield  {journal} {\bibinfo  {journal}
  {Nucl. Phys.}\ }\textbf {\bibinfo {volume} {B637}},\ \bibinfo {pages} {3}
  (\bibinfo {year} {2002})},\ \Eprint {http://arxiv.org/abs/hep-ex/0205076}
  {arXiv:hep-ex/0205076 [hep-ex]} \BibitemShut {NoStop}%
\bibitem [{\citenamefont {Aaron}\ \emph {et~al.}(2010)\citenamefont {Aaron}
  \emph {et~al.}}]{Aaron:2010ab}%
  \BibitemOpen
  \bibfield  {author} {\bibinfo {author} {\bibfnamefont {F.~D.}\ \bibnamefont
  {Aaron}} \emph {et~al.} (\bibinfo {collaboration} {H1}),\ }\href {\doibase
  10.1140/epjc/s10052-010-1369-4} {\bibfield  {journal} {\bibinfo  {journal}
  {Eur. Phys. J.}\ }\textbf {\bibinfo {volume} {C68}},\ \bibinfo {pages} {381}
  (\bibinfo {year} {2010})},\ \Eprint {http://arxiv.org/abs/1001.0532}
  {arXiv:1001.0532 [hep-ex]} \BibitemShut {NoStop}%
\bibitem [{\citenamefont {{NA10 Collaboration}}\ \emph
  {et~al.}(1985)\citenamefont {{NA10 Collaboration}}, \citenamefont {Betev},
  \citenamefont {Blaising}, \citenamefont {Bordalo}, \citenamefont
  {Boumediene}, \citenamefont {Cerrito}, \citenamefont {Degr\'{e}},
  \citenamefont {Ereditato}, \citenamefont {Falciano}, \citenamefont
  {Freudenreich}, \citenamefont {Gorini} \emph {et~al.}}]{Betev:1985pf}%
  \BibitemOpen
  \bibfield  {author} {\bibinfo {author} {\bibnamefont {{NA10 Collaboration}}},
  \bibinfo {author} {\bibfnamefont {B.}~\bibnamefont {Betev}}, \bibinfo
  {author} {\bibfnamefont {J.}~\bibnamefont {Blaising}}, \bibinfo {author}
  {\bibfnamefont {P.}~\bibnamefont {Bordalo}}, \bibinfo {author} {\bibfnamefont
  {A.}~\bibnamefont {Boumediene}}, \bibinfo {author} {\bibfnamefont
  {L.}~\bibnamefont {Cerrito}}, \bibinfo {author} {\bibfnamefont
  {A.}~\bibnamefont {Degr\'{e}}}, \bibinfo {author} {\bibfnamefont
  {A.}~\bibnamefont {Ereditato}}, \bibinfo {author} {\bibfnamefont
  {S.}~\bibnamefont {Falciano}}, \bibinfo {author} {\bibfnamefont
  {K.}~\bibnamefont {Freudenreich}}, \bibinfo {author} {\bibfnamefont
  {E.}~\bibnamefont {Gorini}},  \emph {et~al.},\ }\href {\doibase
  10.1007/BF01550243} {\bibfield  {journal} {\bibinfo  {journal} {Z. Phys.}\
  }\textbf {\bibinfo {volume} {C28}},\ \bibinfo {pages} {9} (\bibinfo {year}
  {1985})}\BibitemShut {NoStop}%
\bibitem [{\citenamefont {Barate}\ \emph {et~al.}(1979)\citenamefont {Barate},
  \citenamefont {Bareyre}, \citenamefont {Bonamy}, \citenamefont {Borgeaud},
  \citenamefont {David}, \citenamefont {Ernwein}, \citenamefont {Gentit},
  \citenamefont {Laurens}, \citenamefont {Lemoigne}, \citenamefont {Roussarie}
  \emph {et~al.}}]{Barate:1979da}%
  \BibitemOpen
  \bibfield  {author} {\bibinfo {author} {\bibfnamefont {R.}~\bibnamefont
  {Barate}}, \bibinfo {author} {\bibfnamefont {P.}~\bibnamefont {Bareyre}},
  \bibinfo {author} {\bibfnamefont {P.}~\bibnamefont {Bonamy}}, \bibinfo
  {author} {\bibfnamefont {P.}~\bibnamefont {Borgeaud}}, \bibinfo {author}
  {\bibfnamefont {M.}~\bibnamefont {David}}, \bibinfo {author} {\bibfnamefont
  {J.}~\bibnamefont {Ernwein}}, \bibinfo {author} {\bibfnamefont {F.~X.}\
  \bibnamefont {Gentit}}, \bibinfo {author} {\bibfnamefont {G.}~\bibnamefont
  {Laurens}}, \bibinfo {author} {\bibfnamefont {Y.}~\bibnamefont {Lemoigne}},
  \bibinfo {author} {\bibfnamefont {A.}~\bibnamefont {Roussarie}},  \emph
  {et~al.},\ }\href {\doibase 10.1103/PhysRevLett.43.1541} {\bibfield
  {journal} {\bibinfo  {journal} {Phys. Rev. Lett.}\ }\textbf {\bibinfo
  {volume} {43}},\ \bibinfo {pages} {1541} (\bibinfo {year}
  {1979})}\BibitemShut {NoStop}%
\bibitem [{\citenamefont {Greenlee}\ \emph {et~al.}(1985)\citenamefont
  {Greenlee}, \citenamefont {Frisch}, \citenamefont {Grosso-Pilcher},
  \citenamefont {Johnson}, \citenamefont {Mestayer}, \citenamefont
  {Schachinger}, \citenamefont {Shochet}, \citenamefont {Swartz}, \citenamefont
  {Pirou\'e}, \citenamefont {Pope}, \citenamefont {Stickland},\ and\
  \citenamefont {Sumner}}]{Greenlee:1985gd}%
  \BibitemOpen
  \bibfield  {author} {\bibinfo {author} {\bibfnamefont {H.~B.}\ \bibnamefont
  {Greenlee}}, \bibinfo {author} {\bibfnamefont {H.~J.}\ \bibnamefont
  {Frisch}}, \bibinfo {author} {\bibfnamefont {C.}~\bibnamefont
  {Grosso-Pilcher}}, \bibinfo {author} {\bibfnamefont {K.~F.}\ \bibnamefont
  {Johnson}}, \bibinfo {author} {\bibfnamefont {M.~D.}\ \bibnamefont
  {Mestayer}}, \bibinfo {author} {\bibfnamefont {L.}~\bibnamefont
  {Schachinger}}, \bibinfo {author} {\bibfnamefont {M.~J.}\ \bibnamefont
  {Shochet}}, \bibinfo {author} {\bibfnamefont {M.~L.}\ \bibnamefont {Swartz}},
  \bibinfo {author} {\bibfnamefont {P.~A.}\ \bibnamefont {Pirou\'e}}, \bibinfo
  {author} {\bibfnamefont {B.~G.}\ \bibnamefont {Pope}}, \bibinfo {author}
  {\bibfnamefont {D.~P.}\ \bibnamefont {Stickland}}, \ and\ \bibinfo {author}
  {\bibfnamefont {R.~L.}\ \bibnamefont {Sumner}},\ }\href {\doibase
  10.1103/PhysRevLett.55.1555} {\bibfield  {journal} {\bibinfo  {journal}
  {Phys. Rev. Lett.}\ }\textbf {\bibinfo {volume} {55}},\ \bibinfo {pages}
  {1555} (\bibinfo {year} {1985})}\BibitemShut {NoStop}%
\bibitem [{\citenamefont {Anderson}\ \emph {et~al.}(1979)\citenamefont
  {Anderson}, \citenamefont {Coleman}, \citenamefont {Hogan}, \citenamefont
  {Karhi}, \citenamefont {McDonald}, \citenamefont {Newman}, \citenamefont
  {Pilcher}, \citenamefont {Rosenberg}, \citenamefont {Sanders}, \citenamefont
  {Smith},\ and\ \citenamefont {Thaler}}]{Anderson:1979tt}%
  \BibitemOpen
  \bibfield  {author} {\bibinfo {author} {\bibfnamefont {K.~J.}\ \bibnamefont
  {Anderson}}, \bibinfo {author} {\bibfnamefont {R.~N.}\ \bibnamefont
  {Coleman}}, \bibinfo {author} {\bibfnamefont {G.~E.}\ \bibnamefont {Hogan}},
  \bibinfo {author} {\bibfnamefont {K.~P.}\ \bibnamefont {Karhi}}, \bibinfo
  {author} {\bibfnamefont {K.~T.}\ \bibnamefont {McDonald}}, \bibinfo {author}
  {\bibfnamefont {C.~B.}\ \bibnamefont {Newman}}, \bibinfo {author}
  {\bibfnamefont {J.~E.}\ \bibnamefont {Pilcher}}, \bibinfo {author}
  {\bibfnamefont {E.~I.}\ \bibnamefont {Rosenberg}}, \bibinfo {author}
  {\bibfnamefont {G.~H.}\ \bibnamefont {Sanders}}, \bibinfo {author}
  {\bibfnamefont {A.~J.~S.}\ \bibnamefont {Smith}}, \ and\ \bibinfo {author}
  {\bibfnamefont {J.~J.}\ \bibnamefont {Thaler}},\ }\href {\doibase
  10.1103/PhysRevLett.42.944} {\bibfield  {journal} {\bibinfo  {journal} {Phys.
  Rev. Lett.}\ }\textbf {\bibinfo {volume} {42}},\ \bibinfo {pages} {944}
  (\bibinfo {year} {1979})}\BibitemShut {NoStop}%
\bibitem [{\citenamefont {Corden}\ \emph {et~al.}(1980)\citenamefont {Corden},
  \citenamefont {Dowell}, \citenamefont {Garvey}, \citenamefont {Homer},
  \citenamefont {Jobes}, \citenamefont {Kenyon}, \citenamefont {McMahon},
  \citenamefont {Owen}, \citenamefont {Sumorok}, \citenamefont {Vallance} \emph
  {et~al.}}]{Corden:1980xf}%
  \BibitemOpen
  \bibfield  {author} {\bibinfo {author} {\bibfnamefont {M.}~\bibnamefont
  {Corden}}, \bibinfo {author} {\bibfnamefont {J.}~\bibnamefont {Dowell}},
  \bibinfo {author} {\bibfnamefont {J.}~\bibnamefont {Garvey}}, \bibinfo
  {author} {\bibfnamefont {R.}~\bibnamefont {Homer}}, \bibinfo {author}
  {\bibfnamefont {M.}~\bibnamefont {Jobes}}, \bibinfo {author} {\bibfnamefont
  {I.}~\bibnamefont {Kenyon}}, \bibinfo {author} {\bibfnamefont
  {T.}~\bibnamefont {McMahon}}, \bibinfo {author} {\bibfnamefont
  {R.}~\bibnamefont {Owen}}, \bibinfo {author} {\bibfnamefont {K.}~\bibnamefont
  {Sumorok}}, \bibinfo {author} {\bibfnamefont {R.}~\bibnamefont {Vallance}},
  \emph {et~al.},\ }\href {\doibase
  https://doi.org/10.1016/0370-2693(80)90800-X} {\bibfield  {journal} {\bibinfo
   {journal} {Phys. Lett. B}\ }\textbf {\bibinfo {volume} {96}},\ \bibinfo
  {pages} {417 } (\bibinfo {year} {1980})}\BibitemShut {NoStop}%
\bibitem [{\citenamefont {Gluck}\ \emph {et~al.}(1992)\citenamefont {Gluck},
  \citenamefont {Reya},\ and\ \citenamefont {Vogt}}]{Gluck:1991ng}%
  \BibitemOpen
  \bibfield  {author} {\bibinfo {author} {\bibfnamefont {M.}~\bibnamefont
  {Gluck}}, \bibinfo {author} {\bibfnamefont {E.}~\bibnamefont {Reya}}, \ and\
  \bibinfo {author} {\bibfnamefont {A.}~\bibnamefont {Vogt}},\ }\href {\doibase
  10.1007/BF01483880} {\bibfield  {journal} {\bibinfo  {journal} {Z. Phys.}\
  }\textbf {\bibinfo {volume} {C53}},\ \bibinfo {pages} {127} (\bibinfo {year}
  {1992})}\BibitemShut {NoStop}%
\bibitem [{\citenamefont {Gluck}\ \emph {et~al.}(1995)\citenamefont {Gluck},
  \citenamefont {Reya},\ and\ \citenamefont {Vogt}}]{Gluck:1994uf}%
  \BibitemOpen
  \bibfield  {author} {\bibinfo {author} {\bibfnamefont {M.}~\bibnamefont
  {Gluck}}, \bibinfo {author} {\bibfnamefont {E.}~\bibnamefont {Reya}}, \ and\
  \bibinfo {author} {\bibfnamefont {A.}~\bibnamefont {Vogt}},\ }\href {\doibase
  10.1007/BF01624586} {\bibfield  {journal} {\bibinfo  {journal} {Z. Phys.}\
  }\textbf {\bibinfo {volume} {C67}},\ \bibinfo {pages} {433} (\bibinfo {year}
  {1995})}\BibitemShut {NoStop}%
\bibitem [{\citenamefont {Zhu}\ \emph {et~al.}(2013)\citenamefont {Zhu},
  \citenamefont {Ruan},\ and\ \citenamefont {Hou}}]{Zhu:2010qa}%
  \BibitemOpen
  \bibfield  {author} {\bibinfo {author} {\bibfnamefont {W.}~\bibnamefont
  {Zhu}}, \bibinfo {author} {\bibfnamefont {J.}~\bibnamefont {Ruan}}, \ and\
  \bibinfo {author} {\bibfnamefont {F.}~\bibnamefont {Hou}},\ }\href {\doibase
  10.1142/S0218301313500134} {\bibfield  {journal} {\bibinfo  {journal} {Int.
  J. Mod. Phys.}\ }\textbf {\bibinfo {volume} {E22}},\ \bibinfo {pages}
  {1350013} (\bibinfo {year} {2013})},\ \Eprint
  {http://arxiv.org/abs/1012.4224} {arXiv:1012.4224 [hep-ph]} \BibitemShut
  {NoStop}%
\bibitem [{\citenamefont {Boroun}(2010{\natexlab{a}})}]{Boroun:2010zza}%
  \BibitemOpen
  \bibfield  {author} {\bibinfo {author} {\bibfnamefont {G.~R.}\ \bibnamefont
  {Boroun}},\ }\href {\doibase 10.1140/epja/i2010-10919-9} {\bibfield
  {journal} {\bibinfo  {journal} {Eur. Phys. J.}\ }\textbf {\bibinfo {volume}
  {A43}},\ \bibinfo {pages} {335} (\bibinfo {year} {2010}{\natexlab{a}})},\
  \Eprint {http://arxiv.org/abs/1402.1186} {arXiv:1402.1186 [hep-ph]}
  \BibitemShut {NoStop}%
\bibitem [{\citenamefont {Boroun}\ and\ \citenamefont
  {Zarrin}(2013)}]{Boroun:2013mgv}%
  \BibitemOpen
  \bibfield  {author} {\bibinfo {author} {\bibfnamefont {G.~R.}\ \bibnamefont
  {Boroun}}\ and\ \bibinfo {author} {\bibfnamefont {S.}~\bibnamefont
  {Zarrin}},\ }\href {\doibase 10.1140/epjp/i2013-13119-8} {\bibfield
  {journal} {\bibinfo  {journal} {Eur. Phys. J. Plus}\ }\textbf {\bibinfo
  {volume} {128}},\ \bibinfo {pages} {119} (\bibinfo {year} {2013})},\ \Eprint
  {http://arxiv.org/abs/1402.0678} {arXiv:1402.0678 [hep-ph]} \BibitemShut
  {NoStop}%
\bibitem [{\citenamefont {Boroun}(2010{\natexlab{b}})}]{Boroun:2010zz}%
  \BibitemOpen
  \bibfield  {author} {\bibinfo {author} {\bibfnamefont {G.~R.}\ \bibnamefont
  {Boroun}},\ }\href {\doibase 10.1134/S1063776110100067} {\bibfield  {journal}
  {\bibinfo  {journal} {J. Exp. Theor. Phys.}\ }\textbf {\bibinfo {volume}
  {111}},\ \bibinfo {pages} {567} (\bibinfo {year} {2010}{\natexlab{b}})},\
  \Eprint {http://arxiv.org/abs/1402.0178} {arXiv:1402.0178 [hep-ph]}
  \BibitemShut {NoStop}%
\bibitem [{\citenamefont {Devee}\ and\ \citenamefont
  {Sarma}(2014)}]{Devee:2014fna}%
  \BibitemOpen
  \bibfield  {author} {\bibinfo {author} {\bibfnamefont {M.}~\bibnamefont
  {Devee}}\ and\ \bibinfo {author} {\bibfnamefont {J.~K.}\ \bibnamefont
  {Sarma}},\ }\href {\doibase 10.1140/epjc/s10052-014-2751-4} {\bibfield
  {journal} {\bibinfo  {journal} {Eur. Phys. J.}\ }\textbf {\bibinfo {volume}
  {C74}},\ \bibinfo {pages} {2751} (\bibinfo {year} {2014})},\ \Eprint
  {http://arxiv.org/abs/1401.0653} {arXiv:1401.0653 [hep-ph]} \BibitemShut
  {NoStop}%
\bibitem [{\citenamefont {Phukan}\ \emph {et~al.}(2017)\citenamefont {Phukan},
  \citenamefont {Lalung},\ and\ \citenamefont {Sarma}}]{Phukan:2017lzp}%
  \BibitemOpen
  \bibfield  {author} {\bibinfo {author} {\bibfnamefont {P.}~\bibnamefont
  {Phukan}}, \bibinfo {author} {\bibfnamefont {M.}~\bibnamefont {Lalung}}, \
  and\ \bibinfo {author} {\bibfnamefont {J.~K.}\ \bibnamefont {Sarma}},\ }\href
  {\doibase 10.1016/j.nuclphysa.2017.09.003} {\bibfield  {journal} {\bibinfo
  {journal} {Nucl. Phys.}\ }\textbf {\bibinfo {volume} {A968}},\ \bibinfo
  {pages} {275} (\bibinfo {year} {2017})},\ \Eprint
  {http://arxiv.org/abs/1705.06092} {arXiv:1705.06092 [hep-ph]} \BibitemShut
  {NoStop}%
\bibitem [{\citenamefont {Rezaei}\ and\ \citenamefont
  {Boroun}(2010)}]{Rezaei:2010zz}%
  \BibitemOpen
  \bibfield  {author} {\bibinfo {author} {\bibfnamefont {B.}~\bibnamefont
  {Rezaei}}\ and\ \bibinfo {author} {\bibfnamefont {G.~R.}\ \bibnamefont
  {Boroun}},\ }\href {\doibase 10.1016/j.physletb.2010.07.047} {\bibfield
  {journal} {\bibinfo  {journal} {Phys. Lett.}\ }\textbf {\bibinfo {volume}
  {B692}},\ \bibinfo {pages} {247} (\bibinfo {year} {2010})},\ \Eprint
  {http://arxiv.org/abs/1402.0752} {arXiv:1402.0752 [hep-ph]} \BibitemShut
  {NoStop}%
\bibitem [{\citenamefont {Boroun}(2009)}]{Boroun:2009zzb}%
  \BibitemOpen
  \bibfield  {author} {\bibinfo {author} {\bibfnamefont {G.~R.}\ \bibnamefont
  {Boroun}},\ }\href {\doibase 10.1140/epja/i2009-10871-9} {\bibfield
  {journal} {\bibinfo  {journal} {Eur. Phys. J.}\ }\textbf {\bibinfo {volume}
  {A42}},\ \bibinfo {pages} {251} (\bibinfo {year} {2009})},\ \Eprint
  {http://arxiv.org/abs/1402.0864} {arXiv:1402.0864 [hep-ph]} \BibitemShut
  {NoStop}%
\bibitem [{\citenamefont {Lalung}\ \emph {et~al.}(2017)\citenamefont {Lalung},
  \citenamefont {Phukan},\ and\ \citenamefont {Sarma}}]{Lalung:2017omk}%
  \BibitemOpen
  \bibfield  {author} {\bibinfo {author} {\bibfnamefont {M.}~\bibnamefont
  {Lalung}}, \bibinfo {author} {\bibfnamefont {P.}~\bibnamefont {Phukan}}, \
  and\ \bibinfo {author} {\bibfnamefont {J.~K.}\ \bibnamefont {Sarma}},\ }\href
  {\doibase 10.1007/s10773-017-3527-z} {\bibfield  {journal} {\bibinfo
  {journal} {Int. J. Theor. Phys.}\ }\textbf {\bibinfo {volume} {56}},\
  \bibinfo {pages} {3625} (\bibinfo {year} {2017})},\ \Eprint
  {http://arxiv.org/abs/1702.05291} {arXiv:1702.05291 [hep-ph]} \BibitemShut
  {NoStop}%
\bibitem [{\citenamefont {Becher}\ \emph {et~al.}(2008)\citenamefont {Becher},
  \citenamefont {Neubert},\ and\ \citenamefont {Xu}}]{Becher:2007ty}%
  \BibitemOpen
  \bibfield  {author} {\bibinfo {author} {\bibfnamefont {T.}~\bibnamefont
  {Becher}}, \bibinfo {author} {\bibfnamefont {M.}~\bibnamefont {Neubert}}, \
  and\ \bibinfo {author} {\bibfnamefont {G.}~\bibnamefont {Xu}},\ }\href
  {\doibase 10.1088/1126-6708/2008/07/030} {\bibfield  {journal} {\bibinfo
  {journal} {JHEP}\ }\textbf {\bibinfo {volume} {07}},\ \bibinfo {pages} {030}
  (\bibinfo {year} {2008})},\ \Eprint {http://arxiv.org/abs/0710.0680}
  {arXiv:0710.0680 [hep-ph]} \BibitemShut {NoStop}%
\bibitem [{\citenamefont {Anastasiou}\ \emph {et~al.}(2003)\citenamefont
  {Anastasiou}, \citenamefont {Dixon}, \citenamefont {Melnikov},\ and\
  \citenamefont {Petriello}}]{Anastasiou:2003yy}%
  \BibitemOpen
  \bibfield  {author} {\bibinfo {author} {\bibfnamefont {C.}~\bibnamefont
  {Anastasiou}}, \bibinfo {author} {\bibfnamefont {L.~J.}\ \bibnamefont
  {Dixon}}, \bibinfo {author} {\bibfnamefont {K.}~\bibnamefont {Melnikov}}, \
  and\ \bibinfo {author} {\bibfnamefont {F.}~\bibnamefont {Petriello}},\ }\href
  {\doibase 10.1103/PhysRevLett.91.182002} {\bibfield  {journal} {\bibinfo
  {journal} {Phys. Rev. Lett.}\ }\textbf {\bibinfo {volume} {91}},\ \bibinfo
  {pages} {182002} (\bibinfo {year} {2003})},\ \Eprint
  {http://arxiv.org/abs/hep-ph/0306192} {arXiv:hep-ph/0306192 [hep-ph]}
  \BibitemShut {NoStop}%
\bibitem [{\citenamefont {Anastasiou}\ \emph {et~al.}(2004)\citenamefont
  {Anastasiou}, \citenamefont {Dixon}, \citenamefont {Melnikov},\ and\
  \citenamefont {Petriello}}]{Anastasiou:2003ds}%
  \BibitemOpen
  \bibfield  {author} {\bibinfo {author} {\bibfnamefont {C.}~\bibnamefont
  {Anastasiou}}, \bibinfo {author} {\bibfnamefont {L.~J.}\ \bibnamefont
  {Dixon}}, \bibinfo {author} {\bibfnamefont {K.}~\bibnamefont {Melnikov}}, \
  and\ \bibinfo {author} {\bibfnamefont {F.}~\bibnamefont {Petriello}},\ }\href
  {\doibase 10.1103/PhysRevD.69.094008} {\bibfield  {journal} {\bibinfo
  {journal} {Phys. Rev.}\ }\textbf {\bibinfo {volume} {D69}},\ \bibinfo {pages}
  {094008} (\bibinfo {year} {2004})},\ \Eprint
  {http://arxiv.org/abs/hep-ph/0312266} {arXiv:hep-ph/0312266 [hep-ph]}
  \BibitemShut {NoStop}%
\bibitem [{\citenamefont {Stirling}\ and\ \citenamefont
  {Whalley}(1993)}]{Stirling:1993gc}%
  \BibitemOpen
  \bibfield  {author} {\bibinfo {author} {\bibfnamefont {W.~J.}\ \bibnamefont
  {Stirling}}\ and\ \bibinfo {author} {\bibfnamefont {M.~R.}\ \bibnamefont
  {Whalley}},\ }\href {\doibase 10.1088/0954-3899/19/d/001} {\bibfield
  {journal} {\bibinfo  {journal} {J. Phys. G}\ }\textbf {\bibinfo {volume}
  {19}},\ \bibinfo {pages} {D1} (\bibinfo {year} {1993})}\BibitemShut {NoStop}%
\end{thebibliography}%
\end{document}